\documentclass[aps,pra,nobalancelastpage,superscriptaddress,twocolumn,reprint]{revtex4-2}
\usepackage{graphicx}
\usepackage{standalone, gensymb, tensor, amsmath,amsfonts,amssymb,amsthm,bbm,bm, braket}
\usepackage[T1]{fontenc}
\usepackage{hyperref}
\usepackage[inline]{enumitem}
\usepackage{scalerel, physics}
\usepackage{wasysym}
\usepackage{comment}
\usepackage{enumerate}
\usepackage{graphicx}
\usepackage{mathtools}
\usepackage{soul}
\usepackage{xparse}
\usepackage{dsfont}
\usepackage{ifthen}
\usepackage{tensor}
\usepackage{tikz}
\usepackage{pgfplots}
\usepackage{tikz-network}
\usepackage{simplewick} 
\usepackage{mathrsfs}

\usetikzlibrary{patterns,decorations.pathreplacing}
\theoremstyle{break}        

\theoremstyle{break}

\usepackage{color}
\definecolor{alessandrored}{RGB}{250, 153, 140}
\definecolor{OliveGreen}{RGB}{85,107,47}
\definecolor{NavyBlue}{RGB}{0,0,128}
\definecolor{alessandrogreen}{RGB}{17, 127, 35}
\definecolor{jonasgreen}{RGB}{81, 160, 37}
\usepackage{tensor}
\usepackage{xifthen}
\usepackage{xargs}

\definecolor{blue1}{RGB}{0, 0, 255}
\definecolor{blue2}{RGB}{0, 150, 255}
\definecolor{blue3}{RGB}{0, 71, 171}
\definecolor{blue4}{RGB}{100, 149, 237}
\definecolor{blue5}{RGB}{93, 63, 211}

\definecolor{red1}{RGB}{238, 75, 43}
\definecolor{red2}{RGB}{233, 116, 81}
\definecolor{red3}{RGB}{222, 49, 99}
\definecolor{red4}{RGB}{250, 160, 160}
\definecolor{red5}{RGB}{236, 88, 0}

\definecolor{green1}{RGB}{80, 180, 152}

\newcommandx{\fineq}[5][1=-.8ex,2=1,3=1,5=0]{
	\begin{tikzpicture}[baseline={([yshift=#1]current  bounding  box.center)}, scale = #2, every node/.style={scale = #3},rotate around={#5:(0,0)},every node/.style={transform shape}]
		#4
	\end{tikzpicture}
}

\definecolor{bertinired}{RGB}{232,102,102}
\definecolor{bertiniblue}{RGB}{101,147,245}
\definecolor{bertinigreyblue}{RGB}{101,147,245}
\definecolor{bertinigreyred}{RGB}{232,102,102}
\definecolor{bertinivioletc}{RGB}{45,130,60}
\definecolor{bertinigreen}{RGB}{166,218,149}
\definecolor{bertiniorange}{RGB}{255, 116, 23}
\definecolor{OliveGreen}{RGB}{85,107,47}
\definecolor{NavyBlue}{RGB}{0,0,128}
\definecolor{bertiniviolet}{RGB}{210,145,178}
\definecolor{bertinigrey1}{RGB}{98,98,98}
\definecolor{bertinigrey2}{RGB}{211,211,211}
\definecolor{bertinigrey3}{RGB}{192,192,192}
\definecolor{bertinigrey4}{RGB}{169,169,169}

\newcommandx{\tikzdiagup}{
	\tikz {\draw[thick] (0,0)--(0.15,0.15); \draw (0,0) rectangle (0.15,0.15);}
}

\newcommandx{\gatecross}[1][1=0.5]{
	\pgfmathparse{#1/2.0}
	\let\x\pgfmathresult
	\draw[thick] (-\x,-\x) -- (\x,\x);
	\draw[thick] (\x,-\x) -- (-\x,\x);
}
\newcommandx{\gatesqu}[2][1=0.25,2=]{
	\pgfmathparse{#1/2.0}
	\let\x\pgfmathresult
	\ifthenelse{\equal{#2}{}}{
		\draw[thick, fill=white, rounded corners=2pt] (-\x,\x) rectangle (\x,-\x);
	}{
		\draw[thick, fill=#2, rounded corners=2pt] (-\x,\x) rectangle (\x,-\x);
	}
}

\newcommandx{\gatemark}[2][1=0.075,2=tr]{
	\pgfmathparse{#1}
	\let\l\pgfmathresult
	\ifthenelse{\equal{#2}{topleft}}{
		\draw[thick] (0,\l) -- ++(-\l,0) --++ (0,-\l);
	}{}
	\ifthenelse{\equal{#2}{topright}}{
		\draw[thick] (0,\l) -- ++(\l,0) --++ (0,-\l);
	}{}
	
	\ifthenelse{\equal{#2}{bottomleft}}{
		\draw[thick] (0,-\l) -- ++(-\l,0) --++ (0,\l);
	}{}
	\ifthenelse{\equal{#2}{bottomright}}{
		\draw[thick] (0,-\l) -- ++(\l,0) --++ (0,\l);
	}{}
	
}

\newcommandx{\squaregate}[3][1=0,2=0,3=white]
{
	\begin{scope}[shift={(#1,#2)},rounded corners= 2pt]
		\draw[thick,fill=#3] (-.13,-.13) rectangle (.15,.15);
	\end{scope}
}

\newcommandx{\roundgate}[6][1=0,2=0,3=1,4=topright,5=white,6=-1]{
	\pgfmathparse{#3}
	\let\l\pgfmathresult
	\begin{scope}[shift={(#1,#2)}]
		\gatecross[\l]
			\pgfmathparse{\l/2.0}
		\let\s\pgfmathresult
		\gatesqu[\s][#5]
		\pgfmathparse{\l*0.15}
		\let\m\pgfmathresult
	\ifthenelse{\equal{#6}{-1}}{		\gatemark[\m][#4]
	}{	\node at ({0},{0}) {\scalebox{1.3}{$#6$}};}
\end{scope}
}

\newcommandx{\wcirc}[2]{\begin{scope}
		\draw[fill=white] (#1,#2) circle (0.15);	\end{scope}} 
\newcommandx{\wcircc}[2]{\begin{scope}
		\draw[fill=white] (#1,#2) circle (0.13);	\end{scope}} 

\newcommandx{\wsqr}[2]{\begin{scope}
		\draw[fill=white,shift={(#1,#2)}] (-.13,.13) rectangle (.13,-.13);	\end{scope}} 
\newcommandx{\wsqrr}[2]{\begin{scope}
		\draw[fill=white,shift={(#1,#2)}] (-.11,.11) rectangle (.11,-.11);	\end{scope}} 
\newcommandx{\bcirc}[2]{\begin{scope}
		\draw[fill=black] (#1,#2) circle (0.15);	\end{scope}} 
\newcommandx{\thetastate}[4][1=0,2=0,3=1,4=]{
	\pgfmathparse{#3/2}
	\let\l\pgfmathresult
	\pgfmathparse{\l*0.15}
	\let\m\pgfmathresult
	\begin{scope}[shift={(#1,#2)}]
		\draw[thick] (0,0)--(\l,\l);
		\draw[thick] (0,0)--(-\l,\l);
		\ifthenelse{\equal{#4}{}}{
			\draw[fill=white] (0,0) circle (0.15);
		}{
			\draw[thick, fill=#4] (0,0) circle (0.15);
		}
	\end{scope}
}

\newcommandx{\thetastateflipped}[4][1=0,2=0,3=1,4=]{
	\pgfmathparse{#3/2}
	\let\l\pgfmathresult
	\pgfmathparse{\l*0.15}
	\let\m\pgfmathresult
	\begin{scope}[shift={(#1,#2)}]
		\draw[thick] (0,0)--(\l,-\l);
		\draw[thick] (0,0)--(-\l,-\l);
		\ifthenelse{\equal{#4}{}}{
			\draw[fill=white] (0,0) circle (0.15);
		}{
			\draw[thick, fill=#4] (0,0) circle (0.15);
		}
	\end{scope}
}

\newcommandx{\vertgate}[5][1=0,2=0,3=4,4=bertiniorange,5=topright]
{
	\begin{scope}[shift={(#1,#2)}]
		\ifthenelse{\equal{#3}{1}}{
			\roundgate[0][0][1][#5][#4]
		}{
			\foreach \n[evaluate=\n as \y using {2*\n-2}] in {1,...,#3}{
				\roundgate[0][\y][1][#5][#4]
			}
		}
	\end{scope}
}

\newcommandx{\tsfmatV}[8][1=0,2=0,3=l,4=4,5=tr,6=init,7=bertiniorange,8=topright]{
	\begin{scope}[shift={(#1,#2)}]
		\ifthenelse{\equal{#3}{l}}{
			\pgfmathsetmacro{\flag}{0}
		}{
			\pgfmathsetmacro{\flag}{1}
		}
		
		\foreach \y[evaluate=\y as \x using {mod(\y+\flag,2)}] in {1,...,#4}{
			\roundgate[\x][\y][1][#8][#7]
		}
		\ifthenelse{\equal{#5}{tr}}{
			\foreach \y[evaluate=\y as \x using {mod(\y+\flag,2)}] in {#4}{
				\draw [fill=white] (\x-0.5,\y+0.5) circle (0.15);
				\draw [fill=white] (\x+0.5,\y+0.5) circle (0.15);
			}
		}{}
		\ifthenelse{\equal{#6}{init}}{
			\thetastate[\flag][0][1][#7]
		}{}
	\end{scope}
}

\newcommandx{\leftriangle}[5][1=0,2=0,3=4,4=bertiniorange,5=topright]{
	\begin{scope}[shift={(#1,#2)}]
		\pgfmathsetmacro{\t}{#3}
		\pgfmathsetmacro{\steps}{ceil(\t/2)}
		\foreach \i[evaluate=\i as \x using -\t+2*\i-1, evaluate=\i as \ylim using \t-2*\i+2] in {1,...,\steps}{
			\foreach \y[evaluate=\y as \thisx using {\x+\y-1}] in {1,...,\ylim}{
				\roundgate[\thisx][\y][1][#5][#4]
			}
		}
	\end{scope}
}

\newcommandx{\rightriangle}[5][1=0,2=0,3=4,4=bertiniorange,5=topright]{
	\begin{scope}[shift={(#1,#2)}]
		\pgfmathsetmacro{\t}{#3}
		\pgfmathsetmacro{\steps}{ceil(\t/2)}
		\foreach \i[evaluate=\i as \x using -\t+2*\i-1, evaluate=\i as \ylim using \t-2*\i+2] in {1,...,\steps}{
			\foreach \y[evaluate=\y as \thisx using {-\x-\y+1}] in {1,...,\ylim}{
				\roundgate[\thisx][\y][1][#5][#4]
			}
		}
	\end{scope}
}

\newcommandx{\eigenVL}[8][1=0,2=0,3=l,4=5,5=tr,6=init,7=bertiniorange,8=topright]{
	\begin{scope}[shift={(#1,#2)}]
		\pgfmathsetmacro{\t}{#4}
		\leftriangle[0][0][\t][#7][#8]
		
		\ifthenelse{\equal{#6}{init}}{
			\drawinitstate[0][0][l][\t][#7]
		}{}
		
		\ifthenelse{\equal{#5}{tr}}{
			\draw[fill=white] \foreach \x in {0,...,\t} {(\x-0.5-\t,0.5+\x) circle (0.15)};
			\ifthenelse{\equal{#3}{r}}{
				\draw[fill=white] (0.5,\t+0.5) circle (0.15);
			}{}
		}{}
		\ifthenelse{\equal{#5}{parttr}}{
			\draw[fill=white] \foreach \x in {0,...,\t} {(\x-0.5-\t,0.5+\x) circle (0.15)};
		}{}
	\end{scope}
}

\newcommandx{\eigenVR}[8][1=0,2=0,3=l,4=5,5=tr,6=init,7=bertiniorange,8=topright]{
	\begin{scope}[shift={(#1,#2)}]
		\pgfmathsetmacro{\t}{#4}
		\rightriangle[0][0][\t][#7][#8]
		
		\ifthenelse{\equal{#6}{init}}{
			\drawinitstate[0][0][r][\t][#7]
		}{}
		
		\ifthenelse{\equal{#5}{tr}}{
			\draw[fill=white] \foreach \x in {0,...,\t}{(-\x+0.5+\t,0.5+\x) circle (0.15)};
			\ifthenelse{\equal{#3}{l}}{
				\draw[fill=white] (-0.5,\t+0.5) circle (0.15);
			}{}
		}{}
	\end{scope}
}

\newcommandx{\tra}[2][1]{\underset{#1}{\text{tr}}\left[#2\right]}

\newcommandx{\tsfmatDgate}[7][1=0,2=0,3=l,4=4,5=tr,6=bertiniorange,7=topright]
{
	\begin{scope}[shift={(#1,#2)}]
		\ifthenelse{\equal{#3}{l}}{
			\pgfmathsetmacro{\flag}{-1}
		}{
			\pgfmathsetmacro{\flag}{1}
		}
		\pgfmathsetmacro{\t}{#4}
		\foreach \i[evaluate=\i as \x using {\flag*\i}, evaluate=\i as \y using \i] in {1,...,\t}{
			\roundgate[\x][\y][1][#7][#6]
		}
		
		\ifthenelse{\equal{#5}{tr}}{
			\foreach \i[evaluate=\i as \x using {\flag*\i}, evaluate=\i as \y using \i] in {\t}{
				\draw [fill=white] (\x-0.5,\y+0.5) circle (0.15);
				\draw [fill=white] (\x+0.5,\y+0.5) circle (0.15);
			}  
		}{}
	\end{scope}
	
}

\newcommandx{\tsfmatD}[8][1=0,2=0,3=l,4=4,5=tr,6=init,7=bertiniorange,8=topright]{
	\begin{scope}[shift={(#1,#2)}]
		\ifthenelse{\equal{#6}{init}}{
			\thetastate[0][0][1][#7]
		}{}
		
		\ifthenelse{\equal{#3}{l}}{
			\pgfmathsetmacro{\flag}{-1}
		}{
			\pgfmathsetmacro{\flag}{1}
		}
		
		\pgfmathsetmacro{\t}{#4}
		\foreach \i[evaluate=\i as \x using {\flag*\i}, evaluate=\i as \y using \i] in {1,...,\t}{
			\roundgate[\x][\y][1][#8][#7]
		}
		
		\ifthenelse{\equal{#5}{tr}}{
			\foreach \i[evaluate=\i as \x using {\flag*\i}, evaluate=\i as \y using \i] in {\t}{
				\draw [fill=white] (\x-0.5,\y+0.5) circle (0.15);
				\draw [fill=white] (\x+0.5,\y+0.5) circle (0.15);
			}  
		}
		\ifthenelse{\equal{#5}{parttr}}{
			\foreach \i[evaluate=\i as \x using {\flag*\i}, evaluate=\i as \y using \i] in {\t}{
				\draw [fill=white] (\x+0.5*\flag,\y+0.5) circle (0.15);
			}  
		}
		{}
	\end{scope}
}

\newcommandx{\drawinitstate}[5][1=0,2=0,3=l,4=4,5=bertiniorange]{
	\pgfmathsetmacro{\t}{#4}
	\begin{scope}[shift={(#1,#2)}]
		\pgfmathsetmacro{\steps}{ceil((\t-1)/2)}
		\ifthenelse{\equal{#3}{l}}{
			\foreach \i[evaluate=\i as \x using -\t+2*\i] in {0,...,\steps}{
				\thetastate[\x][0][1][#5]
			}
		}{
			\foreach \i[evaluate=\i as \x using -\t+2*\i] in {0,...,\steps}{      
				\thetastate[-\x][0][1][#5]
			}
		}
	\end{scope}
}

\newcommandx{\drawinitstateflipped}[5][1=0,2=0,3=l,4=4,5=bertiniorange]{
	\pgfmathsetmacro{\t}{#4}
	\begin{scope}[shift={(#1,#2)}]
		\pgfmathsetmacro{\steps}{ceil((\t-1)/2)}
		\ifthenelse{\equal{#3}{l}}{
			\foreach \i[evaluate=\i as \x using -\t+2*\i] in {0,...,\steps}{
				\thetastateflipped[\x][0][1][#5]
			}
		}{
			\foreach \i[evaluate=\i as \x using -\t+2*\i] in {0,...,\steps}{      
				\thetastateflipped[-\x][0][1][#5]
			}
		}
	\end{scope}
}

\newcommandx{\eigenDL}[6][1=0,2=0,3=l,4=4,5=bertiniorange,6=topright]{
	\begin{scope}[shift={(#1,#2)}]
		\pgfmathsetmacro{\t}{#4}
		\ifthenelse{\equal{#3}{l}}{
			\eigenVL[0][0][l][\t][tr][init][#5][#6]
			\pgfmathsetmacro{\t}{#4-1}
			\rightriangle[1][0][\t][#5][#6]
			\drawinitstate[1][0][r][\t][#5]
		}{
			\begin{scope}[shift={(-0.5,0.5)}]
				\foreach \i[evaluate=\i as \x using \i, evaluate=\i as \y using \i] in {0,...,\t}{      
					\draw (\x,\y)--++(0.5,0);
					\draw[fill=white] (\x,\y) circle (0.15);
				}
			\end{scope}
		}
	\end{scope}
}

\newcommandx{\eigenDR}[6][1=0,2=0,3=l,4=4,5=bertiniorange,6=topright]{
	\begin{scope}[shift={(#1,#2)}]
		\pgfmathsetmacro{\t}{#4}
		\ifthenelse{\equal{#3}{r}}{
			\eigenVR[0][0][r][\t][tr][init][#5][#6]
			\pgfmathsetmacro{\t}{#4-1}
			\leftriangle[-1][0][\t][#5][#6]
			\drawinitstate[-1][0][l][\t][#5]
		}{
			\begin{scope}[shift={(0.5,0.5)}]
				\foreach \i[evaluate=\i as \x using \i, evaluate=\i as \y using \t-\i] in {0,...,\t}{      
					\draw (\x,\y)--++(0.5,0);
					\draw[fill=white] (\x+0.5,\y) circle (0.15);
				}
			\end{scope}
		}
	\end{scope}
}

\newcommandx{\idonpurity}[2][1=0,2=0]
{
	\begin{scope}[shift={(#1,#2)}]
		\draw[thick] (-0.5,0)--++(-0.1,0.1)--++(0,0.2)--++(0.1,-0.1);
		\draw[thick] (-0.5,0.4)--++(-0.1,0.1)--++(0,0.2)--++(0.1,-0.1);
		\draw[thick] (0.5,0)--++(0.1,0.1)--++(0,0.2)--++(-0.1,-0.1);
		\draw[thick] (0.5,0.4)--++(0.1,0.1)--++(0,0.2)--++(-0.1,-0.1);
	\end{scope}
}

\newcommandx{\swaponpurity}[2][1=0,2=0]
{
	\begin{scope}[shift={(#1,#2)}]
		\draw[thick] (-0.5,0)--++(-0.2,0.2)--++(0,0.6)--++(0.2,-0.2);
		\draw[thick] (-0.5,0.2)--++(-0.075,0.075)--++(0,0.2)--++(0.075,-0.075);
		\draw[thick] (+0.5,0)--++(+0.2,0.2)--++(0,0.6)--++(-0.2,-0.2);
		\draw[thick] (+0.5,0.2)--++(+0.075,0.075)--++(0,0.2)--++(-0.075,-0.075);
	\end{scope}
}

\newcommandx{\hook}[4][1=0,2=0,3=t,4=l]{
	\begin{scope}[shift={(#1,#2)}]
		\ifthenelse{\equal{#3}{t}}{
			\ifthenelse{\equal{#4}{l}}{\draw[thick] (0.5,-0.5) arc (45:-90:0.15);}{\draw[thick] (0.5,-0.5) arc (45:270:0.15);}
		}{\ifthenelse{\equal{#4}{l}}{\draw[ thick] (0.5,-0.5) arc (-45:90:0.15);}{\draw[ thick] (0.5,-0.5) arc (315:90:0.15);}
		}
	\end{scope}
}

\newcommandx{\hhook}[4][1=0,2=0,3=t,4=l]{
	\begin{scope}[shift={(#1,#2)}]
		\ifthenelse{\equal{#3}{t}}{
			\ifthenelse{\equal{#4}{l}}{\draw[thick] (0.5,-0.5) arc (-45:175:0.15);}{\draw[thick] (0.5,-0.5) arc (225:0:0.15);}
		}{\ifthenelse{\equal{#4}{l}}{\draw[ thick] (0.5,-0.5) arc (-45:180:-0.15);}{\draw[ thick] (0.5,-0.5) arc (45:-180:0.15);}
		}
	\end{scope}
}

\newcommandx{\Pproj}[3][3=$P_\Lambda$]{
\begin{scope}[shift={(#1-.5,#2-1)}]
\draw[thick,fill=white] (0,0)rectangle (1,2);
\draw[thick] (0,1.5)--(-.5,1.5);
\draw[thick] (1,1.5)--(1.5,1.5);
\draw[thick] (0,.5)--(-.5,.5);
\draw[thick] (1,.5)--(1.5,.5);
\node[scale=2] at (.5,1) {#3};
\end{scope}}

\definecolor{FcolU}{rgb}{0.71,0.78,0.91}
\definecolor{colLines}{rgb}{0.31,0.31,0.31}
\definecolor{colVMPSLines}{rgb}{0.11,0.11,0.11}
\definecolor{IcolUc}{rgb}{0.71,0.41,0.42}
\definecolor{IcolU}{rgb}{0.71,0.8,0.76}
\definecolor{IcolVMPSc}{rgb}{0.73,0.69,0.7}
\definecolor{IcolVMPS}{rgb}{0.81,0.77,0.78}
\definecolor{colObs}{rgb}{1.,1.,1.}

\renewcommand{\comment}[1]{\ifdef{0}{}{#1}}

\newcommandx{\eightlegs}[2][1=0,2=0]{
	\begin{scope}[shift={(#1,#2)}]
		\foreach \x in {1,...,8}{
			\draw (\x, 0)--++(0,0.25);
			\draw[fill] (\x,0) circle (0.05);
		}
		\foreach \x in {1,3}{
			\pgfmathsetmacro\result{2*\x-1} 
			\node () at (\result,-0.5) {$i_{\x}$};
			\pgfmathsetmacro\result{2*\x}
			\node () at (\result,-0.5) {$j_{\x}$};	
		}
		\foreach \x in {2,4}{
			\pgfmathsetmacro\result{2*\x} 
			\node () at (\result,-0.5) {$i_{\x}$};
			\pgfmathsetmacro\result{2*\x-1}
			\node () at (\result,-0.5) {$j_{\x}$};	
		}
	\end{scope}
}

\newcommandx{\MPSinitialstate}[5][1=0,2=0,3=bertiniorange,4=topright,5=-1]{
\begin{scope}[shift={(#1,#2)},rounded corners=1.5pt]
	\draw[black,thick,fill=#3] 
	(-0.25,.25)--++(.5,0)--++(0,-.3)--++(-.5,0)--cycle;
	\draw[thick] (-.25,.25)--++(-.25,.25);
	\draw[thick] (.25,.25)--++(.25,.25);
	\draw[very thick] (-1,.-.05)--++(2,0);
\ifthenelse{\equal{#5}{-1}}{
	\ifthenelse{\equal{#4}{topright}}{\draw[thick,rounded corners=0.3]
	(-.1,.15)--++(.2,0)--++(0,-0.1); }{}
	\ifthenelse{\equal{#4}{topleft}}{\draw[thick,rounded corners=0.3]
	(.1,.15)--++(-.2,0)--++(0,-0.1); }{}
	\ifthenelse{\equal{#4}{bottomleft}}{\draw[thick,rounded corners=0.3]
	(.1,-.15)--++(-.2,0)--++(0,0.1); }{}	\ifthenelse{\equal{#4}{bottomright}}{\draw[thick,rounded corners=0.3]
	(-.1,-.15)--++(.2,0)--++(0,0.1); }{}}{			\node at ({0},{0.085}) {\scalebox{1.}{{$#5$}}};}
\end{scope}
}

\newcommandx{\Cmatrix}[6][1=0,2=0,3=2,4=bertiniorange,5=,6=topright]{
	\pgfmathsetmacro\result{#3-1} 
	\begin{scope}[shift={(#1,#2)}]
		\foreach \i in {0,...,\result}
		{\foreach \j in {0,...,\i}
			{\roundgate[\i+\j][\i-\j][1][#6][#4]}
		}
		\ifthenelse{\equal{#5}{init}}{
			\foreach \i in {0,...,#3}
			{
				\MPSinitialstate[-1+2*\i][-1][#4]
			}
		}{}
	\end{scope}
}

\newcommand{\msqr}{\fineq[-0.8ex][0.7][1]{\sqrstate[0][0][]}}

\newcommand{\mcirc}{\fineq[-0.8ex][0.7][1]{\cstate[0][0][]}}
\renewcommand{\bcirc}{\fineq[-0.8ex][0.7][1]{\cstate[0][0][][black]}}

\newcommandx{\cstate}[4][1=0,2=0,3= ,4=white]{
	\begin{scope}[shift={(0,0)}]
				\draw[fill=#4,thick] (#1,#2) circle (0.13);
				\node[scale=1.1] at (#1,#2) {$#3$};
\end{scope}
}
\newcommandx{\sqrstate}[4][1=0,2=0,3= ,4=white]{
	\begin{scope}[shift={(#1,#2)}]
		\draw[thick,fill=#4] (-0.13,-0.13) rectangle (0.13,0.13) ;
		\node[scale=1.1] at (0,0) {$#3$};
	\end{scope}
}
\newcommandx{\pairproduct}[2][1=0,2=0]
{\begin{scope}[shift={(#1 ,#2)}]
\draw[thick] (-.5,.5) arc(-135:-45:1/1.414);
\sqrstate[0][.5-.1414][][black]	
\end{scope}
}
\newcommandx{\bellpair}[2][1=0,2=0]
{\begin{scope}[shift={(#1 ,#2)}]
		\draw[thick] (-.5,.5) arc(-135:-45:1/1.414);
	\end{scope}
}
\newcommandx{\charge}[3][1=0,2=0,3=black]
{
	\ifthenelse{\equal{#3}{blue}}{\def \chargecolor{bertinigreyblue}
	}{
	\ifthenelse{\equal{#3}{red}}{\def \chargecolor{bertinigreyred}}{\def \chargecolor {#3}}}
\begin{scope}[shift={(#1 ,#2)}]
	\draw[ fill=\chargecolor] circle (0.08);        
\end{scope}
}
\newcommandx{\trianglediag}[7][1=0,2=0,3=1,4=bertiniorange,5= ,6=-1,7=topright]
{\begin{scope}[shift={(#1 ,#2)}]
	\foreach \i in {0,...,#3}
	{	\foreach \j in {0,...,\i}
		{	\roundgate[-\j+2*\i][\j][1][topright][#4][#6]
		}
	}
	\foreach \i in {-1,...,#3}
	{\ifthenelse{\equal{#5}{bellpair}}{\bellpair[\i*2+1][-1]}{\ifthenelse{\equal{#5}{pairproduct}}{\pairproduct[\i*2+1][-1]}{\MPSinitialstate[\i*2+1][-1][#4][#7][#6]}}}
\end{scope}
}
\newcommandx{\projectorleg}[4][1=0,2=0,3=R,4=left]
{
	\begin{scope}[shift={(#1 ,#2)}]
		{\ifthenelse{\equal{#3}{R}}{		\draw[thick] (-.25,-.25)--++(.5,.5);}{\ifthenelse{\equal{#3}{L}}{		\draw[thick] (-.25,.25)--++(.5,-.5);}{\draw[thick] (-.25,0)--++(.5,0);}}
		}
	\draw[thick, fill=white] circle (0.13);
	\ifthenelse{\equal{#4}{right}}{
		\draw[thick] (.0,.07)--++(.07,0)--++(0,-.07);
		\node[scale=0.5] at (0,-.02) {$\alpha$};
	}{}
	\ifthenelse{\equal{#4}{left}}{	
	\draw[thick] (0,.07)--++(-.07,0)--++(0,-.07);
	\node[scale=0.5] at (0,-.03) {$\beta$};
	}{}
	\end{scope}
}
\usetikzlibrary{patterns,decorations.pathreplacing}

\newcommand{\be}{\begin{equation}}
	\newcommand{\ee}{\end{equation}}

\pgfplotsset{
	colormap={bright}{rgb255=(251,51,255) rgb255=(255,128,0) rgb255=(0,0,255)
		rgb255=(255,0,0) rgb255=(128,255,0) rgb255=(204,204,0) rgb255=(127,0,255)
		rgb255=(0,0,0)}
}
\pgfplotsset{
	colormap={lightbluepalette}{ rgb255=(200, 240, 250) rgb255=(135, 206, 250) rgb255=(115, 194, 251)
		rgb255=(124,158,217) rgb255=(96, 130, 182)}
}

\theoremstyle{plain}

\newtheorem{theorem}{Theorem}
\newtheorem{definition}{Definition}

\newcommand{\lt}[0]{{({\ell})}}
\newcommand{\rt}[0]{{({r})}}
\newcommand{\ltb}[0]{\ell}
\newcommand{\rtb}[0]{r}
\pgfplotsset{
    colormap={springpastels}{ rgb255=(253, 127, 111) rgb255=(126, 176, 213) rgb255=(178, 224, 97) rgb255=(189, 126, 190) rgb255=(255, 181, 90)  rgb255=(190, 185, 219) rgb255=(253, 204, 229) rgb255=(139, 211, 199)}
    }

\definecolorseries{foo}{hsb}{step}[hsb]{0,1,1}[hsb]{.618,0,0}

\hypersetup{
	pdftitle={Non-equilibrium dynamics of  symmetric dual-unitary circuits},
	pdfsubject={Many-body quantum physics},
	pdfauthor={Alessandro, Pasquale, and Bruno},
	pdfkeywords={Many-body quantum physics, Entanglement, OTOC, Information spreading}
}

\renewcommand{\vec}[1]{\boldsymbol{#1}}
\begin{document}

\title{Non-equilibrium dynamics of charged dual-unitary circuits}
	
\author{Alessandro Foligno}
\affiliation{School of Physics and Astronomy, University of Nottingham, Nottingham, NG7 2RD, UK} 
\affiliation{Centre for the Mathematics and Theoretical Physics of Quantum Non-Equilibrium Systems, University of Nottingham, Nottingham, NG7 2RD, UK}
	
\author{Pasquale Calabrese}
\affiliation{SISSA and INFN Sezione di Trieste, via Bonomea 265, 34136 Trieste, Italy}
\affiliation{International Centre for Theoretical Physics (ICTP), Strada Costiera 11, 34151 Trieste, Italy}

\author{Bruno Bertini}
\affiliation{School of Physics and Astronomy, University of Birmingham, Edgbaston, Birmingham, B15 2TT, UK}
	
\begin{abstract}
The interplay between symmetries and entanglement in out-of-equilibrium quantum systems is currently at the centre of an intense multidisciplinary research effort. Here we introduce a setting where these questions can be characterised exactly by considering dual-unitary circuits with an arbitrary number of $U(1)$ charges. After providing a complete characterisation of these systems we show that one can introduce a class of solvable states, which extends that of generic dual unitary circuits, for which the non-equilibrium dynamics can be solved exactly. In contrast to the known class of solvable states, which relax to the infinite temperature state, these states relax to a family of non-trivial generalised Gibbs ensembles. The relaxation process of these states can be simply described by a linear growth of the entanglement entropy followed by saturation to a non-maximal value but with maximal entanglement velocity. We then move on to consider the dynamics from non-solvable states, combining exact results with the entanglement membrane picture we argue that the entanglement dynamics from these states is qualitatively different from that of the solvable ones. It shows two different growth regimes characterised by two distinct slopes, both corresponding to sub-maximal entanglement velocities. Moreover, we show that non-solvable initial states can give rise to the quantum Mpemba effect, where less symmetric initial states restore the symmetry faster than more symmetric ones.
\end{abstract}
	
\maketitle

\tableofcontents

\section{Introduction}

Understanding the non-equilibrium dynamics of interacting many-particle systems, classical or quantum, has been a key open question in theoretical physics for almost two centuries. Since the turn of the millennium, however, an interdisciplinary research effort involving experts from condensed matter physics, quantum information, and particle physics has started to crack the nut. 

For instance, one aspect that made the study of many-body systems out of equilibrium particularly hard is that the standard probes used to analyse systems at equilibrium, like Hamiltonian gaps or correlation functions of local observables, are either ill-defined or return obscure non-universal information in the non-equilibrium setting. Recent research, however, has finally identified a convenient probe in \emph{quantum entanglement}~\cite{amico2008entanglement, calabrese2009entanglement, santos2011entropy, gurarie2013global, laflorencie2016quantum, calabrese2018entanglement, bertini2022growth}. The latter gives an observable-independent characterisation of the correlations between the different finite parts of the system and displays universal time evolution even in out-of-equilibrium settings. For example, under very general conditions on the microscopic dynamics, after a quantum quench the entanglement between a finite subsystem and the rest shows an irreversible linear growth followed by saturation~\cite{calabrese2005evolution, fagotti2008evolution, alba2018entanglement, alba2017entanglement, alba2018entanglement,lagnese2021entanglement, liu2014entanglement, asplund2015entanglement, laeuchli2008spreading, kim2013ballistic, pal2018entangling, bertini2019entanglement, piroli2020exact, gopalakrishnan2019unitary}.

At the same time researchers began to find minimal interacting models where the entanglement dynamics can be characterised exactly~\cite{nahum2018operator, vonKeyserlingk2018operator, bertini2019entanglement, gopalakrishnan2019unitary, piroli2020exact, jonay2021triunitary, foligno2024quantum}. These can be considered as the out-of-equilibrium analogues of the solved models of statistical mechanics~\cite{baxter1982exactly}, and, just like their equilibrium counterparts, they can be used to define and characterise suitable universality classes for dynamical behaviours. The aforementioned minimal models have been found in the context of \emph{quantum circuits}, i.e., many-body systems in discrete space-time~\cite{hosur2016chaos, nahum2017quantum, bertini2018exact, gopalakrishnan2019unitary,piroli2020exact,nahum2018operator,khemani2018operator, chan2018solution,vonKeyserlingk2018operator,bertini2019entanglement,bertini2019exact,friedman2019spectral,li2019measurement,skinner2019measurement,rakovszky2019sub, zabalo2020critical, claeys2021ergodic}, owing to the more symmetric treatment of space and time provided by this setting. Specifically, there have been two main strategies to find solvable quantum circuits. The first is to introduce noise~\cite{nahum2017quantum, khemani2018operator, chan2018solution, bertini2024quantum} (see also Ref.~\cite{fisher2022random} and references therein). In this case one can compute the average of relevant physical quantities by mapping them to classical partition functions, which can sometimes be solved exactly. The second is to impose additional constraints, or symmetries, on the dynamics~\cite{bertini2019exact, klobas2021exact, prosen2021many, kos2021correlations, claeys2022exact, bertini2022exact, kos2023circuitsofspacetime, bertini2023exact}. This second approach has the advantage of being more fundamental --- it does not require any external influence --- but the additional constraints can make these minimal models non-generic. Remarkably, however, among the constrained minimal models there are some exhibiting, ergodic i.e., generic, dynamics. The simplest class with this property are dual unitary circuits~\cite{bertini2019exact}, which are constrained to generate unitary dynamics also when the roles of space and time are exchanged.    

Having obtained a blueprint of the thermalisation process through the entanglement the research has now shifted to seek more refined probes. A natural next step has been to investigate the fluctuations of globally conserved quantities within local subsystems, as they should also show a universal dynamics, and their interplay with the entanglement~\cite{laflorencie2014spin, goldstein2018symmetry, xavier2018equipartition, bonsignori2019symmetry, murciano2020entanglement, ares2022entanglement, bertini2022nonequilibrium, bertini2024dynamics,parez2021quasiparticle,mrc-20,Vitale-22}. For example, a simple but instructive problem studied in this context is that of symmetry restoration when a symmetric system is prepared in a non-equilibrium initial state that explicitly breaks the symmetry. This process can be characterised in a universal way by measuring the distance between the time-evolving state and its symmetrised counterpart via the so called \emph{entanglement asymmetry}~\cite{ares2022entanglement} (see also Refs.~\cite{ares2023lack,murciano2024entanglement,khor2023confinement,ferro2023nonequilibrium,capizzi2023entanglement,capizzi2023universal,bertini2024dynamics,rylands2023microscopic,chen2024renyi,fossati2024entanglement,caceffo2024entangled,ares2023entanglement,liu2024symmetry,yamashika2024entanglement,ares2024quantum,chalas2024multiple,turkeshi2024quantum, klobas2024translation,lastres2024entanglementasymmetrycriticalxxz,benini2024entanglementasymmetryconformalfield}). This revealed, for instance, that a symmetry can sometimes be restored more rapidly  when it is broken more by the initial state, providing a quantum analogue of the famous Mpemba effect~\cite{mpemba1969cool}.  

To sustain this effort it is again crucial to identify minimal models where these questions can be studied exactly. Some progress in this direction has been achieved in the context of noisy quantum circuits~\cite{liu2024symmetry,turkeshi2024quantum, klobas2024translation}, however, for the case of clean systems the results are scarce. Essentially, the only existing partial results in this direction ~\cite{giudice2021temporal} are for a class of Yang-Baxter integrable dual unitary circuits~\cite{bertini2020operator2,holdendye2023fundamental}. Here we introduce and characterise a class of symmetric dual-unitary circuits, which are chaotic in each symmetry sector. We then use these systems to provide exact results on the interplay between entanglement and charge fluctuations. We do so by first introducing a class of initial states with exactly solvable quench dynamics. This extends the known class of solvable states of dual-unitary circuits~\cite{piroli2020exact}. We then study the dynamics of states that, instead, are not solvable, by combining exact bounds with the entanglement membrane theory. We prove that these two classes of states show a qualitatively different entanglement dynamics and that the latter can display the quantum Mpemba effect.

The rest of the paper is laid out as follows. In Sec.~\ref{sec:setting} we present a summary and a discussion of our main results. In Sec.~\ref{sec:characterisation} we report a systematic characterisation of dual unitary circuits with $U(1)$ charges for arbitrary local Hilbert space dimension. In Sec.~\ref{sec:solvablestates} we introduce the class of solvable states for charged dual-unitary circuits, which we dub ``charged solvable states''. In Sec.~\ref{sec:solvablestatesquench} we present an exact solution of the quench dynamics of charged solvable states. In Sec.~\ref{sec:genericinitialstates} we instead study the dynamics of non-solvable states while in Sec.~\ref{sec:asymmetrysec} we discuss the dynamics of entanglement asymmetry. Our conclusions are reported in Sec.~\ref{sec:conclusions}. The main text is complemented by a number of appendices containing the proofs of some of the theorems presented in the main text and some explicit calculations.

\section{Setting and Summary of the Results}
\label{sec:setting}

We consider brickwork quantum circuits, i.e., quantum many-body systems defined on a lattice where the time evolution is discrete and the interactions are local. A step of time evolution is implemented by the many-body unitary operator $\mathbb{U}$ constructed in terms of a two-qudit gate $U$ according to the following staggered pattern  
\be
\begin{aligned}
&\mathbb{U} =\mathbb{U}_e\mathbb{U}_o,\\
&\mathbb{U}_e=\bigotimes_{x=0}^{L-1} U_{x,x+1/2},\quad \mathbb{U}_o=\bigotimes_{x=0}^{L-1} U_{x+1/2,x+1}\,.
\end{aligned}
\label{eq:U}
\ee
Here the sites are labelled by half integers, $U_{a,b}$ acts on the qudits located at $a$ and $b$, and we denoted by $L$ the length of the system (which contains $2L$ qubits). Moreover, we take periodic boundary conditions, so that sites $0$ and $L$ coincide, and indicate by $d$ the local dimension (number of states of each qudit) on each site. 

We take our local gate $U$ be \emph{dual-unitary}~\cite{bertini2019exact}, meaning that both $U$ and $\widetilde{U}$ --- the gate corresponding to $U$ upon exchange of space and time --- are unitary matrices. More formally, the spacetime swapped gate $\widetilde{U}$ is defined through the following matrix element reshuffling 
\be
\bra{ij}\widetilde{U}\ket{kl}\equiv \mel{jl}{U}{ik},
\ee	
and dual unitarity corresponds to requiring 	
\begin{align}
&UU^\dagger = U^\dagger U=\mathds{1}_{d^2}\qquad 
 \widetilde{U}\widetilde{U}^\dagger
= \widetilde{U}^\dagger\widetilde{U}=\mathds{1}_{d^2},
\label{eq:DU}
\end{align}
where $\mathds{1}_{x}$ represents the identity operator in $\mathbb C^{x}$. A quantum circuit composed of gates with this property is typically referred to as dual-unitary circuit. Specifically, in this work we consider dual unitary circuits with $n$ independent $U(1)$ symmetries (hence having $n$ independent conservation laws with 1-local density). In fact, since left and right moving charges are independent in dual unitary circuits~\cite{bertini2020operator} (see also the review in Appendix~\ref{app:chargepropagation}), it is natural to consider independent symmetry groups for left $\lt$ and right $\rt$ movers. Therefore, we consider circuits with a symmetry group given by 
\be
\mathcal G  = U(1)_\ltb^{n_\ltb} \times U(1)_\rtb^{n_\rtb},
\label{eq:symmetrygroup}
\ee
where $n_{\ltb/\rtb}$ can take any value from $0$ to $d-1$ (since we require our charges to commute with each other). The lower bound corresponds to circuits with no symmetry, while the upper bound to circuits with the maximal number of commuting 1-local charge densities. 
\begin{figure}
\includegraphics{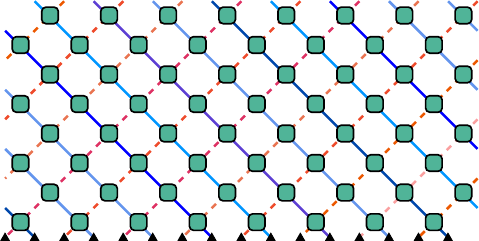}
\caption{Graphical representation of the state of a $\mathcal G$-symmetric dual-unitary circuit in a given charge sector. Left-moving (blue shades) and right-moving (red shades) legs have generically different local dimensions. See Sec.~\ref{sec:diagrams} for a detailed explanation of the diagrammatic notation.}
\label{fig:schemechargesector}
\end{figure}

In Sec.~\ref{sec:characterisation} we show that in $\mathcal G$-symmetric dual unitary circuits the local gate $U$ can be decomposed in different charge sectors as 
\be
U = \bigoplus_{\alpha=1}^{n_\rtb+1}  \bigoplus_{\beta=1}^{n_\ltb+1} U^{(\alpha,\beta)},
\label{eq:directsum}
\ee
where the blocks $U^{(\alpha,\beta)}$ have dimension $d_\alpha^{{\rt}} d_\beta^{{\lt}}$ and $d_\alpha^{(\ltb/\rtb)}> 0$ are such that  
\be
\sum_{\alpha=1}^{n_\rtb+1} d_\alpha^{{\rt}} = \sum_{\beta=1}^{n_\ltb+1} d_\beta^{{\lt}}  = d\,. 
\ee
Note that $U^{(\alpha,\beta)}$ implements unitary transformations with different domains and codomains, i.e., 
\be
U^{(\alpha,\beta)}: \mathbb C^{d_\alpha^{{\rt}}}\otimes  \mathbb C^{d_\beta^{\lt}} \rightarrow \mathbb C^{d_\beta^{\lt}} \otimes  \mathbb C^{d_\alpha^{{\rt}}}.
\ee
Since the domain and codomain of $U^{(\alpha,\beta)}$ are isomorphic, however, one can still think of it as a square unitary matrix. In fact, whenever $U$ is dual unitary, also the blocks $U^{(\alpha,\beta)}$ are dual-unitary, meaning that both $U^{(\alpha,\beta)}$ and its space-time swapped counterpart are unitary.

In passing we note that unitary transformations with different (but isomorphic) domain and codomain can be used to define quantum cellular automata generalising the concept of brickwork quantum circuits~\cite{Farrelly2020reviewofquantum}. These systems arise at the boundary of certain Floquet systems displaying many-body localisation in the bulk~\cite{po2016chiral, po2017radical, harper2017floquet, fidkowski2019interacting}, and can be shown to generate a qualitatively different quantum information spreading~\cite{gong2022coarse, zadnik2024quantum}. Here we find that the dual unitary version of these systems emerge naturally in the charge sectors of $\mathcal G$-symmetric dual unitary circuits. A schematic representation of the state of the system in a given charge sector is presented in Fig.~\ref{fig:schemechargesector}.

Next, in Sec.~\ref{sec:solvablestates}, we show that for $\mathcal G$-symmetric dual unitary circuits one can extend the family of solvable states introduced in Ref.~\cite{piroli2020exact} (a minimal example is provided at the end of the section). We dub this larger family \emph{charged solvable} states and, in Sec.~\ref{sec:solvablestatesquench}, we present the exact solution of their many body dynamics. We show that in contrast with the solvable states of Ref.~\cite{piroli2020exact}, which always relax to the infinite temperature state, charged solvable states relax to non-trivial generalised Gibbs ensembles. Their entanglement velocity, however, is still maximal. Specifically, at the leading order the entanglement entropy of an interval $A$ of length $L_A$ is described by the standard linear-growth/sharp relaxation form  
\be
S_A(t)= 2 s \min(2t, L_A), 
\label{eq:entropysolvable}
\ee
where $2 s$ is the entropy density of the generalized Gibbs ensemble (GGE). In fact, we find that $S_A(\infty)$ can be split into a ``number'' and a ``configurational'' part, where the former measures the average of the entanglement in the charge sectors and the latter the fluctuations of the charge~\cite{lukin2019probing, parez2021quasiparticle, parez2021exact}. The emergence of this splitting at infinite times is a typical feature of quantum many-body systems with conserved charges~\cite{bertini2022nonequilibrium}, however, in our case it takes a special form: the number part of $S_A(\infty)$ is extensive rather than logarithmic in $L_A$. One can understand this by noting that $\mathcal G$-symmetric dual unitary circuits have exponentially many (in $L_A$) charge sectors as opposed to the polynomially many occurring in generic systems with conservation laws. 

\begin{figure}
\includegraphics{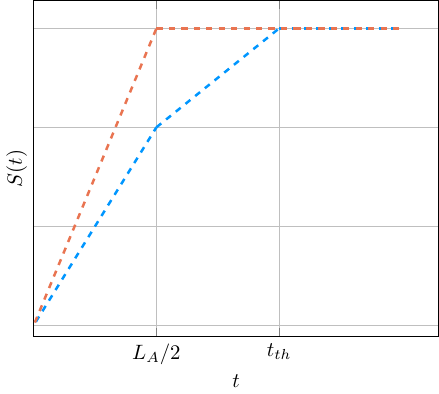}
\caption{Schematic representation of entanglement growth of a finite, connected interval of size $L_A$ from a charged solvable state (orange) and a generic one (blue). Solvable charged states behave as those studied in \cite{piroli2020exact}, meaning they thermalise at the fastest possible rate and the entanglement saturates at $t=L_A/2$. In the generic case, instead, there is a secondary growth phase which is always slower than the first at times between the thermalization time $t_{th}$ and $L_A/2$.}
\label{fig:entgrowth}
\end{figure}

Then, in Sec.~\ref{sec:genericinitialstates}, we consider the dynamics of $\mathcal G$-symmetric dual unitary circuits from states that are \emph{not} charged solvable. Combining exact results with the entanglement membrane approach~\cite{jonay2018coarsegrained, zhou2020entanglement} we find that the entanglement dynamics is \emph{qualitatively different} from the one described by Eq.~\eqref{eq:entropysolvable} as it shows a \emph{two step relaxation}. More precisely, we find that left and right movers give different contributions to the entanglement growth (they contribute equally for charged solvable states). Calling their respective contributions $s^{{\rt}}$ and $s^{{\lt}}$ and considering $s^{{\rt}}<s^{{\lt}}$ we have the following scaling form of the entanglement entropy for large times and sub-system sizes 
\be
\!\!S_A(t) \simeq \begin{cases}
		4t  s^{{\rt}} &t\le {L_A}/{2}
		\\
		4t s^{{\rt}}_{\phantom{{\rm num}}}-(2t-L_A) s^{{\rt}}_{\rm num} & {L_A}/{2}< t\le t_{\rm th}\\
		L_A (s^{{\rt}}+s^{{\lt}})  & t > t_{\rm th}
	\end{cases}\!\!
\label{eq:entropynonsolvable}
\ee
where $\simeq$ means equality at leading order in time, $s^{{\rt}}_{\rm num}$ is the number entropy density of right movers and the thermalisation time $t_{\rm th}$ is fixed by continuity to be
\be
t_{\rm th}=\frac{L_A}{2}\left(1+\frac{s^\lt-s^\rt}{2s^\rt-s^\rt_{\rm num}}\right).
\label{eq:thermalisationtime}
\ee
Eq.~\eqref{eq:entropynonsolvable} implies that, for a given entropy density of the GGE  to which the state thermalizes, i.e. $2s=s^{{\rt}}+s^{{\lt}}$ non-solvable states show a slower (state dependent) entanglement growth compared to the solvable ones. Namely their entanglement velocity is not 1 despite the circuit being dual unitary. Moreover, the slope in the second phase is always lower than the one in the first as long as $s^{{\rt}}_{\rm num}\neq 0$. This behaviour arises because the charge structure acts as a constraint that can create an imbalance on the amount of entanglement available for left- and right-moving charges: While solvable states are forced to have no imbalance, non-solvable ones are generically imbalanced and their thermalisation process is slower. The occurrence of the second stage of thermalisation is another interesting consequence of the charge structure, this time in concurrence with the non-Yang--Baxter-integrable nature of the system. Indeed, this phase corresponds to a time regime where a simple quasiparticle picture would predict the entanglement to be already saturated. The growth is only observed because the creation and destruction of quasiparticles generates a scrambling of quantum information. This scrambling, however, cannot be maximal because all the creation/annihilation processes have to conserve the charge. Therefore, the slope in the second phase is generically smaller than in the first. A schematic representation of the entanglement growth from solvable and non-solvable states is reported in Fig.~\ref{fig:entgrowth}, while a numerical simulation confirming this picture is presented in Fig.~\ref{fig:entgrowthnumerics} (see Sec.~\ref{sec:genericstatesent}).  

We remark that the linear growth discussed above for the entanglement entropy is also observed for all R\'enyi entropies $S_A^{(n)}(t)$
, including those of order $n>1$, for both charge solvable and non-charged solvable states. This is not in contradiction with the results of Refs.~\cite{rakovszky2019sub, huang2020dynamics, znidaric2020entanglement}, which found sub-ballistic growth of $S_A^{(n>1)}(t)$ in generic quantum circuits with conservation laws. Indeed, a key ingredient for the onset of sub-ballistic growth is the presence of diffusive charge transport whereas charge-transport in dual-unitary circuits is always ballistic (see Sec.~\ref{sec:characterisation}).

Finally, in Sec.~\ref{sec:asymmetrysec}, we study symmetry restoration in $\mathcal G$-symmetric dual-unitary circuits. Namely, we quantify how a non-symmetric initial state becomes gradually more symmetric in the course of the time evolution. We do so by computing the entanglement asymmetry $\Delta S_A(t)$ defined as the relative entropy between the reduced state of $A$ and its symmetrised version (see Sec.~\ref{sec:asymmetrysec} for details).

An interesting aspect of this process is the possible occurrence of the \emph{quantum Mpemba effect}, which means that a symmetry is restored faster for states breaking it more at $t=0$~\cite{ares2022entanglement}. Its name is due to the conceptual similarity that this process has with the ``classical'' Mpemba effect, arising when hot water freezes faster than cold water~\cite{mpemba1969cool}. Since its first observation in Ref.~\cite{ares2022entanglement} this effect has been observed in a range of different physical settings~\cite{ares2022entanglement,ares2023lack,murciano2024entanglement,khor2023confinement,ferro2023nonequilibrium,capizzi2023entanglement,capizzi2023universal,bertini2024dynamics,rylands2023microscopic,chen2024renyi,fossati2024entanglement,caceffo2024entangled,ares2023entanglement,liu2024symmetry,yamashika2024entanglement,ares2024quantum,chalas2024multiple,turkeshi2024quantum, klobas2024translation}, including trapped ion quantum simulators~\cite{joshi2024observing}. Here we find that non-charged solvable initial states in $\mathcal G$-symmetric dual-unitary circuits show the quantum Mpemba effect owing to their two step relaxation process. In particular, we provide an explicit example where the quantum Mpemba effect is controlled by the number entropy of right movers: larger number entropy implies larger initial asymmetry and faster relaxation.

\section{$\mathcal G$-Symmetric Dual Unitary Circuits}
\label{sec:characterisation}

In this work we are interested in dual-unitary circuits with continuous internal symmetries. In other words, we require the existence of extensive operators 
\be
Q=\sum_{x\in \mathds{Z}_{2L}/2} q_x,
\label{eq:Q}
\ee
with $q_x$ acting nontrivially only at site $x$, that commute with $\mathbb{U}$ in Eq.~\eqref{eq:U}.

In dual-unitary circuits, conserved charges as the one in Eq.~\eqref{eq:Q} split into two independent components supported respectively on integer and half-odd-integer sites, i.e.
\be
Q^{\lt} = \sum_{x\in \mathds{Z}_{L}}q_{x+1/2},\qquad Q^{\rt} = \sum_{x\in \mathds{Z}_{L}} q_{x}.
\label{eq:QLQR}
\ee
Note that these operators are two-site shift invariant like the time evolution operator. Moreover, we denoted the two charges in Eq.~\eqref{eq:QLQR} by the labels $\ltb$ and $\rtb$ as their densities are respectively left- and right-moving~\cite{bertini2020operator}. 

To see how this occurs we recall that, as a consequence of charge conservation, the charge density obeys an operatorial continuity equation of the form
\begin{align}
\mathbb{U}^\dag q_x \mathbb{U}= q_x - J_{x+1/2} + J_{x-1/2},
\label{eq:continuityequation2}
\end{align}
where $J_{x}$ is a local operator with support at most on two sites starting at site $x$. For dual unitary circuits, however, this operator takes the following special form~\cite{bertini2020operator} (for completeness, we report a proof of the relevant facts in Appendix~\ref{app:chargepropagation})
\be
J_{x}=
\begin{cases}
  - q_{x+1/2}, & x\in\mathbb Z\\
\phantom{-}  q_{x-1/2}, & x\in\mathbb Z+1/2\\
\end{cases},
\ee 
meaning that transport of charge is \emph{always ballistic} in dual-unitary circuits. Moreover, the charge densities are simply shifted by the time evolution
\begin{align}
\mathbb{U} q_x\mathbb{U}^\dagger= \begin{cases}
q_{x+1}, & x\in\mathbb Z\\
q_{x-1}, & x\in\mathbb Z+1/2\\
\end{cases}\,,
\label{eq:chargeconservationrelation}
\end{align}
and the charges in Eq.~\eqref{eq:QLQR} are independently conserved.

Operators evolving as in Eq.~\eqref{eq:chargeconservationrelation} have been referred to as ``solitons''~\cite{bertini2020operatorII} or ``gliders''~\cite{borsi2022construction} in the literature and their presence immediately implies that the circuit has far more local conserved operators than those in Eq.~\eqref{eq:QLQR}. Indeed if the circuit admits a non-trivial soliton, for any given $\ell \geq 1$ one can produce $2^{\ell+1}-2$ independent charge densities  with support on at most $2\ell$ sites (as shown in  App.~\ref{app:chargepropagation}), each of which obeys \eqref{eq:chargeconservationrelation}. Physically, this happens because the solitons are effectively ``fragmenting'' the dynamics of the circuit in different charge sectors as in Refs.~\cite{fragmentation,PhysRevX.10.011047}. This structure has been sometimes referred to as \emph{super-integrability}, see e.g., Ref.~\cite{gombor2022superintegrable, gombor2024integrable}. Despite this, $\mathcal G$-symmetric dual-unitary circuits are generically not Yang-Baxter integrable. In fact, the charges are generically unable to constraint the dynamics within each given charge sector and the latter are chaotic. The only exception to this is when all charge sectors have dimension one, which is realised, for instance, for $d=2$~\cite{bertini2020operator2}. This is conceptually similar to what happens in holographic conformal field theories (see, e.g.,~\cite{heemskerk2009holography}), which, despite the huge amount of conserved charges provided by the conformal structure, behave chaotically in each charge sector. The goal of this paper is to characterise the conserved charge structure of $\mathcal G$-symmetric dual-unitary circuits and investigate its effects on the many-body dynamics. In the rest, we will use the word ``soliton'' to designate the fundamental building blocks of the charge densities, that is those with support on a single site.

Considering circuits with the symmetry group $\mathcal G$ (as defined in Eq.~\eqref{eq:symmetrygroup}) we then have $m_{\ltb}=1+ n_{\ltb}$ independent left moving solitons and $m_{\ltb}=1+ n_{\ltb}$ solitons. Indeed, Eq.~\eqref{eq:chargeconservationrelation} is fulfilled by the charges associated to the $U(1)$ symmetries and, trivially, by the identity operator. Noting then that Eq.~\eqref{eq:chargeconservationrelation} is linear in the charge density, we have that the space of solitons on each site is a vector space. Therefore we introduce the following bases of solitons  
\be
\mathcal S^{(\rtb/\ltb)} = \{\sigma_1^{(\rtb/\ltb)}, \ldots,\sigma_{m_{{\ltb}/{\rtb}}}^{(\rtb/\ltb)}\},
\label{eq:solitonbasis}
\ee
where the subscript denotes the element of the basis and we suppressed the position index $x$: unless explicitly stated position indices for solitons will be suppressed from now on. This basis can be taken to be Hilbert-Schmidt orthogonal. 

The bases of solitons in Eq.~\eqref{eq:solitonbasis} are characterised by means of the following theorem, proven in Appendix~\ref{app:chargeclassifications}. 

\begin{theorem}
\label{th:chargeclassification}
In a dual unitary circuit with  commuting solitons supported on a \emph{single site} one can always choose a basis where the solitons are \emph{mutually orthogonal projectors}. In addition, the bases in Eq.~\eqref{eq:solitonbasis} can be taken to be  
\be
\mathcal S^{(\rtb/\ltb)} =\{\{\Pi^{(\rtb/\ltb)}_\alpha\}, \{P_{+,\beta}^{(\rtb/\ltb)}, P_{-,\beta}^{(\rtb/\ltb)}\}\},
\label{eq:solitonbasisproj}
\ee  
where the two types of solitons, $\Pi_\alpha$ and $P_{+/-,\beta}$,  respectively fulfil 
\be
\begin{aligned}
&U	\left(\Pi^{{\rt}}_\alpha \otimes \mathds{1}_d\right) U^\dagger=  \mathds{1}_d \otimes \Pi^{{\rt}}_\alpha,\\
&U	\left( \mathds{1}_d \otimes \Pi^{\lt}_\alpha \right) U^\dagger=  \Pi^{\lt}_\alpha \otimes \mathds{1}_d,
\end{aligned}
\label{eq:firstgroupPi}
\ee
and
\be		
\begin{aligned}
&U\left(P^{{\rt}}_{\pm,\alpha} \otimes \mathds{1}_d\right) U^\dagger=  \mathds{1}_d \otimes P^{{\rt}}_{\mp,\alpha},\\
&U	\left( \mathds{1}_d\otimes P^{\lt}_{\pm,\alpha}\right) U^\dagger=   P^{\lt}_{\mp,\alpha}\otimes\mathds{1}_d .
\end{aligned}
\label{eq:secondgroupP}
\ee
\end{theorem}
In a nutshell, Eq.~\eqref{eq:solitonbasisproj} involves three different families of projectors because Eq.~\eqref{eq:chargeconservationrelation} implies that there are two different kinds of solitons: those that are shifted by one site every half time-step and those that are shifted and multiplied by $-1$. Solitons of the first kind originate the family $\Pi^{{\lt}/{\rt}}_\alpha$ and those of the second originate $P^{{\rt}}_{+,\alpha}$ and $P^{{\rt}}_{-,\alpha}$ (see Appendix~\ref{app:chargeclassifications} for details). 
We also remark that, since the set of solitons includes the identity operator, the basis in Eq.~\eqref{eq:solitonbasisproj} fulfils the following completeness relation 
\begin{align}
\sum_\alpha \Pi^{(\rtb/\ltb)}_\alpha+\sum_\beta P^{(\rtb/\ltb)}_{+,\beta}+P^{(\rtb/\ltb)}_{-,\beta}=\mathds{1}_d\,.
\label{eq:completeness}
\end{align}

For the sake of simplicity here we only consider the case of commuting solitons, i.e., circuits where the symmetry group is abelian (this excludes, for example, circuits made of SWAP gates as they have non-abelian symmetry groups). Moreover, we restrict to solitons of the type $\Pi_\alpha$. The latter is not a real restriction as the solitons of type $P_{\pm,\alpha}$ are mapped into $\Pi_\alpha$ by ``renormalising'' the quantum circuit, i.e., combining together two subsequent time steps and two neighbouring qudits (see the discussion at end of App.~\ref{app:transferM} for an explicit example).

Under these conditions Theorem~\ref{th:chargeclassification} implies that one  can decompose the local gates in a number of blocks corresponding to the charge sectors. Namely, using the completeness relations in Eq.~\eqref{eq:completeness}, the strong soliton condition in Eq.~\eqref{eq:firstgroupPi}, and the projector property of $\Pi_\alpha^{(\ltb/\rtb)}$ one can write
\be
 U = \sum_{\alpha,\beta}  (\Pi^{{\lt}}_\beta \otimes \Pi^{{\rt}}_\alpha) U = \sum_{\alpha,\beta} U^{(\alpha,\beta)},
\label{eq:DUdecomposiiton}
\ee
where we introduced
\be
\label{eq:Ualphabeta}
U^{(\alpha,\beta)} \equiv (\Pi^{{\lt}}_\beta \otimes \Pi^{{\rt}}_\alpha) U (\Pi^{{\rt}}_\alpha\otimes\Pi^{{\lt}}_\beta)\,. 
\ee
Note that domain and codomain of this transformation are different. Specifically, we have  
\begin{align}
&U^{(\alpha,\beta)}: \mathbb C^{d_\alpha^{{\rt}}}\otimes  \mathbb C^{d_\beta^{\lt}} \rightarrow \mathbb C^{d_\beta^{\lt}} \otimes  \mathbb C^{d_\alpha^{{\rt}}},
\label{eq:map}
\end{align}
where we set 
 \be
d^{(\rtb/\ltb)}_\alpha=\tra{\Pi_\alpha^{(\rtb/\ltb)}}\,.
\label{eq:dalphalr}
\ee 
These spaces, however, are isomorphic. Moreover, using Eq.~\eqref{eq:firstgroupPi} and the projector property of $\Pi_\alpha^{(\rtb/\ltb)}$ one can show that $U^{(\alpha,\beta)}$ is a unitary transformation between the two spaces, i.e. 
\be
\begin{aligned}
& (U^{(\alpha,\beta)})^\dag U^{(\alpha,\beta)}=\Pi^{{\rt}}_\alpha\otimes\Pi^{{\lt}}_\beta, \\
&U^{(\alpha,\beta)}(U^{(\alpha,\beta)})^\dag = \Pi^{{\rt}}_\beta \otimes \Pi^{{\lt}}_\alpha. 
\end{aligned}
\label{eq:unitaritytimered}
\ee
In fact, noting that 
\be
\begin{aligned}
\widetilde{U^{(\alpha,\beta)}}&=(\Pi^{{\lt}}_\alpha \otimes \Pi^{{\rt}}_\beta) \widetilde U\\
&= (\Pi^{{\lt}}_\alpha \otimes \Pi^{{\rt}}_\beta) \widetilde U (\Pi^{{\rt}}_\beta\otimes\Pi^{{\lt}}_\alpha),
\end{aligned}
\ee
and using again Eq.~\eqref{eq:firstgroupPi} and the projector property of $\Pi_\alpha^{{\lt}/{\rt}}$ one can show that also $\widetilde{U^{(\alpha,\beta)}}$ is unitary, i.e.    
\be
\begin{aligned}
& (\widetilde{U^{(\alpha,\beta)}})^\dag \widetilde{U^{(\alpha,\beta)}} = \Pi^{{\rt}}_\beta\otimes\Pi^{{\lt}}_\alpha, \\
&\widetilde{U^{(\alpha,\beta)}} (\widetilde{U^{(\alpha,\beta)}})^\dag= \Pi^{{\lt}}_\alpha \otimes \Pi^{{\rt}}_\beta. 
\end{aligned}
\label{eq:unitarityspacered}
\ee
Therefore, each one of the blocks on the r.h.s.\ of Eq.~\eqref{eq:DUdecomposiiton} is a \emph{dual-unitary transformation}. This provides a full characterisation of $\mathcal G$-symmetric dual-unitary circuits: they are circuits in which the gate is decomposed in a number of dual unitary blocks. The precise decomposition depends on the soliton content of the circuit, and the local spaces on which the blocks act can have different local dimensions ($d_\alpha^{{\rt}}\neq d_\alpha^{\lt}$). 

Before proceeding with the study of the properties of these circuits, in the two upcoming subsections we present an example of explicit parameterisation of such dual-unitary transformations and introduce a useful diagrammatic representation that will be extensively used in the rest of the paper.

\subsection{Explicit parameterisation of dual-unitary transformations}

A simple class of dual-unitary transformations can be constructed following Ref.~\cite{prosen2021many, borsi2022construction}. Namely, one can set 
\begin{align}
U \left(\ket{a}_{{\rt}}\otimes\ket{b}_{\lt}\right)\equiv  u^{(a)}\ket{b}_{\ltb}\otimes v\ket{a}_{{\rtb}}, 
\label{eq:example2}
\end{align}
where $\ket{a}_\rtb$ and $\ket{b}_\ltb$ are bases of $\mathbb C^{d_\alpha^{{\rt}}}$ and $\mathbb C^{d_\beta^{\lt}}$,  $\{u^{(1)}, \ldots, u^{(d_\alpha^{{\rt}})}\} \in U(d^{{\rt}}_\alpha)$, and $v\in U(d^{\lt}_\beta)$. It is immediate to see that the map in Eq.~\eqref{eq:example2} is of the type in Eq.~\eqref{eq:map} and fulfils both Eqs.~\eqref{eq:unitaritytimered} and \eqref{eq:unitarityspacered}. Although this parameterisation is not exhaustive, for generic choices of $\{u^{(a)}\}$ and $v$ the resulting gate will not admit non-trivial solitons nor will fulfil any Yang--Baxter integrability condition. We will assume such irreducibility throughout this work.


\subsection{Diagrammatic representation}	
\label{sec:diagrams}
The study of quantum circuits is significantly simplified by the introduction of the following diagrammatic representation borrowed from the theory of tensor networks~\cite{cirac2021matrix}. The matrix elements of the local gate $U$ and its complex conjugate are represented as 	
\begin{align}
	\mel{kl}{U}{ij}=\fineq[-0.8ex][1][1]{
		\tsfmatV[0][-0.5][r][1][][][bertinired]
		\node at (-0.5,-0.2) {$i$};
		\node at (0.5,-0.15) {$j$};
		\node at (-0.5,1.2) {$k$};
		\node at (0.5,1.25) {$l$};},\quad \mel{kl}{U^*}{ij}=\fineq[-0.8ex][1][1]{
		\tsfmatV[0][-0.5][r][1][][][bertiniblue]		\node at (-0.5,-0.2) {$i$};
		\node at (0.5,-0.15) {$j$};
		\node at (-0.5,1.2) {$k$};
		\node at (0.5,1.25) {$l$};}.\label{eq:singlegate}
\end{align}
Matrix multiplication is represented connecting the legs of indices summed over and acts bottom to top. For example, the unitary conditions for $U$ and $\tilde U$ (see Eq.~\eqref{eq:DU}) are expressed as 	
\be
\begin{aligned}
\fineq[-0.8ex][1][1]{
\tsfmatV[0][1.25][r][1][][][bertinired][topright]
\tsfmatV[0][0][r][1][][][bertiniblue][bottomright]
\draw[ thick] (-0.5,1.5) -- (-0.5,1.75);
\draw[ thick] (0.5,1.5) -- (0.5,1.75);}
& = 
\fineq[-0.8ex][1][1]{
\tsfmatV[0][1.25][r][1][][][bertiniblue][bottomright]
\tsfmatV[0][0][r][1][][][bertinired][topright]
\draw[ thick] (-0.5,1.5) -- (-0.5,1.75);
\draw[ thick] (0.5,1.5) -- (0.5,1.75);}=
		\fineq[-0.8ex][.65][1]{
			\draw[ thick] (-0.45,0.15) -- (-0.45,2.35);
			\draw[ thick] (0.45,0.15) -- (0.45,2.35);
		}\,, \\	
\fineq[-0.8ex][1][1]{
		\tsfmatV[0][0][r][1][][][bertinired][topright]
		\tsfmatV[1.25][0][r][1][][][bertiniblue][topleft]
		\draw[ thick] (.5,1.5) -- (.75,1.5);
		\draw[ thick] (.5,0.5) -- (.75,0.5);
	}
	& = 
	\fineq[-0.8ex][1][1]{
		\tsfmatV[0][0][r][1][][][bertiniblue][topleft]
		\tsfmatV[1.25][0][r][1][][][bertinired][topright]
		\draw[ thick] (.5,1.5) -- (.75,1.5);
		\draw[ thick] (.5,0.5) -- (.75,0.5);
	}
	=
	\fineq[-0.8ex][.65][1]{
		\draw[ thick] (0,1.5) -- (1,1.5);
		\draw[ thick] (0,0.5) -- (1,0.5);}\,,
\end{aligned}
\label{eq:dualunitaritynonfolded}
\ee
where we removed the indices to represent the full matrix, rather than its elements. Also note that in the second line we rotated two of the diagrams to arrange them along a horizontal line. 

Many body operators are obtained by composing the above building blocks via the above rules. For example, the time evolution operator  in Eq.~\eqref{eq:U} for a system of $2L=10$ qudits is represented as
\begin{align}
	\mathds{U}=\fineq[-0.8ex][.65][1]{	
		\foreach \i in {0,...,4}
		{
			\roundgate[2*\i][0][1][topright][bertinired][-1]
			\roundgate[2*\i+1][1][1][topright][bertinired][-1]
		}	
	}.
\end{align}

We also introduce the so called ``folded representation'', which helps representing multi-replica quantities needed, e.g., for entanglement calculations. To this end we introduce symbols for the replicated local gate
\begin{align}
\left(U\otimes U^*\right)^{\otimes n}=\fineq{
\foreach \i in {0,...,-2}
{\roundgate[\i*0.3][\i*0.15][1][topright][bertiniblue]
\roundgate[\i*0.3-.15][\i*0.15-.075][1][topright][bertinired]
}
\draw[decorate, decoration={brace}] (-1.5,.2)--++(1,0.5);
\node[scale=0.7,rotate=30] at (-1.1,.7) {$2n$ gates};}=\fineq{\roundgate[0][0][1][topright][bertiniorange][n]},\label{eq:foldedgatepicture}
\end{align}
and the replicated solitons
\be
\bigotimes_{i=1}^n(q^{{\rt}}_{i}  \otimes \mathds{1}_d) = \fineq[-0.8ex][1.4][1]{	
			\draw[thick] (0,-.5)--++(0.5,0);
			\charge[0.25][-.5][red]}, \quad \bigotimes_{i=1}^n(q^{\lt}_{i} \otimes \mathds{1}_d) =  \fineq[-0.8ex][1.4][1]{	
			\draw[thick] (0,.5)--++(.5,0);
			\charge[0.25][.5][blue]},
\ee
where we used the symbol $q_{i}$ for the solition in each replica $i$ as it can be any linear combination of the basis of $\Pi_{\alpha}$ (we will specify to what it refers in each case) and we assumed to have in general different solitons on each replica (also this will be specified in each case).

To represent contractions between replicas it is useful to introduce symbols for two special states in the multi replica space. The first one is the ``circle state''
\begin{align}
\ket{\mcirc_{d, n}}=\left(\sum_{i=1}^d \ket{i,i}\right)^{\otimes n}\equiv\fineq{\draw[thick] (0,0)--++(-.5,0);\cstate[0][0]}\,.
\label{eq:circlestate}
\end{align}
The second one is the ``square state'' $\ket{\msqr_{d,n}}$. Labelling the $2n$ replicas as $(1,1^*, 2, 2^*,\ldots, n, n^*)$ we can represent it as 
\begin{align}
	\ket{\msqr_{d, n}}=\bigotimes_{a=1}^n \left( \sum_{i=1}^d \ket{i}_a\ket{i}_{(a+1)^*}\right)\equiv \fineq{\draw[thick] (0,0)--++(-.5,0);\sqrstate[0][0]}\,,
\label{eq:squarestate}
\end{align}
where the replica indexes are understood to be cyclic, so that $n+1\equiv 1$. 

In the folded representation the unitarity conditions for $U$ and $\tilde U$ are represented as
\begin{align}
	\fineq[-0.8ex][1][1]{
	\roundgate[1][1][1][topright][bertiniorange][n]
	\cstate[1.5][1.5] 
	\cstate[.5][1.5] 
}&=
\fineq[-0.8ex][1][1]{
	\draw (.5,1.5)--++(0,-1);
	\draw (1.5,1.5)--++(0,-1);
	\cstate[1.5][1.5] 
	\cstate[.5][1.5] },\qquad
\fineq[-0.8ex][1][1]{
	\roundgate[1][1][1][topright][bertiniorange][n]
	\cstate[1.5][.5] 
	\cstate[.5][.5] }=\fineq[-0.8ex][1][1]{
	\draw (.5,1.5)--++(0,-1);
	\draw (1.5,1.5)--++(0,-1);
	\cstate[.5][.5] 
	\cstate[1.5][.5] },\label{eq:unitarityfoldeddiagram}\\
\fineq[-0.8ex][1][1]{
	\roundgate[1][1][1][topright][bertiniorange][n]
	\cstate[1.5][1.5] 
	\cstate[1.5][.5] 
}&=
\fineq[-0.8ex][1][1]{
	\draw (1.5,1.5)--++(-1,0);
	\draw (1.5,.5)--++(-1,0);
	\cstate[1.5][1.5] 
	\cstate[1.5][.5] },\qquad
\fineq[-0.8ex][1][1]{
	\roundgate[1][1][1][topright][bertiniorange][n]		
	\cstate[.5][1.5] 
	\cstate[.5][.5] }=\fineq[-0.8ex][1][1]{
	\draw (1.5,.5)--++(-1,0);
	\draw (1.5,1.5)--++(-1,0);
	\cstate[.5][1.5] 
	\cstate[.5][.5] },
	\label{eq:DUgraph}
\end{align}
\begin{align}
	\fineq[-0.8ex][1][1]{
		\roundgate[1][1][1][topright][bertiniorange][n]
		\sqrstate[1.5][1.5] 
		\sqrstate[.5][1.5] 
	}&=
	\fineq[-0.8ex][1][1]{
		\draw (.5,1.5)--++(0,-1);
		\draw (1.5,1.5)--++(0,-1);
		\sqrstate[1.5][1.5] 
		\sqrstate[.5][1.5] },\qquad
	\fineq[-0.8ex][1][1]{
		\roundgate[1][1][1][topright][bertiniorange][n]
		\sqrstate[1.5][.5] 
		\sqrstate[.5][.5] }=\fineq[-0.8ex][1][1]{
		\draw (.5,1.5)--++(0,-1);
		\draw (1.5,1.5)--++(0,-1);
		\sqrstate[.5][.5] 
		\sqrstate[1.5][.5] },\label{eq:unitarityfoldeddiagramsquare}\\
	\fineq[-0.8ex][1][1]{
		\roundgate[1][1][1][topright][bertiniorange][n]
		\sqrstate[1.5][1.5] 
		\sqrstate[1.5][.5] 
	}&=
	\fineq[-0.8ex][1][1]{
		\draw (1.5,1.5)--++(-1,0);
		\draw (1.5,.5)--++(-1,0);
		\sqrstate[1.5][1.5] 
		\sqrstate[1.5][.5] },\qquad
	\fineq[-0.8ex][1][1]{
		\roundgate[1][1][1][topright][bertiniorange][n]		
		\sqrstate[.5][1.5] 
		\sqrstate[.5][.5] }=\fineq[-0.8ex][1][1]{
		\draw (1.5,.5)--++(-1,0);
		\draw (1.5,1.5)--++(-1,0);
		\sqrstate[.5][1.5] 
		\sqrstate[.5][.5] },
	\label{eq:DUgraphsquare}
\end{align}
while, e.g., the soliton ballistic propagation can be expressed as
\begin{align}
\fineq[-0.8ex][1][1]{
\roundgate[0][0][1][topright][bertiniorange][n]
					\charge[-.35][-.35][red]
				}=\fineq[-0.8ex][1][1]{
				\roundgate[0][0][1][topright][bertiniorange][n]
				\charge[.35][.35][red]
			}, \qquad \quad\fineq[-0.8ex][1][1]{
					\roundgate[0][0][1][topright][bertiniorange][n]
					\charge[.35][-.35][blue]
				}=\fineq[-0.8ex][1][1]{
				\roundgate[0][0][1][topright][bertiniorange][n]
				\charge[-.35][.35][blue]
			}.
\label{eq:Picharges}
\end{align}

\section{Charged solvable states}
\label{sec:solvablestates}

A key concept in the study of the dynamics of dual-unitary circuits is that of \emph{solvable} states~\cite{bertini2019entanglement, piroli2020exact}. This is a class of non-equilibrium states compatible with the dual unitarity condition and allowing for exact calculations of dynamical properties. In particular, all solvable states can be shown to relax to the infinite temperature state~\cite{piroli2020exact}. Here we extend this concept to the case of $U(1)$ symmetric dual-unitary circuits. We show that, because of the block structure of the local gate (see Eq.~\eqref{eq:DUdecomposiiton}), one can introduce a larger family of solvable states, which we dub \emph{charged solvable} states. In Sec.~\ref{sec:solvablestatesquench} we show that charged solvable states generically relax to non-trivial generalised Gibbs ensembles.

	We begin by considering states in matrix product state (MPS) form, defined in terms of $d^2$, $\chi\times \chi$ matrices $\mathcal{M}^{i,j}$ as
\begin{align}
\!\!\ket{\Psi_0(\mathcal{M})}	\!=\!\!\!\!\!\! \sum_{i_1,i_2,\ldots=1}^d \!\!\!\!\tra{\mathcal{M}^{i_1,i_2}\mathcal{M}^{i_3,i_4}\!\!\!\!\ldots}\!\!\ket{i_1,i_2,\ldots, i_{2L}}.
\label{eq:twositeMPS}
\end{align}

The tensor generating the MPS has a convenient graphical representation 
\begin{align}
\fineq{\MPSinitialstate[0][0][bertiniorange][topright][n]}=\left(\mathcal{M}\otimes\mathcal{M}^*\right)^{\otimes n},
\label{eq:MPSstatedef}
\end{align}
where the thick line represents the replicated auxiliary space, i.e. the vector space of dimension $\chi$ on which the  matrices $\mathcal{M}^{i,j}$ act, while the thin ones represent the replicated physical space. It is useful to reshuffle the indices of $\mathcal{M}^{i,j}$ and view them as matrices $\Gamma_\mathcal{M}$ acting on $\mathds{C}^d\otimes\mathds{C}^\chi$ (whose input/output indices are the right/left legs of the tensor in Eq.~\eqref{eq:MPSstatedef})
\begin{align}
\Gamma_\mathcal{M}=\sum_{i,j=1}^d\ketbra{i}{j}\otimes \mathcal{M}^{ij}, 
\end{align}
and to define the transfer matrix of the MPS 
\begin{align}
		\tau(\mathcal{M})=\sum_{i,j} \mathcal{M}^{i,j} \otimes \left(\mathcal{M}^{i,j}\right)^*=		\fineq{				\MPSinitialstate[0][0][bertiniorange][topright][1]
		\cstate[.5][.5]\cstate[-.5][.5]}\,.
\label{eq:transfermatrix}
\end{align}

Having established the notation we are now in a position to introduce charged solvable states
\begin{definition}[Charged Solvable States]
\label{def:chargedsolvability}
An MPS $\ket{\Psi_0(\mathcal{M})}$ is \emph{left charged solvable} if the transfer matrix $\tau(\mathcal{M})$ (see Eq.~\eqref{eq:transfermatrix}) has a unique maximal eigenvalue with absolute value $1$ and there exists an operator $S$ acting on the auxiliary space $\mathds{C}^\chi$ such that
\be
\Gamma^\dagger_{\mathcal{M}} (\mathds{1}_d \otimes S) \Gamma_{\mathcal{M}}= \sum_\alpha \frac{c^{{\rt}}_\alpha}{d_\alpha^{{\rt}}} \Pi^{{\rt}}_{\alpha} \otimes S,
\label{eq:leftsolvabilityformula} 
\ee
where $d_\alpha^{{\rt}}$ is defined in Eq.~\eqref{eq:dalphalr} and we introduced  the expectation value of a soliton 
\be
c^{{\rt}}_\alpha = \expval{(\Pi^{\rt}_\alpha)_{x}}{\Psi_0(\mathcal{M})}, \qquad x\in \mathbb Z\,.  
\ee
By definition we then have $c_\alpha^{{\rt}}\ge 0$ and $\sum_{\alpha} c_\alpha^{{\rt}}=1$. Instead, $\ket{\Psi_0(\mathcal{M})}$ is \emph{right charged solvable} if the transfer matrix has a unique maximal eigenvalue and there exists a $S'$ such that
\begin{align}
\Gamma_{\mathcal{M}} (\mathds{1}_d \otimes S') \Gamma_{\mathcal{M}}^\dagger= \sum_\alpha \frac{c^{{\lt}}_\alpha}{d_\alpha^{{\lt}}} \Pi^{{\lt}}_{\alpha} \otimes S' ,	
\label{eq:rightsolvabilityformula}
\end{align} 
where 
\be
c^{{\lt}}_\alpha = \expval{(\Pi^{\lt}_\alpha)_{x+1/2}}{\Psi_0(\mathcal{M})}, \qquad x\in \mathbb Z\,.  
\ee
Finally, it is \emph{charged solvable} if it is both left and right charged solvable.
\end{definition} 
Note that these definitions also imply that for left/right charged solvable states the left/right leading eigenvectors of $\tau(\mathcal{M})$ correspond to the matrices $S$ and $S'$ in Eqs.~\eqref{eq:leftsolvabilityformula} and~\eqref{eq:rightsolvabilityformula} upon matrix to vector mapping (see Appendix~\ref{app:solvablestates})
\be
\bra{S}=\sum_{i,j=1}^d \left(S\right)_{i,j}\bra{i,j},\quad \ket{S'}=\sum_{i,j=1}^d \left(S'\right)_{i,j}\ket{i,j}.
\label{eq:vectorSSp}
\ee

Recalling that two MPS states $\ket{\Psi_0(\mathcal{M})},\ket{\Psi_0(\mathcal{M'})}$ are called \emph{equivalent} if, for every local operator $O_R$ (i.e. operator with a finite support)
\begin{align}
\lim_{L\rightarrow\infty}	\mel{\Psi_0(\mathcal{M'})}{O_R}{\Psi_0(\mathcal{M'})} =\notag\\
=\lim_{L\rightarrow\infty}	\mel{\Psi_0(\mathcal{M})}{O_R}{\Psi_0(\mathcal{M})},
\end{align}
we characterise charged solvable states by means of the following theorem (proven in Appendix~\ref{app:solvablestates}).
\begin{theorem}[Equivalent MPS]
\label{th:solvabilityclassification}
A left charged solvable MPS state is always equivalent, in the thermodynamic limit, to an MPS state $\mathcal{M'}$ with bond dimension $\chi$ such that 
\begin{align}
\Gamma^\dagger_{\mathcal{M}'} \Gamma_{\mathcal{M}'}= \sum_i \frac{c^{\rt}_\alpha}{d_\alpha^{\rt}} \Pi^{\rt}_{\alpha} \otimes \mathds{1}_\chi , 
\label{eq:leftsolv}
\end{align}
 whose unique  maximal right eigenvector of the transfer matrix $\tau(\mathcal{M}) $ is
\begin{align}
\ket{\mcirc_{\chi,1}}=\sum_{i=1}^{\chi} \ket{i} \otimes \ket{i}.
\label{eq:identitystate}
\end{align}
Analogously, a right charged solvable state is equivalent to an MPS such that  
\begin{align}
\Gamma_{\mathcal{M}'} \Gamma_{\mathcal{M}'}^\dagger= \sum_i \frac{c^{{\lt}}_\alpha}{d_\alpha^{{\lt}}} \Pi^{{\lt}}_{\alpha} \otimes \mathds{1}_\chi , 
\label{eq:rightsolv}
\end{align}
 whose transfer matrix has unique maximal left eigenvector 
\begin{align}
\bra{\mcirc_{\chi,1}}=\sum_{i=1}^{\chi} \bra{i} \otimes \bra{i}.
\end{align}
\end{theorem}
In essence, Theorem~\eqref{th:solvabilityclassification} guarantees that one can always replace left/right charged solvable states with equivalent MPS where the matrix $S$ in Definition~\ref{def:chargedsolvability} is the identity. 
	
The equivalent MPS states of Theorem~\ref{th:solvabilityclassification} can be further characterised by specialising the local basis of the quantum circuit. To show this we consider the singular value decomposition of $\Gamma_\mathcal{M}$, i.e.\ $\Gamma_\mathcal{M}= V D W$, and rewrite condition for left charged solvability \eqref{eq:leftsolv} as
\begin{align}
 \Gamma_\mathcal{M}^\dagger\Gamma_\mathcal{M}=W D^2 W^\dagger=\sum_i \frac{c_\alpha^{{\rt}}}{d_\alpha^{\rt}} \Pi_\alpha^{{\rt}}\otimes \mathds{1}_\chi. 
\label{eq:SVD_MPSstate}
\end{align}
The matrix $D^2$ is diagonal by construction, and its spectrum needs to match the one of the matrix on the right (the unitary $W$ cannot alter the spectrum). If we choose a basis for each qudit such that all projectors $\Pi^{{\rt}/\lt}_\alpha$ are diagonal (always possible as they commute), we have that the right hand side of Eq.~\eqref{eq:SVD_MPSstate} is also diagonal. In this case, upon reordering of the basis, we can take $D^2$ to be equal to the righthand side. With these choices we find
\begin{align}
		W \sum_\alpha \frac{c_\alpha^{{\rt}}}{d^{{\rt}}_\alpha} \Pi_\alpha^{{\rt}}\otimes \mathds{1}_\chi W^\dagger=\sum_\alpha \frac{c_\alpha^{{\rt}}}{d^{{\rt}}_\alpha} \Pi_\alpha^{{\rt}}\otimes\mathds{1}_\chi,
\end{align}
so we can move $W$ past $D$ in the SVD of $\Gamma_\mathcal{M}$ and include it in the definition of $V$. This means that most generic $\Gamma_\mathcal{M}$ representing a left-solvable state can be written as
\begin{align}
\Gamma_\mathcal{M}=\sum_\alpha\sqrt{\frac{c_\alpha^{{\rt}}}{d^{{\rt}}_\alpha}} V \left(\Pi_\alpha^{{\rt}}\otimes \mathds{1}_\chi\right), \qquad V\in U(\chi \cdot d),
\label{eq:leftsolvablestateconstruction}
\end{align}
and an analogous discussion holds for right charged solvable states. Finally, using this characterisation, in Appendix~\ref{app:proofsolvabilitybothsides} we prove the following compatibility condition for charged solvable states
\begin{theorem}[Compatibility Condition]
\label{th:characterizationsolvMPS}
Given the bases for left and right solitons 
\be
\mathcal S^{{\rt}} =\{\Pi^{{\rt}}_\alpha\}_{\alpha=1,\ldots,m_{\rtb}},\quad 
\mathcal S^{{\lt}} =\{\Pi^{{\lt}}_\alpha\}_{\alpha=1,\ldots,m_{\ltb}},
\ee
and a charged solvable state fulfilling Eqs.~\eqref{eq:leftsolvabilityformula}-\eqref{eq:rightsolvabilityformula}, than there exists a partition $\mathcal P^{{\rt}}=\{\mathcal P^{{\rt}}_1,\ldots,\mathcal P^{{\rt}}_{k'}\}$ of $\{1,\ldots,m_{\rtb}\}$ and $\mathcal P^{{\lt}}=\{\mathcal P^{{\lt}}_1,\ldots,\mathcal P^{{\lt}}_{k''}\}$ of $\{1,\ldots,m_{\ltb}\}$, such that $k'=k''=k$ and for all $i=1,\ldots,k$ we have 
\be  
\begin{aligned}
&\sum_{\alpha\in \mathcal P^{{\rt}}_i} d^{\rt}_\alpha=\sum_{\beta\in \mathcal P^{{\lt}}_i} d^{{{\lt}}}_\beta\equiv d_i\\
&\frac{c^{\rt}_\alpha}{d^{\rt}_\alpha}=\frac{c^{{{\lt}}}_\beta}{d^{{{\lt}}}_\beta}\equiv \frac{c_i}{d_i}, \qquad\qquad\quad \forall \alpha\in  \mathcal P^{{\rt}}_i, \beta \in  \mathcal P^{{\lt}}_i\,.
\end{aligned}
\label{eq:blockingcondition}
\ee
Moreover, the left/right fixed points of $\tau(\mathcal{M})$ can be taken to be $\bra{\mcirc_{\chi,1}}, \ket{S}$, where $S$ is a strictly positive matrix.
 \end{theorem}
The essence of this theorem is that to have a state that is charged solvable both directions, there must be a compatible blocking structure for solitons moving left and right. Namely, the two matrices 
\be
\sum_\alpha \frac{c_\alpha^{{\rt}}}{d^{{\rt}}_\alpha} \Pi_\alpha^{{\rt}}, \qquad \sum_\alpha \frac{c_\alpha^{\lt}}{d^{\lt}_\alpha} \Pi_\alpha^{\lt},
\ee
should have the same spectrum. Moreover, in contrast to what happens for the standard solvability condition (see Ref.~\cite{piroli2020exact}), it is generically not possible to choose an equivalent MPS such that \emph{both} left and right eigenvectors of $\tau(\mathcal{M})$ are the (vectorised) identity matrix.

In fact, we make two remarks concerning Theorem~\ref{th:characterizationsolvMPS}. First, we emphasise that the charged solvable states corresponding to the trivial partitions $\mathcal P^{{\rt}}=\{\{1,\ldots,m_{\rtb}\}\}$ and $\mathcal P^{{\lt}}=\{\{1,\ldots,m_{\ltb}\}\}$ (i.e.\ partitions with a single element corresponding to the full set) are those without a charge structure, and correspond to the standard solvable states. In this particular case, it can be shown that the fixed point of the transfer matrix on both sides can be taken to be $\ket{\mcirc_{\chi,1}}$~\cite{piroli2020exact}. Second, we note that the compatibility condition is relatively easy to achieve for gates that have the same dimension for every charge subspace (which also means that they have the same number of left- and right-moving solitons), i.e.\ $d_\alpha^{\rt}=d_\alpha^{\lt}=d/m_{\rtb}=d/m_{\ltb}$. In this case one just needs to match each right-moving charge with a left-moving one.

Finally, let us conclude by illustrating the results of this and the previous section in a simple example. Consider the $U(1)$ symmetric dual-unitary circuit defined for qudits of local dimension $d = 4$ and with local gate 
\be
U = \begin{pmatrix}
U_A &&&\\
&&U_B&\\
&U_C&&\\
&&&U_D 
\end{pmatrix},
\label{eq:newexample}
\ee
where $U_A, \ldots, U_D$ generic $4 \times 4$ dual-unitary matrices while all the other blocks $4 \times 4$ are 0. Despite this system is not Yang-Baxter integrable, this system admits $m_{\ltb}=2$ left moving solitons and $m_{\rtb}=2$ right moving solitons that \emph{take the same form}, i.e.   
\be
\{\Pi^{{\rt}}_1,\Pi^{{\rt}}_2\} =\{\Pi^{{\lt}}_1,\Pi^{{\lt}}_2\} = \{P , \bar P\},  
\ee
where $P= {\rm diag}(1,1,0,0)$ and $\bar P= {\rm diag}(0,0,1,1)$. In this case a simple class of solvable states are the pair product states of the form
\begin{align}
	\ket{\Psi_0}= 	\left(\sum_{i,j=1}^4 (m)_{i,j}\ket{i,j}\right)^{\otimes L},
\label{eq:stateexample}
\end{align}
where the matrix $m$ is written as 
\be
m= \sqrt{\frac{\alpha}{4}} P + \sqrt{\frac{1-\alpha}{4}} \bar P\,, \quad \alpha\in[0,1]\,.
\label{eq:stateexample2}
\ee
This construction provides a direct generalisation of the Bell pair initial state~\cite{piroli2020exact}, which is the simplest solvable state in a dual-unitary circuit. The state in Eqs.~\eqref{eq:stateexample}--\eqref{eq:stateexample2}, however, is not solvable in a non-symmetric circuit. Note that one can immediately extend this example to any ${d= 2k > 4}$ by taking $U_A, \ldots, U_D$ to be $k^2 \times k^2$ dual-unitary matrices.

\section{Entanglement dynamics from charged solvable states}
\label{sec:solvablestatesquench}

In this section, we consider the entanglement dynamics after a quantum quench from a charged solvable state. Our calculations are carried out in full generality, without specifying a particular $U(1)$ symmetric dual-unitary circuit or charged solvable state, but for concreteness one can think of the example given at the end of the previous subsection. Specifically, we characterise the growth of entanglement between a subsystem $A$, composed by $2 L_A$ contiguous qudits, and the rest of the system by computing the R\'enyi entropies
\be
\label{eq:Renyi}
S_A^{(n)}(t)=\frac{1}{1-n}\log\tra{\rho_A^n(t)}, \qquad n \in \mathbb N\,,
\ee 
where $\rho_A^n(t)$ is the state of $A$ at time $t$. Note that $\lim_{n\to1}S_A^{(n)}(t) = S_A(t)$ where $S_A(t)$ is the standard entanglement entropy. 

We split the calculation in two steps: In Sec.~\ref{Sec:initialgrowth} we consider the early time regime $t< {L_A}/{2}$ while in Sec.~\ref{Sec:GGE} the late time regime $t\geq {L_A}/{2}$. 

\subsection{Early-time Regime}
\label{Sec:initialgrowth}

Let us begin considering the early time regime, i.e., we restrict to times $t< {L_A}/{2}$. In this regime the two edges of $A$ are causally disconnected (the speed of light is one in our units) and entanglement is generated independently at both edges with equal rate. The contribution of the two edges, however, cannot be completely separated as the initial MPS can encode non-trivial correlations. 

Specifically, the diagram giving the $n$-th moment of the reduced density matrix reads as 
\begin{widetext} 
\begin{align}
\tra{\rho_A^n(t\leq {L_A}/{2})}
=\fineq[-0.8ex][.65][1]{
	\draw[very thick,dashed] (-1,-1.05)--++(20,0);
	\foreach \i in {0,...,8}
	{\MPSinitialstate[2*\i+1][-1][bertiniorange][topright][n]
					\roundgate[\i*2][0][1][topright][bertiniorange][n]				\roundgate[\i*2+1][1][1][topright][bertiniorange][n]
	}
	\foreach \i in {0,...,4}
	{
		\cstate[\i+.5][1.5]
	}
	\foreach \i in {5,...,12}
	{
		\sqrstate[\i+.5][1.5]
	}
	\foreach \i in {13,...,17}
	{
		\cstate[\i+.5][1.5]
	}
	\draw [decorate, decoration = {brace,mirror}]   (13,2)--++(-8,0) ;
	\node[scale=1.25] at (9,2.5) {$2L_A$};
}&\notag\\
=\fineq[-0.8ex][.7][1]{			
\draw[very thick,dashed] (-7,-1.05)--++(20,0);
\foreach \i in {-3,...,5}{
\MPSinitialstate[2*\i+1][-1][bertiniorange][topright][n]}
\trianglediag[0][0][1][bertiniorange][MPS][n][d]	
\trianglediag[8][0][1][bertiniorange][MPS][n][d]	
\foreach \i in {2,...,4}{
\cstate[-\i+2.5][-\i+3.5]	
\sqrstate[\i-.5][-\i+3.5]	
\cstate[\i+7.5][-\i+3.5]	
\sqrstate[-\i+10.5][-\i+3.5]}
\sqrstate[5.5][-.5]	
\sqrstate[4.5][-.5]	
\cstate[-2.5][-.5]	
\cstate[-3.5][-.5]	
\cstate[-4.5][-.5]	
\cstate[-5.5][-.5]	
\draw [decorate, decoration = {brace}]   (6,2)--++(3.5,0) ;
\node[scale=1.25] at (7.8,2.6) {$2t$};
	\draw [decorate, decoration = {brace}]   (4.15,2)--++(1.7,0) ;
\node[scale=1.25] at (5,2.6) {$2L_A-4t$};
\draw [decorate, decoration = {brace}]   (-6.,2)--++(4,0);
\node[scale=1.25] at (-4,2.6) {$2L-4t-2L_A$};
}&,
\label{eq:rdmentgrowth}
\end{align}
where, in going from the first to the second line, we repeatedly used the unitarity conditions in Eqs.~\eqref{eq:unitarityfoldeddiagram} and \eqref{eq:unitarityfoldeddiagramsquare}. 

To proceed, we note that on the left of this diagram we have $L-2t-L_A$ powers of the MPS transfer matrix $\tau(\mathcal{M})$ (see Eq.~\eqref{eq:transfermatrix}) replicated $n$ times, i.e., 
\be
(\tau(\mathcal{M})^{\otimes n})^{L-2t-L_A}. 
\label{eq:replicatedTMLpower}
\ee
As we now show this allows us to compute the thermodynamic limit value of this diagram whenever the initial MPS is charged solvable. 

By definition, a charged solvable MPS has a transfer matrix $\tau(\mathcal{M})$ with unique fixed points given by $\ket{S}$ and $\bra{\mcirc_{\chi,1}}$ (see Definition~\ref{def:chargedsolvability}), Eq.~\eqref{eq:vectorSSp}, and Theorem~\ref{th:solvabilityclassification}), where we choose the left fixed point to be the identity. This means that
\be
\lim_{L \rightarrow \infty }	\left(\fineq[-0.8ex][1][1]{\MPSinitialstate[0][0][bertiniorange][topright][n]
\cstate[-.5][.5]\cstate[.5][.5]}\right)^{L-2t-L_A} =\fineq[-2ex][1][1]{\phantom{\MPSinitialstate[0][0][bertiniorange][topright][n]}	
\draw[very thick] (-1,-.05)--++(.5,0);
\draw[very thick] (.5,-.05)--++(-.5,0);
\cstate[-.5][-.05][][black]
\cstate[0][-.05][][white]}\label{eq:leadingMPSTmat}
\ee
where the black circle denotes  
\be
\fineq[-0.8ex][1][1]
{
	\draw[very thick] (.5,-.05)--++(-.5,0);
	\cstate[0.5][-.05][][black]
}=	\prod_{a=1}^n \left[\sum_{i,j=1}^d \left(S\right)_{i,j}\ket{i}_a\ket{j}_{(a)^*}\right].
\ee
It is then immediate to see that, in the thermodynamic limit, the diagram in Eq.~\eqref{eq:rdmentgrowth} is reduced to the following 
\begin{align}
\fineq[-0.8ex][.7][1]{			
	\foreach \i in {-1,...,5}
	{
		\MPSinitialstate[2*\i+1][-1][bertiniorange][topright][n]
	}
	\trianglediag[0][0][1][bertiniorange][MPS][n][d]	
	\trianglediag[8][0][1][bertiniorange][MPS][n][d]	
	\foreach \i in {2,...,4}{
		\cstate[-\i+2.5][-\i+3.5]	
		\sqrstate[\i-.5][-\i+3.5]	
		\cstate[\i+7.5][-\i+3.5]	
		\sqrstate[-\i+10.5][-\i+3.5]}
	\sqrstate[5.5][-.5]	
	\sqrstate[4.5][-.5]	
	\cstate[-2.][-1.05]	
	\cstate[12][-1.05][][black]	
}. 
\label{eq:rdmentgrowthTL}	
\end{align}
This diagram can be fully contracted using the charged solvability conditions \eqref{eq:leftsolv} and \eqref{eq:rightsolvabilityformula}. Indeed the latter implies the following diagrammatic relations for the replicated state
\begin{align}
	\fineq{				
		\MPSinitialstate[0][0][bertiniorange][topright][n]
		\cstate[-1][-.05][][white]
		\cstate[-.5][.5]
	}=\fineq{		
		\draw[thick] (0,0)--++(1,0);		
		\draw[thick] (0.5,.5)--++(.5,.5);		
		\charge[0.75][.75][red]
		\cstate[0][0][][white]
		\cstate[.5][.5]
	},\label{eq:leftsolvgraph}
	\\
	\fineq{				
		\MPSinitialstate[0][0][bertiniorange][topright][n]
		\cstate[1][-.05][][black]
		\cstate[.5][.5]
	}=\fineq{		
		\draw[thick] (0,0)--++(-1,0);		
		\draw[thick] (-0.5,.5)--++(-.5,.5);		
		\charge[-0.75][.75][blue]
		\cstate[0][0][][black]
		\cstate[-.5][.5]},
		\label{eq:rightsolvgraph}
\end{align}
where the solitons represented in red/blue correspond to the following linear combination of projectors
\begin{align}
 	\left(\sum_{\alpha=1}^{m_{\rtb/\ltb}} \frac{c^{{{\rt/\lt}}}_\alpha}{d^{{{\rt/\lt}}}_\alpha}\Pi^{{{\rt/\lt}}}_\alpha\right).
\end{align}
We can move each soliton to the top, using \eqref{eq:Picharges}, obtaining 
\begin{align}
	\fineq[-0.8ex][.7][1]{			
		\foreach \i in {0,...,4}
		{
			\MPSinitialstate[2*\i+1][-1][bertiniorange][topright][n]
		}
		\trianglediag[2][0][0][bertiniorange][MPS][n][d]	
		\trianglediag[8][0][0][bertiniorange][MPS][n][d]	
		\roundgate[1][1][1][topright][bertiniorange][n]				
		\roundgate[0][0][1][topright][bertiniorange][n]				
		\roundgate[9][1][1][topright][bertiniorange][n]				
		\roundgate[10][0][1][topright][bertiniorange][n]				
		\foreach \i in {2,...,3}{
			\cstate[-\i+2.5][-\i+3.5]		
			\cstate[\i+7.5][-\i+3.5]	}
		\foreach \i in {2,...,4}{
		\sqrstate[\i-.5][-\i+3.5]	
		\sqrstate[-\i+10.5][-\i+3.5]}
		\sqrstate[5.5][-.5]	
		\sqrstate[4.5][-.5]	
		\cstate[0.][-1.05]	
		\cstate[10][-1.05][][black]	
		\cstate[-0.5][-.5]	
	\cstate[10.5][-.5]
	\charge[10.3-1.6][-.3+1.6][blue]	
	\charge[-0.5+1.8][-.5+1.8][red]	
	\draw [decorate, decoration = {brace,mirror}]   (8.5,2)--++(-7,0) ;
	\node[scale=1.5] at (5,2.5) {$2 L_A$ sites};
	},
\end{align}
\end{widetext}
and then use dual unitarity \ref{eq:DUgraph} to simplify both outer diagonals of gates, and reiterate the procedure until the triangle at both edges are fully simplified.
The final result reads as 
\begin{align}
	\tra{\rho_A^n(t)}= \left(\sum_\alpha \frac{{c^{\rt}_\alpha}^n}{{d^{\rt}_\alpha}^{n-1}}\right)^{4t} \Xi_n(L_A-2t),
\end{align}
where the extra contribution $\Xi_n(x)$ comes from the initial MPS entanglement, and ${\ln(\Xi_n(x))}/{(1-n)}$ can be thought as the R\'enyi entropy, on the initial state, of an interval including $2x$ sites . Namely, it is written as  
\begin{align}
	\Xi_n(x)=\fineq[-1.6ex][1][1]{
	\def \size {2}
	\draw[very thick] (0,0)--++(.5,0);
	\foreach \i in {1,...,\size}
	{
		\MPSinitialstate[2*\i-1][0.05][bertiniorange][topright][n]
		\sqrstate[2*\i-.5][.55]
		\sqrstate[2*\i-1.5][.55]
	}
	\draw[decorate,decoration={brace}] (0,.75)--++(4,0);
	\node[scale=1] at (1.75,1) {$x$};
	\cstate[0][0]
	\cstate[2*\size][0][][black]
}.
\end{align}
 Noting that this term is $O(t^0)$ we then have
\begin{align}
	{S^{(n)}_A(t)}=\frac{4t}{1-n}\log(\sum_{i=1}^k \frac{c_i^n}{d_i^{n-1}})+o(t),
\label{eq:entgrowthsolvable}
\end{align}
where we used the block matching condition in Eq.~\eqref{eq:blockingcondition} to eliminate the labels $\ltb/\rtb$ from $c_i$ and $d_i$ (and $k$ is defined above Eq.~\eqref{eq:blockingcondition}). Indeed, using the latter condition one has 
\begin{align}
&\sum_\alpha \frac{{c_\alpha^{\rt}}^n}{{d^{\rt}_\alpha}^{n-1}}= \sum_{i=1}^k \sum_{\alpha\in P^{{\rt}}_i}\frac{{c_\alpha^{\rt}}^n}{{d^{\rt}_\alpha}^{n-1}}=\sum_{i=1}^k \frac{c_i^n}{d_i^n} \sum_{\alpha\in P^{{\rt}}_i}d_\alpha\notag\\
&=\sum_{i=1}^k \frac{ c_i^n}{d_i^{n-1}}=\sum_\alpha \frac{{c_\alpha^{{{\lt}}}}^n}{{d^{{{\lt}}}_\alpha}^{n-1}}\,.
\end{align}
In particular, in the limit $n\rightarrow1$ we obtain 
\be
{S_A(t)}= 4t (s_{\rm num}+s_{\rm conf}) + o(t),
\label{eq:entropygrowthexpressionsolvable}
\ee
where introduced the quantities 
\be
s_{\rm num}\equiv -\sum_{i=1}^k  c_i\log(c_i),\quad s_{\rm conf}\equiv \sum_{i=1}^k c_i \log(d_i), 
\label{eq:snumsconf}
\ee
whose physical meaning will become clear in the upcoming subsection.

\subsection{Late-time Regime}
\label{Sec:GGE}
We now consider times $t\ge L_A/2$. In this regime, after taking the thermodynamic limit, the reduced density matrix $\rho_A(t)$ has the following graphical representation
\be
\rho_A(t) =	\hspace{-1cm}\fineq[-0.8ex][0.50001][1]{
	\draw [decorate, decoration = {brace,mirror}]   (6.5,5)--++(-3,0) ;
	\node[scale=2] at (5,5.5) {$2 L_A$ sites};
			\trianglediag[0][0][4][bertiniorange][MPS][1][1]
			\trianglediag[2][0][4][bertiniorange][MPS][1][1]
			\foreach \i in {0,...,4}
			{
				\cstate[\i-1.5][\i-.5]
				\cstate[-\i+11.5][\i-.5]
			}
			\cstate[-2][-1.05][][white]
			\cstate[12][-1.05][][black]
			\draw[thick,color=red,dashed] (2.5,4.5)--++(2.5,-2.5)--++(2.5,2.5);}.
\label{eq:rdm}
\ee

Making now repeated use of the charged solvability conditions in Eqs.~\eqref{eq:leftsolvgraph} and \eqref{eq:rightsolvgraph}, the soliton relation \eqref{eq:Picharges} (to move solitons at the top open legs) and dual unitarity  in Eq.~\eqref{eq:DUgraph}, this diagram can be simplified as follows  
\begin{align}
\rho_A(t) =\!\!\!\!\!\!\!\!\!\!\!\fineq[-0.8ex][0.50001][1]{
					\roundgate[0][0][1][topright][bertiniorange][1]
					\roundgate[-1][1][1][topright][bertiniorange][1]
					\roundgate[1][1][1][topright][bertiniorange][1]
					\draw[thick,color=red,dashed] (-2.5,1.5)--++(2.5,-2.5)--++(2.5,2.5);
					\foreach \i in {0,1}
					{
						\charge[1.35-2*\i][1.35][red]
						\charge[0.65-2*\i][1.35][blue]			
					\cstate[-.5-\i][-.5+\i]
					\cstate[+.5+\i][-.5+\i]
					}
\draw [decorate, decoration = {brace,mirror}]   (1.5,1.65)--++(-3,0) ;
\node[scale=2] at (0,2.25) {$2 L_A$ sites};
}	
=\fineq[-0.8ex][1][1]{
\foreach \i in {0,1}{
				\draw[thick](4.+2*\i,4)--++(0,.5);
				\charge[4.+2*\i][4.3][blue]
				\cstate[4.+2*\i][4]
				\draw[thick](5+2*\i,4)--++(0,.5);
				\charge[5+2*\i][4.3][red]
				\cstate[5+2*\i][4]
			}
		},	\label{eq:GGE}
\end{align}

where in the second step we used the unitarity condition in Eq.~\eqref{eq:Picharges} to fully simplify the diagram. Note that this can only be done if the dashed red line in \eqref{eq:GGE} does not cross the bottom line of initial states, that is for $t\ge L_A/2$.
Using the simplified form in Eq.~\eqref{eq:GGE}, we can write explicit expressions for  all R\'enyi entropies of $A$. The result reads as   
\begin{align}
		S^{(n)}_A(t)=\frac{2 L_A}{1-n}\log(\sum_{i=1}^{k} \frac{\left(c_i\right)^n}{\left(d_i\right)^{n-1}}).\label{eq:ententropysaturation}
\end{align}
In the limit $n\to1$ the result can be written as  	
\begin{align}
S_A(t)={2 L_A} (s_{\rm num}+s_{\rm conf})\,,
\label{eq:VNexpressionsolvable}
\end{align}
where $s_{\rm num}$ and $s_{\rm conf}$ are defined in Eq.~\eqref{eq:snumsconf}. To unveil the physical meaning of these quantities we introduce configurational, $S_{\rm conf}(t)$, and number, $S_{\rm num}(t)$, entropies defined as~\cite{lukin2019probing, parez2021quasiparticle, parez2021exact} 
\be
\begin{aligned}
\!\!\!\!\!\!S_{\rm conf}(t) &\!\equiv \!- \!\!\sum_{\boldsymbol \alpha, \boldsymbol \beta}p_{\boldsymbol \alpha, \boldsymbol \beta}(t)  (\rho_{A,\boldsymbol \alpha, \boldsymbol \beta}(t) \log \rho_{A,\boldsymbol \alpha, \boldsymbol \beta}(t) ),\\
\!\!\!\!\!\!S_{\rm num}(t) &\!\equiv\!-\!\!\sum_{\boldsymbol \alpha, \boldsymbol \beta} p_{\boldsymbol \alpha, \boldsymbol \beta}(t)  \log p_{\boldsymbol \alpha, \boldsymbol \beta}(t),
\end{aligned}
\ee
where we introduced the state reduced to a symmetry block  
\be
\begin{aligned}
\rho_{A,\boldsymbol \alpha, \boldsymbol \beta}(t) &\equiv  \frac{\Pi^{{\rt}}_{\boldsymbol {\alpha}}\Pi^{{\lt}}_{\boldsymbol {\beta}} \rho_A(t) \Pi^{{\rt}}_{\boldsymbol {\alpha}}\Pi^{{\lt}}_{\boldsymbol {\beta}}}{p_{\boldsymbol \alpha, \boldsymbol \beta}(t) },\\ p_{\boldsymbol \alpha, \boldsymbol \beta}(t) &\equiv \tr[\Pi^{{\rt}}_{\boldsymbol {\alpha}}\Pi^{{\lt}}_{\boldsymbol {\beta}} \rho_A(t) \Pi^{{\rt}}_{\boldsymbol {\alpha}}\Pi^{{\lt}}_{\boldsymbol {\beta}}],
\end{aligned}
\label{eq:projectedstate}
\ee
with the projectors in the charge sector with $\boldsymbol \alpha = (\alpha_1, \ldots, \alpha_{m_\rtb})$ right moving solitons and $\boldsymbol \beta = (\beta_1, \ldots, \beta_{m_\ltb})$ left moving solitons defined as 
\be
\Pi^{{\rt}}_{\boldsymbol {\alpha}}=\bigotimes_{i=1}^{L_A} \Pi^{{\rt}}_{\alpha_i} \otimes \mathds{1}_d,\qquad
\Pi^{{\lt}}_{\boldsymbol {\beta}}=\bigotimes_{i=1}^{L_A}  \mathds{1}_d\otimes \Pi^{{\lt}}_{\beta_i}.
\label{eq:projectors}
\ee
 From their definition, we see that $S_{\rm conf}(t)$ measures the average of the entanglement in the charge sectors, while $S_{\rm num}(t)$ the fluctuations of the charge~\cite{lukin2019probing, parez2021quasiparticle, parez2021exact}. Using Eq.~\eqref{eq:GGE} one can readily show 
\be
S_{\rm conf}(\infty) = 2 L_A s_{\rm conf}, \qquad S_{\rm num}(\infty) = 2 L_A s_{\rm num}.
\label{eq:relationwithnumberandconf}
\ee
Therefore we see that $s_{\rm num}$ and $s_{\rm conf}$ are respectively the densities of number and configurational entropy in the stationary state. 

Three comments are in order at this point. First, we see that in our case the number entropy is extensive, as opposed to the $\log L_A$ scaling that it displays in the generic situation~\cite{parez2021exact, bertini2022nonequilibrium}. This is because in our setting the number of non-trivial charge sectors is exponentially large in $L_A$, which should be contrasted with the polynomial scaling of the number of charge sectors with $L_A$ that one has in the generic case. Second, even though Eqs.~\eqref{eq:VNexpressionsolvable} and \eqref{eq:relationwithnumberandconf} imply that the entanglement entropy at infinite times is the sum of number and configurational entropy, this does not mean that such a decomposition holds for all times. Indeed, from Eq.~\eqref{eq:entropygrowthexpressionsolvable} one generically has 
\be
{S_A(t)} \geq S_{\rm conf}(t) + S_{\rm num}(t). 
\ee
The equality holds only if the initial state does not have initial entanglement (which corresponds to the case $\chi=1$). Third, we stress that Eq.~\eqref{eq:entgrowthsolvable} implies linear growth of \emph{all} R\'enyi entropies at early times. This includes also ${S^{(n>1)}_A(t)}$, which have been found to grow sub-ballistically in generic quantum circuits with conservation laws~\cite{rakovszky2019sub, huang2020dynamics, znidaric2020entanglement}. Our result is not in contradiction with the aforementioned references, however, because dual-unitary circuits lack a key ingredient for their arguments to apply, i.e., diffusive charge transport: as discussed in Sec.~\ref{sec:characterisation} the charge transport in dual unitary circuits is always ballistic. In fact, as we show in the next section, the linear growth of R\'enyi entropies persists also for non-charged solvable states.

\section{Entanglement dynamics from non-charged-solvable states}
\label{sec:genericinitialstates}

In this section we look again at the entanglement dynamics after a quench from an MPS state but, crucially, we lift the assumption of full charged solvability. Interestingly, we find that in this case the dynamics of entanglement shows a qualitative change: while it agrees with the charged solvable state result in the early time regime, it does not show immediate saturation for $t\geq L_A/2$. Instead, it continues to grow with a different slope set by the value of the number entropy. 

To deal with the great complication of characterising the growth of entanglement without any form of solvability we attack the problem in a gradual fashion. First, we lift charged solvability only on one side. In this case the phenomenology described above can be rigorously shown under a genericity assumption on the blocks $U^{(\alpha,\beta)}$ in Eq.~\eqref{eq:DUdecomposiiton}. Then, we move to consider fully generic initial states and adapt the entanglement membrane approach~\cite{jonay2018coarsegrained, zhou2020entanglement} to argue that the above features are stable.

\subsection{Entanglement dynamics from left charged solvable states}

Here we consider the entanglement dynamics from MPS states \eqref{eq:twositeMPS} that are solvable only on one side: for definiteness we choose left solvability. 

In the early time regime, partial solvability means that the diagram in Eq.~\eqref{eq:rdmentgrowthTL} can be simplified only partially. Specifically, using Eq.~\eqref{eq:leftsolvgraph} and the dual unitarity condition in Eq.~\eqref{eq:DUgraph} we can simplify the left triangle representing entanglement growth at the left edge of the interval, obtaining
\begin{align}
	\fineq[-0.8ex][.7][1]{			
		\foreach \i in {2,5}
		{
			\MPSinitialstate[2*\i+1][-1][bertiniorange][topright][n]
		}
		\trianglediag[8][0][1][bertiniorange][MPS][n][d]	
		\foreach \i in {2,...,4}{
			\cstate[\i+7.5][-\i+3.5]	
			\sqrstate[-\i+10.5][-\i+3.5]}
		\sqrstate[5.5][-.5]	
		\sqrstate[4.5][-.5]	
		\cstate[4.][-1.05]	
		\cstate[12][-1.05][][black]	
\draw[decorate,decoration={brace}] (4,0)--++(2,0);	
\node[scale=1.5] at (5,0.5) {$L_A-2t$};
}.
	\label{eq:rdmentgrowthLS}	
\end{align}
Although this diagram cannot be further simplified for finite $L_A$ and $t$, a simplification emerges in the scaling limit 
\be
L_A, t\to \infty, \qquad\qquad \frac{t}{L_A} = \zeta < \frac{1}{2}.  
\ee
Indeed, in this limit the horizontal line of replicated transfer matrices  in \eqref{eq:rdmentgrowthLS} (which are $L_A-2t$) becomes a projector on the fixed points, i.e.,
\begin{align}
	\lim_{L \rightarrow \infty }	\left(\fineq[-0.8ex][1][1]{\MPSinitialstate[0][0][bertiniorange][topright][n]
		\sqrstate[-.5][.5]\sqrstate[.5][.5]}\right)^{L_A-2t} =\fineq[-2ex][1][1]{
		\phantom{\MPSinitialstate[0][0][bertiniorange][topright][n]}	
		\draw[very thick] (-1,-.05)--++(.5,0);
		\draw[very thick] (.5,-.05)--++(-.5,0);
		\sqrstate[-.5][-.05][][black]
		\sqrstate[0][-.05][][white]},
	\label{eq:TMsquares}
\end{align}
where we introduced the replicated fixed point 
\begin{align}
\fineq[-0.8ex][1][1]
{
	\draw[very thick] (.5,-.05)--++(-.5,0);
	\sqrstate[0.5][-.05][][black]
}=	\prod_{a=1}^n \left[\sum_{i,j=1}^d \left(S\right)_{i,j}\ket{i}_a\ket{j}_{(a+1)^*}\right].
\end{align}
Eq.~\eqref{eq:TMsquares} allows us to eliminate the line of transfer matrices and contract the remaining part of the diagram by means of the following multireplica version of the left charged solvability condition 
\be
\fineq{				
		\MPSinitialstate[0][0][bertiniorange][topright][n]
		\sqrstate[-1][-.05][][white]
		\sqrstate[-.5][.5]
	}=\fineq{		
		\draw[thick] (0,0)--++(1,0);		
		\draw[thick] (0.5,.5)--++(.5,.5);		
		\charge[0.75][.75][red]
		\sqrstate[0][0][][white]
		\sqrstate[.5][.5]
	},
\label{eq:leftsolvgraphsquare}
\ee
which involves a different pairing of replicas compared to Eq.~\eqref{eq:leftsolvgraph}. The final result reads as 
\be
\tra{\rho_A^n(t)} \simeq \left(\sum_\alpha \frac{\left(c^{\rt}_\alpha\right)^n}{\left(d^{\rt}_\alpha\right)^{n-1}}\right)^{4t} \tr[S^n]^2\,,
\label{eq:earlytimeasyLS}
\ee
where $\simeq$ denotes leading order in the scaling limit and the last term comes from the scalar product of the fixed points of replicated transfer matrices with different pairings, i.e., 
\be
\fineq[-0.8ex][1][1]
	{
		\draw[very thick] (.5,-.05)--++(-.75,0);
		\sqrstate[0.5][-.05][][black]	\cstate[-0.25][-.05][][white]
	}= \tr[S^n]\,.
\ee
Eq.~\eqref{eq:earlytimeasyLS} coincides with the scaling limit of Eq.~\eqref{eq:entgrowthsolvable}. Therefore, at leading order, lifting solvability on one side leads to the same entanglement growth in the early time regime, and can be written as \begin{align}
	S(t\le L_A/2)=4t(s^{\rt}_{\rm num}+s^{\rt}_{\rm  conf})\equiv 4t s^{\rt},
	\label{eq:earlytimeasyLSexplicit}
\end{align}
where $s^{\rt}_{\rm num/conf}$ are defined as in \eqref{eq:snumsconf}, using the $c^{\rt}_\alpha$ that define left solvability in  \eqref{eq:leftsolvabilityformula} instead of $c_i$.
As we now show, beyond this regime we instead have qualitative differences. Let us begin considering the large time limit. The relevant diagram in this case is the one in Eq.~\eqref{eq:rdm}. Once again, for left charged solvable states the latter can be simplified only partially and the minimal expression reads as  
\begin{align}
	\rho_A(t)	=\!\!\!\!\!\!\fineq[-0.8ex][0.6][1]{
\MPSinitialstate[1][-1][bertiniorange][topright][1]
\roundgate[2][0][1][topright][bertiniorange][1]
\MPSinitialstate[3][-1][bertiniorange][topright][1]
	\foreach\i in {0,...,4}
		{
			\roundgate[-\i][\i][1][topright][bertiniorange][1]
			\roundgate[-\i+1][\i+1][1][topright][bertiniorange][1]
		}
	\roundgate[-5][5][1][topright][bertiniorange][1]
	\roundgate[-4][6][1][topright][bertiniorange][1]
	\roundgate[-6][6][1][topright][bertiniorange][1]
		\cstate[0][-1.05][][white]
		\cstate[4][-1.05][][black]
		\draw[dashed, color=red, rounded corners=5pt](-4.5,2.5)--++(3,3)--++(-1,1)--++(-3,-3)--cycle;
		\foreach \i in {0,...,6}
{
	\cstate[-\i+3.5][\i-.5]
	{
		\cstate[-\i-.55][\i-.55]}{}
	\charge[-\i-.3][\i-.3][red]
}
	\draw [decorate, decoration = {brace,mirror}]   (2.2,1.8)--++(-4.5,4.5) ;
	\node[scale=1.5] at (1.5,4.5){$2t-L_A$};
	\node[scale=1.5, red] at (-5,2.5){$\mathcal{T}$};
	}.\label{eq:rdmls}
\end{align}
We now note that this diagram contains $2t-L_A$ transfer matrices $\mathcal{T}$ as the one circled in red . Therefore, the large time limit can be computed by replacing $\mathcal{T}$ with a projector on its leading eigenvectors, which can be proven to correspond to eigenvalues of unit magnitude (see Appendix~\ref{app:transferM}). As we discuss in Appendix~\ref{app:transferM}, these eigenvectors are in one-to-one correspondence with the conserved, ballistic charges in the circuit so, under the assumption of starting with a complete set of solitons, we can find all of them. 

From now on we assume this to be the case, namely, we assume that the circuit has no independent conserved-charge densities of support $\ell>1$: all charges with densities of larger support are written as products of solitons on different sites. In essence, this is a chaoticity assumption for the blocks $U^{(\alpha,\beta)}$ in Eq.~\eqref{eq:DUdecomposiiton}: for generic blocks there should be no independent conserved charges apart from those enforced by the block structure. Note that the proof in Ref.~\cite{folignoinprep} guarantees that whenever ${d_\alpha^{{\rt}}}={d_\beta^{\lt}}=2$ (see, e.g., the gate in Eq.~\eqref{eq:newexample}) any randomly selected dual unitary block fulfils this assumption with probability one.

Under this assumption, the large time limit is obtained by the following substitution 
\begin{align}
		\fineq[-0.8ex][0.7][1]{	\roundgate[0][0][1][topright][bertiniorange][1]
			\roundgate[1][1][1][topright][bertiniorange][1]
			\cstate[1.5][1.5]
			\cstate[-.6][-.6]
		\charge[-.3][-.3][red]		}\rightarrow
		\sum_{\vec{\alpha}}\fineq[-0.8ex][0.5][1]{
			\draw[thick] (-.5,.5)--++(-.5,.5);
			\charge[-.75][.75][blue];
			\draw [->] (-1,-1)--(-.8,.6);
			\draw[thick] (.5,1.5)--++(-.5,.5);
			\draw [->] (1.5,2.5)--(.35,1.85);
			\charge[.25][1.75][blue];
			\draw[thick] (1,1)--++(.5,-.5);
			\charge[1.25][.75][blue];
			\draw [->] (1.75,2.5)--(1.35,.9);
			\draw[thick] (0,0)--++(.5,-.5);
			\draw [->] (-.8,-1)--(.25,-.35);
			\node[scale=2] at (-.9,-1.5) {$\Pi^{{\lt}}_{\alpha^{{\lt}}_1}$};
			\node[scale=2] at (1.9,3) {$\Pi^{{\lt}}_{\alpha^{{\lt}}_2}$};
			\charge[.25][-.25][blue];
			\cstate[1.][1]
			\cstate[-.5][.5]
			\cstate[0][0]
			\cstate[.5][1.5]
	} \frac{1}{d^{{\lt}}_{\alpha_1^{{\lt}}}d^{{\lt}}_{\alpha_2^{{\lt}}}\ldots},\label{eq:transfermatrsubstitution}
\end{align}
where the sum is all over the strings $\{\alpha^{{\lt}}_1, \ldots, \alpha^{{\lt}}_{L_A}\}$ with $\alpha^{{\lt}}_i=1,\ldots, m_{\ltb}$, and $m_{\ltb}$ being the number of left moving solitons. Plugging this form into Eq.~\eqref{eq:rdmls} we then obtain  
\begin{align}
\rho_A(\infty) = \sum_{\vec{\alpha}}	\fineq[-0.8ex][.7][1]{
		\MPSinitialstate[1][-1][bertiniorange][topright][1]
		\MPSinitialstate[3][-1][bertiniorange][topright][1]
		\cstate[.4][-.4]
		\draw(0,0)--++(-.5,.5);
		\cstate[0][0]
		\draw(0.5,0)--++(.5,.5);
		\cstate[.5][0]
		\charge[.8][.3][red];
		\charge[-.3][.3][blue]
		\cstate[1.6][-.4]
		\cstate[2.4][-.4]
		\draw(2,0)--++(-.5,.5);
		\draw(2.5,0)--++(.5,.5);
		\charge[2.8][.3][red];
		\cstate[2][0]
		\cstate[2.5][0]
		\cstate[3.6][-.4]
		\charge[.65][-.65][blue]
		\charge[1.7][.3][blue]
		\charge[2.65][-.65][blue]
		\cstate[0][-1.05][][white]
		\cstate[4][-1.05][][black]
	}.
\label{eq:GGEforMPS}
\end{align}
Note that the sum in this expression is over left-moving solitons and, therefore, this state has correlations only between its odd sites. This means that, unlike the one in Eq.~\eqref{eq:GGE}, the GGE in Eq.~\eqref{eq:GGEforMPS} has non-zero chemical potentials also for left moving charges of larger support (which are built using solitons). 

To have and explicit, simple expression, we focus on the case $\chi=1$. In this way, all the charges of support larger than $1$ site have a $0$ expectation value. Defining   $c^{\lt}_\alpha/c^\rt_\alpha$ as the expectation value of the $1$-site left/right moving solitons $\Pi^{\lt/\rt}_\alpha$ (notice that this definition agrees with Eq. \eqref{eq:leftsolv}),  we can fully characterize the asymptotic thermal ensemble and write the asymptotic value of the entanglement entropy as 
\be
S_A(\infty)\simeq L_A(s^{{\lt}}+s^{{\rt}}),
\label{eq:ententropysolvabilityonesideasymptotic}
\ee
where we introduced 
\begin{align}
	s^{(\ltb/\rtb)}_n= \frac{1}{1-n}\log(\sum_{\alpha=1}^{m_{\ltb/\rtb}} \frac{\left(c^{(\ltb/\rtb)}_\alpha\right)^n}{\left(d^{(\ltb/\rtb)}_\alpha\right)^{n-1}}),
\end{align}
and set $s^{(\ltb/\rtb)}_1=s^{(\ltb/\rtb)}$. This expression has a very interesting implication: we first note that Eq.~\eqref{eq:ententropysolvabilityonesideasymptotic} does not coincide with the value reached at the end of the early time regime. Indeed, considering Eq.~\eqref{eq:earlytimeasyLSexplicit} we find 
\be
S_A(L_A/2) \simeq 2 L_A s^{{\rt}}. 
\label{eq:ententropysolvabilityonesideearly}
\ee
Next we note that whenever the state is left charged solvable one has $s^{{\lt}}_n\ge s^{{\rt}}_n$ (see Appendix \ref{app:boundsRL}). This means that Eq.~\eqref{eq:ententropysolvabilityonesideasymptotic} is always \emph{larger} than Eq.~\eqref{eq:ententropysolvabilityonesideearly} and \emph{there must be an additional phase of growth after the early time regime}. 
This is true also without the $\chi=1$ assumption on the MPS: In Appendix \ref{app:boundsRL} we show that, even though we cannot find an explicit expression for $S^\lt$ in terms of few parameters we have 
\begin{align}
	S^\lt\ge L_A s^\rt,\label{eq:bound2}
\end{align} 
showing the existence of an additional growth phase.

To characterise this second phase we use the following rigorous bound (proven in Appendix \ref{app:boundsecondphase} using the data processing inequality)  
\be
S_A (t\!>\!L_A/2)-\!S_A(L_A/2)\!\le\! (2t-L_A) (2s^{{\rt}}\!-\!s^{{\rt}}_{\rm num}).
\label{eq:slopesecondphase}
\ee

This ensures that, whenever the number entropy density is non-zero, the slope of entanglement growth  in the second phase is smaller than the initial one (which is equal to $2s^\rt$) and we have a qualitatively different phase of thermalisation. Note that Eq.~\eqref{eq:slopesecondphase} also gives the following lower bound for the thermalisation time 

\begin{align}
	t_{\rm th} \ge \frac{L_A}{2}\left(\frac{(s^{{\lt}}+s^{{\rt}}-s^{{\rt}}_{\rm num})/2}{s^{{\rt}}-(s^{{\rt}}_{\rm num}/2)}\right),
\end{align}
obtained imposing the r.h.s. of the inequality in Eq.~\eqref{eq:slopesecondphase} to be smaller than $S_A(\infty)-S_A(L_A/2)$. 

In the upcoming subsection we use the entanglement membrane picture to argue that the bound in Eq.~\eqref{eq:slopesecondphase} is in fact saturated at leading order in the scaling limit, justifying our expression for the entanglement velocity at times $t>L_A$ \eqref{eq:entanglementvelocities}. 

\subsection{Entanglement dynamics from generic states}
\label{sec:genericstatesent}

Let us now move to initial states with no solvability properties. To treat the problem, we decompose the relevant quantities as sums over the charge sectors and treat each of the sectors using the entanglement membrane picture~\cite{jonay2018coarsegrained, zhou2020entanglement}. This amounts to assuming that the reduced blocks $U^{(\alpha,\beta)}$ (see Eq.~\eqref{eq:DUdecomposiiton}) are chaotic. \\
In order for this approach to be valid, we request that the dimension of each charge block is $d_\alpha^{(\ltb/\rtb)}\ge2$ (the membrane picture cannot be applied if the local dimension is one). 

For the sake of simplicity, we consider a simple initial state: a normalised pair product state (a special case of the MPS in Eq.~\eqref{eq:twositeMPS} with bond dimension $1$), defined by a $d\times d$ matrix $m$
\begin{align}
	\ket{\Psi_0}= 	\left(\sum_{i,j=1}^d  (m)_{i,j}\ket{i,j}\right)^{\otimes L} \qquad \tra{m m^\dagger}=1,
\label{eq:statedef}
\end{align}
where $2L$ is the number of sites of the system (which we always take to be arbitrarily large as to ignore boundary effects).

The graphical representation we use for the matrix $m$ (in the replicated space) is 
\begin{align}
	\left(m\otimes m^*\right)^{\otimes n}=\fineq[-0.8ex][1.4]{\pairproduct[0][0][n]}.
	\end{align}
The expectation values $c^{(\ltb/\rtb)}_{\alpha}$ are then given by 
\begin{align}
	c^{{\rt}}_\alpha=\tra{ m \Pi_{\alpha}^{{\rt}} m^\dagger}=
	\fineq[-0.8ex][1.4]{\pairproduct[0][0][1] \cstate[.5][.5]\cstate[-.5][.5] \charge[.27][.35][red]}\notag\\
	c^{{\lt}}_\alpha=\tra{\Pi_{\alpha}^{\lt}m   m^\dagger}=	\fineq[-0.8ex][1.4]{\pairproduct[0][0][1] \cstate[.5][.5]\cstate[-.5][.5] \charge[-.27][.35][blue]}.
	\label{eq:chargesvaluedef}
\end{align}
and obey 
\begin{align}
	c^{(\ltb/\rtb)}_\alpha\ge 0 \qquad \sum_{\alpha} c^{(\ltb/\rtb)}_\alpha=1. 
\label{eq:normalconditioncharges}
\end{align}
Similarly we can defined joint expectation values for left/right solitons as
\begin{align}
c_{\alpha,\beta}\equiv \tra{\Pi^{\lt}_\alpha m \Pi^{{\rt}}_\beta m^\dagger}=\fineq[-0.8ex][1.4]{\pairproduct[0][0][1] \cstate[.5][.5]\cstate[-.5][.5] \charge[-.27][.35][blue]\charge[ .27][.35][red]}\ge 0.\label{eq:fulldistribution}
\end{align}
The values of $c_{\alpha,\beta}$ define a joint classical probability distribution corresponding to the probability of observing simultaneously charges $\alpha$ and $\beta$ on two neighbouring sites of the initial state.
	The values $c^{(\ltb/\rtb)}_\alpha$ are the marginals of this distribution. Indeed, it is easy to see that
	\begin{align}
	c^{{\rt}}_\beta=\sum_{\alpha=1}^{m_\ltb} c_{\alpha,\beta}, \qquad 
	c^{{\lt}}_\alpha=\sum_{\beta=1}^{m_\rtb} c_{\alpha,\beta}.
\label{eq:marginal}
\end{align}

\subsubsection{Early time growth}
\label{Sec:firstphase}

The rate of entanglement production at early times ($t < L_A/2$) can be found by evaluating the diagram in Eq.~\eqref{eq:rdmentgrowthTL}. In fact, since we are now focussing on pair product states, the contributions at the two edges factorise and we are left to evaluate 
\begin{align}
	\tra[A]{\rho_A^n(t)}=	\left(\fineq[-0.8ex][0.445][1]{		
		\trianglediag[0][0][3][bertiniorange][pairproduct][n][d]	\foreach \i in {0,...,4}{
			\cstate[-\i+2.5][-\i+3.5]	
			\sqrstate[\i+3.5][-\i+3.5]	
		}
		\draw [decorate, decoration = {brace}]   (3.8,4)--++(4,-4) ;
		\phantom{\draw [decorate, decoration = {brace}]   (8.8,4)--++(0,-4) ;}
		\node[text width=15,scale=2] at (7.5,2.3) {$2t$ sites};  
	}\right)^2
\label{eq:Sgrowthnonsolvable}.
\end{align}
We then decompose the $2t$ legs on the left of this diagram using the following resolution of the identity
\begin{align}
\left(\mathds{1}_{n\times d}\right)^{\otimes 2t}=\sum_{\vec{\alpha}}\bigotimes_{i=1}^{2t}\left[\bigotimes_{a=1}^n \left(\Pi^{{\lt}}_{\alpha_{i,a}}\otimes \mathds{1}_d\right)\right].
\label{eq:identitydecompositionl}
\end{align}
Analogously, for the $2t$ legs on the right diagonal we have 
\begin{align}
	\left(\mathds{1}_{n\times d}\right)^{\otimes 2t}=\sum_{\vec{\beta}}\bigotimes_{i=1}^{2t}\left[\bigotimes_{a=1}^n \left(\Pi^{{\rt}}_{\beta_{i,a}}\otimes \mathds{1}_d\right)\right]. 
\label{eq:identitydecompositionr}
\end{align}
The sum is over all possible strings of $\alpha$s of size $2tn$, since in general we need a different projector on each replica. 

In diagrams we then rewrite the square root of Eq.~\eqref{eq:Sgrowthnonsolvable} as 
\begin{align}
\sqrt{\tra[A]{\rho_A^n(t)}} = \sum_{\vec{\alpha},\vec{\beta}}	\!\!\!\!\!\!\!\!\! \fineq[-0.8ex][0.5][1]{		
		\trianglediag[0][0][3][bertiniorange][pairproduct][n][d]	\foreach \i in {0,...,4}{
			\draw[thick](-\i+1.5,-\i+4.5)--++(1,-1);
			\cstate[-\i+1.5][-\i+4.5]	
			\draw[thick](\i+4.5,-\i+4.5)--++(-1,-1);
			\sqrstate[\i+4.5][-\i+4.5]	
			\charge[-\i+2.25][-\i+3.75][blue]
			\charge[\i+3.75][-\i+3.75][red]
		}
	\begin{scope}[shift={(-2.2,0)},rotate around={-45:(0,0)},dashed,blue]
		\draw[thick] (0,0)--++(0,6)--++(1,0)--++(0,-6)--cycle;
	\end{scope}
	\begin{scope}[shift={(7.5,-.7)},rotate around={45:(0,0)},dashed,red]
	\draw[thick] (0,0)--++(0,6)--++(1,0)--++(0,-6)--cycle;
\end{scope}
\node[scale=2.5] at (0.4,5) {$\vec{\alpha}$};
\node[scale=2.5] at (+5.5,5) {$\vec{\beta}$};
\draw[thick,->] (0.4,4.6)--++(.6,-1);\draw[thick,->] (5.4,4.6)--++(-.6,-1);}\!\!\!\!\!\!\!\!\!\! . 
=\label{eq:Sgrowth}
\end{align}\begin{align}
\sum_{\vec{\alpha},\vec{\beta}}	\!\!\!\!\!\!\!\!\! \fineq[-0.8ex][0.5][1]{		
	\trianglediag[0][0][3][bertiniorange][pairproduct][n][d]	\foreach \i in {0,...,4}{
		\draw[thick](-\i+2,-\i+4)--++(.5,-.5);
		\draw[thick](\i+4,-\i+4)--++(-.5,-.5);
		\cstate[-\i+2][-\i+4]	
		\sqrstate[\i+4][-\i+4]	
	}
	\foreach \i in {0,...,4}
	{		
	\foreach \j in {\i,...,4}
	{
		\charge[-\j+2.5+2*\i][-\j+3.5][blue]
		\charge[\j+3.5-2*\i][-\j+3.5][red]		
	}
	}
	\node[scale=2.5] at (0.4,5) {$\vec{\alpha}$};
	\node[scale=2.5] at (+5.5,5) {$\vec{\beta}$};
	\draw[thick,->] (0.4,4.6)--++(.6,-1);\draw[thick,->] (5.4,4.6)--++(-.6,-1);}\!\!\!\!\!\!\!\!\!\!,
\label{eq:Sgrowth2}
\end{align}
where, to go from \eqref{eq:Sgrowth} to \eqref{eq:Sgrowth2} we used the fact that the solitons in the diagram are projectors, thus they obey\begin{align}
\left(	\Pi^{(\ltb/\rtb)}_\alpha\right)^m=	\Pi^{(\ltb/\rtb)}_\alpha
\end{align}
and then we used the ballistic property \eqref{eq:Picharges} to move them around in the diagram showing explicitly that each gate is in a chaotic block.
Keeping fixed the values of $\vec{\alpha},\vec{\beta}$ so that each local gate is now chaotic we are enabled to apply the entanglement membrane theory~\cite{jonay2018coarsegrained, zhou2020entanglement}. The upshot of this approach is to argue that the leading contribution to the diagram is found when the internal legs are divided in two domains; in one domain the internal legs are all set to the state $\ket{\mcirc_{d, n}}$ and in the other to the state $\ket{\msqr_{d, n}}$. 

To find the leading contribution we then consider all possible separations between the two domains that start at the top of the triangle and end at position $x$, and then sum over the possible values of $x$.
Thanks to dual unitarity, all paths separating the domains that start at the top and end at $x$, which are causally connected with the top and bottom point are equivalent (indeed the line tension for a dual unitary circuit is constant) and correspond to only one diagram, so the result does not depend on the path chosen.
Projecting the internal legs on the two different domains, the diagram becomes the following:
\begin{widetext}
	\begin{align}
		\!\!\!\!\!\sqrt{\tra[A]{\rho_A^n(t)}}\!=\!\!\!\sum_{\vec{\alpha},\vec{\beta},x}\frac{1}{\mathcal{N}_{\vec{\alpha},\vec{\beta}}}\!\!\!\!\!\!\!\!	\!\!\!\!\!\!\!\!\!\!\!\!\!\!\!\!		\fineq[-0.8ex][0.7][1]{	
			\trianglediag[0][0][3][bertiniorange][pairproduct][n][d]
			\foreach \i in {0,...,4}{
				\cstate[+\i-1.5][\i-0.5]	
			}\foreach \i in {0,...,3}
			{		
				\cstate[+\i+3.5][-\i+3.5]	
				\charge[3.3+\i][3.3-\i][red]
				\charge[2.7-\i][3.3-\i][blue]
			}
		\cstate[4+3.5][-4+3.5]	
		\charge[3.3+4][3.35-4][red]
		\charge[2.7-4][3.35-4][blue]
		\foreach \i in {0,...,4}
			{		
				\draw[thick] (\i+4,-\i+4)--++(.5,.5);
				\charge[4.25+\i][4.25-\i][red]
				\cstate[+\i+4][-\i+4]	
				\sqrstate[+\i+4.5][-\i+4.5]	
			}	
			\begin{scope}[shift={(-1.5,1.5)}]
				\trianglediag[13][0][0][bertiniorange][pairproduct][n]
				\sqrstate[14.6][-.4][]
				\draw[thick](6.5-1,4.5+1)--++(.5,-.5);
				\draw[thick](7.5-1,5.5+1)--++(.5,-.5);
				\def \trasl {-1.}
				\charge[7.75+\trasl][5.25-\trasl][blue]
				\charge[6.75+\trasl][4.25-\trasl][blue]
				\sqrstate[7.4-1][3.6+1]
				\sqrstate[8.4-1][4.6+1]
				\sqrstate[7-1][4+1]
				\sqrstate[8-1][5+1]
				\cstate[6.5-1][4.5+1]
				\cstate[7.5-1][5.5+1]
				\foreach \i in {0,...,5}{
					\roundgate[13-\i][+\i][1][topright][bertiniorange][n]
					\sqrstate[14.15-\i-.5][+\i-.35+1]
					\ifnumequal{\i}{5}{}{
						\roundgate[11-\i][+\i][1][topright][bertiniorange][n]
						\sqrstate[10.85-\i-.5][+\i-.65]
						\charge[11.15-\i-.5][+\i-.35][red]}
					\charge[13.35-\i][+\i+.15+.2][red]
				}
			\charge[7.65][5.25+.1][blue]
				\charge[6.75-.1][4.25+.14][blue]
				\charge[13.35+1][-1+.2+.2][red]
			\end{scope}	
			\draw[red,dashed,very thick] (6.8,7.2)--++(4.4,-4.4)--++(-3.8,-3.8)--++(-4.4,4.4)--cycle;
			\draw[blue,dashed,very thick] (6.8,7.2)--++(-1.7,1.7)--++(-2.2,-2.2)--++(1.7,-1.7)--cycle;
			\draw [thick,decorate,decoration={mirror,brace}] (-1,-1.5)--++(8,0);
			\draw [thick,decorate,decoration={mirror,brace}] (9,-.5)--++(4,0);
			\node[scale=1.5] at (3.5,-2) {$x$};
			\node[scale=1.5] at (11,-1) {$2t-x$};
	\node[scale=2.5] at (0.4,7) {$\vec{\alpha^{(1)}}$};	\node[scale=2.5] at (0.6,5) {$\vec{\alpha^{(2)}}$};
	\node[scale=2.5] at (+11,6.1) {$\vec{\beta}^{(1)}$};
	\node[scale=2.5] at (+13.5,4.5) {$\vec{\beta}^{(2)}$};
	\draw[thick,->] (.3,4.5)--++(1,-1);
	\draw[thick,->] (1.2,7)--++(2,0.4);
	\draw[thick,->] (+10.2,6)--++(-1,-.3);		
	\draw[thick,->] (+13.,4)--++(-.5,-2);	
	}\!\!\!\!\!\!\!\!\!\!,
		\label{eq:picture2}
	\end{align}
\end{widetext}
where the normalisation factor $\mathcal{N}_{\vec{\alpha},\vec{\beta}}$ comes from the fact that the states $\ket{\mcirc_{d, n}}$ and $\ket{\msqr_{d, n}}$ are not normalised (see~Eqs.~\eqref{eq:circlestate} and \eqref{eq:squarestate}).

\begin{figure}
\includegraphics{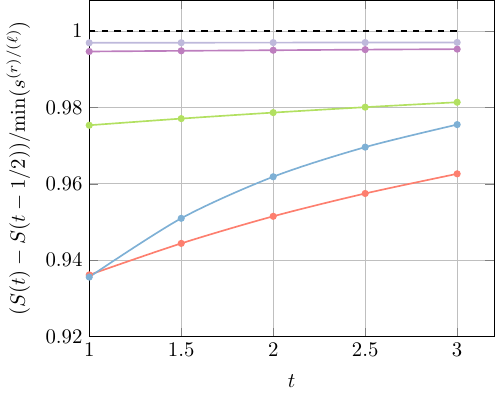}
\caption{Entanglement growth at a single edge (meaning that we evaluate $\lim_{n\rightarrow 1} \frac{1}{1-n}\log\smash{(\sqrt{\tr_A{\rho_A^n(t)}})}$) for randomly generated initial states (coloured dots) divided by the analytic prediction \eqref{eq:entgrowthmembrane} obtained measuring the values of $c^\rt/c^\lt$ randomly chosen. More details on the numerics is presented in App. \ref{app:numericsdetails}.}
\label{plot:entanglementgrowth}
\end{figure}

To evaluate the diagram, we begin by noting that the matrix element of a replicated projector between a bullet and a square state is only non-zero if the the components on each replica are the same, i.e., 
\be
\begin{aligned}
	\fineq{
		\draw[thick](0,0)--++(.5,.5);
		\cstate[0][0] 
		\sqrstate[.5][.5]
		\charge[.25][.25][red]}&=\tra{\Pi^{{\rt}}_{\beta_{i,1}}\Pi^{{\rt}}_{\beta_{i,2}},\ldots\Pi_{\beta_{i,n}}}\\
		&=d^{{\rt}}_{\beta_i} (\delta_{\beta_i,\beta_{i,1}} \ldots \delta_{\beta_i,\beta_{i,n}}). 
\end{aligned}
\label{eq:picture}
\ee
Looking at  Eq.~\eqref{eq:picture2} we see that this observation applies to the first  $x$ right moving legs and the last $2t-x$ left moving legs (the ones encircled in respectively red and blue dashed lines). Therefore, it is convenient to write the string $\vec{\alpha}= \vec{\alpha}^{(1)} \circ \vec{\alpha}^{(2)}$, where $\vec{\alpha}^{(1)}$ and $\vec{\alpha}^{(2)}$ are respectively of sizes $2t-x$ and $x$ (see Eq. \eqref{eq:picture2}). $\alpha^{(1)}$ is therefore constrained to have the same entries on each replica layer. Similarly, we write $\vec{\beta}=\vec{\beta}^{(1)}\circ \vec{\beta}^{(2)}$, where $\vec{\beta}^{(1)}$ has size $x$ and the same entries on each replica layer. In this language we can write the normalisation factor as 
\begin{align}
\mathcal{N}_{\vec{\alpha},\vec{\beta}}=\left(\prod_{i=1}^{2t-x} {d^{{\lt}}_{\alpha^{(1)}_i}}\prod_{j=1}^{x} {d^{{\rt}}_{\beta^{(1)}_j}}\right)^n.
\end{align}
We now proceed to evaluate the diagram \eqref{eq:picture2} using Eqs.~\eqref{eq:Picharges}, \eqref{eq:unitarityfoldeddiagram}, and \eqref{eq:unitarityfoldeddiagramsquare} to simplify all the gates. In particular, using the definitions in Eqs.~\eqref{eq:fulldistribution}, \eqref{eq:marginal} we find 
\begin{align}
&\sqrt{\tra[A]{\rho_A^n(t)}} =	\notag\\&
=\sum_{\substack{x,\\
		\vec{\alpha}^{(1)},\vec{\beta}^{(1)}\\
		\vec{\alpha}^{(2)},\vec{\beta}^{(2)}}}\prod_{i=1}^x\frac{\left[\prod_{a=1}^n {c_{\alpha^{(2)}_{i,a} \:\beta^{(1)}_{i}}}\right]}{{d^{{\rt}}_{\beta^{(1)}_i}}^{n-1}}{}\prod_{j=1}^{2t-x}\frac{\left[\prod_{a=1}^n {c_{\alpha^{(1)}_{j} \:\beta^{(2)}_{j,a}}}\right]}{{d^{{\lt}}_{\alpha^{(1)}_j}}^{n-1}}\notag\\
	&=\sum_{\substack{x,\\{\vec{\alpha}^{(1)},\vec{\beta}^{(1)}}}}\prod_{i=1}^x\frac{{c^{{\rt}}_{\beta^{(1)}_{i}}}^n}{{d^{{\rt}}_{\beta^{(1)}_i}}^{n-1}}\prod_{j=1}^{2t-x} \frac{{c^{\lt}_{\alpha^{(1)}_{j}}}^n}{{d^{{\lt}}_{\alpha^{(1)}_j}}^{n-1}}\notag\\
	&=\sum_x \left(\sum_{\alpha} \frac{{c^{{\rt}}_\alpha}^n}{{d^{{\rt}}_\alpha}^{n-1}}\right)^{2t-x}\left(\sum_{\beta} \frac{{c^{{\lt}}_\beta}^n}{{d^{{\lt}}_\beta}^{n-1}}\right)^x.\label{eq:resum}
\end{align}
The remaining sum over $x$ is a geometric sum and can be evaluated explicitly. Ignoring $o(1)$ factors, we can write the final result as
\be
\sqrt{\tra[A]{\rho_A^n(t)}} =\exp(t(1-n) \min(s_n^{{\rt}},s_n^{{\lt}})),
\ee
where we introduced
\be
s_n^{(\ltb/\rtb)}=\frac{1}{1-n} \log(\sum_\alpha \frac{{c^{(\ltb/\rtb)}_\alpha}^n}{{d^{(\ltb/\rtb)}_\alpha}^{n-1}}).
\ee
Putting all together this finally gives 
\be
S^{(n)}_A(t< L_A/2)=4t \min(s_n^{{\rt}},s_n^{{\lt}}). 
\label{eq:entgrowthmembrane}
\ee
This result is consistent with Eq.~\eqref{eq:earlytimeasyLS}, as one can see recalling that for a left charged solvable state $s^{{\rt}}_n\le s^{{\lt}}_n$ (see~App.~\ref{app:boundsRL}) and can be analytically continued to all values of $n$, in particular $n=1$ \footnote{This is correct even if the  $\min$ function, in \eqref{eq:entgrowthmembrane}, is not analytic, as it is just an approximation, in the scaling limit, of the smooth function obtained summing \eqref{eq:resum}}. In fact, Eq.~\eqref{eq:entgrowthmembrane} can also be tested numerically. For instance, in Fig.~\ref{plot:entanglementgrowth} we compare its prediction for the entanglement entropy with exact numerical results obtained for randomly generated initial states with two solitons of dimensions $d^{\rt}_\alpha= d^{\lt}_\alpha =2$ on each leg. The agreement observed is convincing (with corrections of the order of a few percentage points) even for the very short times accessible by the exact numerics.

Note that an interesting prediction of Eq.~\eqref{eq:entgrowthmembrane} is the possible occurrence of a non-analyticity in $n$ of the R\'enyi entropies in the limit $L_A\gg t \gg 1$. The latter takes place, if there exist two $n\ne m$ such that 
\begin{align}
	s_n^{{\rt}}<s_n^{{\lt}}, \qquad s_{m}^{{\rt}}>s_{m}^{{\lt}}. 
\end{align}

\subsubsection{Asymptotic entanglement value}
\label{GGEentanglement}
At asymptotic times $t\gg L_A$, the subsystem $A$ is at equilibrium with the rest and we expect the entanglement entropy to relax to the thermodynamic entropy~\cite{calabrese2020entanglement}. 

\begin{figure}[h!]
\includegraphics{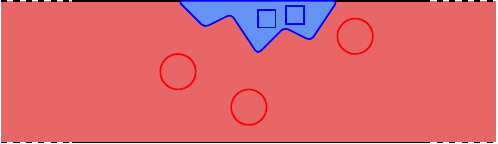}
\caption{Pictorial representation of the domain-wall configuration giving the leading contribution to Eq.~\eqref{eq:rdm2}. The domain of squares does not reach the bottom part of the diagram (where the initial states are located) as the contribution would otherwise be suppressed as $a^t$, for some $a<1$. Instead, if the square-state domain is confined to the top part as in this picture, then the suppression does not scale with $t$. We stress that this domain wall approach works once the gates are projected in a sector where they are ``chaotic'' and \emph{then} one looks for the dominant domain configuration in each of these sectors. }
\label{fig:domainconfig}
\end{figure}

In the domain wall language this can be seen by noting that the diagram for the $n$-th moment of $\rho_A(t)$, i.e.  
\be
\!\!\!\!\tra{\rho_A^n}= \fineq[-0.8ex][0.5][1]{
		\foreach \i in {0,...,2}
		{	
			\foreach \j in {0,...,5}
			{
				\roundgate[2*\j][2*\i][1][topright][bertiniorange][n]
				\roundgate[2*\j+1][2*\i+1][1][topright][bertiniorange][n]
				\pairproduct[2*\j+1][-1]
			}
		}
		\foreach \j in {0,1,2,3,8,9,10,11}
		{
			\cstate[\j+.5][5.5]
		}
	\foreach \j in {4,5,6,7}
	{
		\sqrstate[\j+.5][5.5]
	}\draw[decorate,decoration={brace}] (4,5.8)--++(4,0);
	\node[scale=2] at (6,6.4)  {$A$};},
	\label{eq:rdm2}
\ee
is dominated by the configuration reported in Fig.~\ref{fig:domainconfig}.

To evaluate this contribution, we first decompose the identities along the cuts separating the domain of squares from the rest, proceeding as in Eqs.~\eqref{eq:identitydecompositionl} and \eqref{eq:identitydecompositionr}. This gives \be
	\sum_{ \boldsymbol{\alpha},\boldsymbol{\beta}}
	\fineq[-0.8ex][0.445001][1]{
	\trianglediag[0][0][4][bertiniorange][pairproduct][n]
	\trianglediag[4][0][4][bertiniorange][pairproduct][n]
	\foreach \i in {0,...,5}
	{
		\cstate[\i-1.5][\i-.5]
		\cstate[-\i+13.5][\i-.5]
	}
	\def \diff {0.1}
	\sqrstate[7.5-\diff][4.5+\diff]
	\sqrstate[6.5-\diff][3.5+\diff]
	\sqrstate[5.5+\diff][3.5+\diff]
	\sqrstate[4.5+\diff][4.5+\diff]
	\charge[7.65][4.35][blue]
	\charge[6.65][3.35][blue]
	\charge[4.35][4.35][red]
	\charge[5.35][3.35][red]
},\label{eq:6}
\ee
where 
\be
\boldsymbol {\alpha}=\left(\alpha_{i,a}\right)_{{i=1,\ldots,L_A; a=1,\ldots,n}}, 
\ee
represents strings of projectors on the right-moving solitons, and $\vec{\beta}$ the left-moving ones.
Again, we can pin the domain of bullets by projecting the legs across a cut on bullet states  (for a fixed value of $\vec{\alpha}$ and $\vec{\beta}$).
Thanks to dual unitarity, all cuts separating the two domains correspond to only one diagram to evaluate (meaning that they can be deformed into one another), so we can just choose the simplest one:
\begin{align}
	\frac{1}{\mathcal{N}_{\vec{\alpha},\vec{\beta}}}\hspace{-.75cm}
	\fineq[-0.8ex][0.50001][1]{
		\trianglediag[0][0][4][bertiniorange][pairproduct][n]
		\trianglediag[4][0][4][bertiniorange][pairproduct][n]
		\foreach \i in {0,...,5}
		{
			\cstate[\i-1.5][\i-.5]
			\cstate[-\i+13.5][\i-.5]
		}
		\def \diff {0.1}
		\cstate[7.5-\diff][4.5+\diff]
		\cstate[6.5-\diff][3.5+\diff]
		\cstate[5.5+\diff][3.5+\diff]
		\cstate[4.5+\diff][4.5+\diff]
		\charge[7.65][4.35][blue]
		\charge[6.65][3.35][blue]
		\charge[4.35][4.35][red]
		\charge[5.35][3.35][red]
		\def \difff {0.15}
		\def \heig {-1}
		\def \dif {-0.45}
		\draw[thick] (7.5-\difff,6.5+\difff+\heig)--(7.5-\dif,6.5+\dif+\heig);
		\draw[thick] (6.5-\difff,5.5+\difff+\heig)--(6.5-\dif,5.5+\dif+\heig);
		\draw[thick] (5.5+\difff,5.5+\difff+\heig)--(5.5+\dif,5.5+\dif+\heig);
		\draw[thick] (4.5+\difff,6.5+\difff+\heig)--(4.5+\dif,6.5+\dif+\heig);
		\charge[7.5+0.15][6.5-0.15+\heig][blue];
		\charge[6.5+0.15][5.5-0.15+\heig][blue];
		\charge[4.5-0.15][6.5-0.15+\heig][red];
		\charge[5.5-0.15][5.5-0.15+\heig][red];
		\sqrstate[7.5-\difff][6.5+\difff+\heig]
		\sqrstate[6.5-\difff][5.5+\difff+\heig]
		\sqrstate[5.5+\difff][5.5+\difff+\heig]
		\sqrstate[4.5+\difff][6.5+\difff+\heig]
		\cstate[7.5-\dif][6.5+\dif+\heig]
		\cstate[6.5-\dif][5.5+\dif+\heig]
		\cstate[5.5+\dif][5.5+\dif+\heig]
		\cstate[4.5+\dif][6.5+\dif+\heig]}\hspace{-.75cm},
\end{align}
where $\mathcal{N}_{\vec{\alpha},\vec{\beta}}$ is a normalization factor.
Eq. \eqref{eq:picture} forces then the solitons $\vec{\beta},\vec{\alpha}$ to be the same on each replica; dropping the indices of replicas on the string, we can write the normalization factor as\begin{align}
	\mathcal{N}_{\vec{\alpha},\vec{\beta}}=\prod_{i=1}^{L_A}\left(d^\rt_{\alpha_i}d^\lt_{\beta_i}\right)^n.
\end{align}
Finally, using the charge conservation relations in Eq.~\eqref{eq:Picharges} to move all the solitons to the bottom part of the diagram, and then the unitarity relation in Eq.~\eqref{eq:unitarityfoldeddiagram}, we can reduce the diagram to 
\be
\tra{\rho^n_A(t\gg L_A)}=	\sum_{\boldsymbol {\alpha},\boldsymbol {\beta}}\prod_{i=1}^{L_A} \frac{\left(c_{\alpha_i}^{{\rt}}c_{\beta_i}^{{\lt}}\right)^n}{\left(d^{{\rt}}_{\alpha_i}d^{{\lt}}_{\beta_i}\right)^{n-1}}.
\ee
The sum over strings can be carried out by exchanging it with the product, giving again the result obtained in Eq.~\eqref{eq:ententropysolvabilityonesideasymptotic}, i.e., 
\be
S_A^{(n)}=L_A(s_n^{{\rt}}+s_n^{{\lt}}).
\ee

\subsubsection{Second phase of thermalisation}

In Sec.~\ref{Sec:firstphase} we found that, for generic initial states, the R\'enyi-entropy growth at times $t < L_A/2$ is, at leading order in $t$ 
\be
S^{(n)}_A(t)=4t\min(s^{{\rt}}_n,s^{{\lt}}_n)+o(t),
\ee
while in Sec.~\ref{GGEentanglement} we determined its asymptotic value to be  
\be
S^{(n)}_A(t\gg L_A)=L_A(s^{{\rt}}_n+s^{{\lt}}_n). 
\ee
These expressions coincide at $t=L_A/{2}$ \emph{only if} $s^{{\rt}}_n=s^{{\lt}}_n$ (e.g.\ for a charged solvable state), where we find a fast thermalisation process, concluded at the minimal possible time allowed by causality. If this is not the case, we expect a second non-trivial phase of thermalisation.  Let us now characterise this phase using the entanglement membrane approach:  in the rest of the section we set $s^{{\rt}}_n<s^{{\lt}}_n$ for the calculations.

\begin{figure}[t!]
	\includegraphics{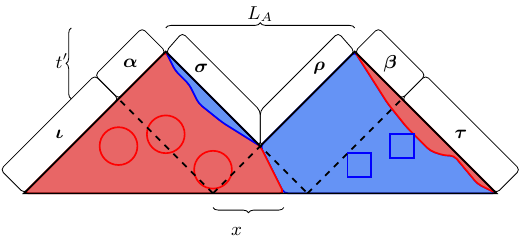}
\caption{Pictorial representation of the dominant domain contributions to evaluate Eq.~\eqref{eq:6} for small values of $t'=t-L_A/2$. We used the fact that, since $s^{{\rt}}_n<s^{{\lt}}_n$, the dominant domain at the left/right borders is always the one on the left for small values of $t'$, and we sum over all possible choices of cuts in between the triangles (dashed lines in Fig.~\ref{fig:domainconfig2ndphase}), which we label as $x=1,\ldots 2t'$. 
}\label{fig:domainconfig2ndphase}
\end{figure}

\begin{figure}
\includegraphics{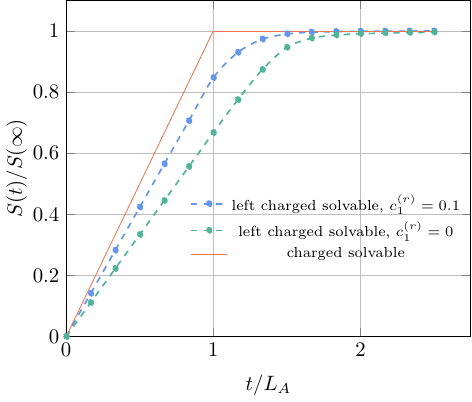}
\caption{Entanglement growth for a charged solvable state (red) and a left charged solvable state (blue and green), normalised on the asymptotic entanglement value and on the interval size $L_A$ (chosen to be $L_A = 5$ for the numerics). In order to reduce the finite size effects, we choose a pair product initial state, i.e.\ a state as in Eq.~\eqref{eq:twositeMPS} with $\chi=1$.More details on the numerics is presented in App. \ref{app:numericsdetails}.}
\label{fig:entgrowthnumerics}
\end{figure}

We again consider the diagram in Eq.~\eqref{eq:rdm2} but now focus on the case of finite $t>{L}_A/{2}$ and set $t'=t-{L}_A/{2}$. We again insert a resolution of the identity in order to decompose the gates into chaotic blocks and then apply the membrane picture (see Eq.~\eqref{eq:6}). Next we sum over several different strings, as indicated in Fig.~\ref{fig:domainconfig2ndphase}. Referring to the figure we have that only the $\boldsymbol {\alpha}$ and $\boldsymbol {\beta}$ strings contribute to a growth of entanglement, while the contribution from the others is constant and equal to the entanglement value at the end of the early time regime (see Sec.~\ref{Sec:firstphase}). The calculation is similar to those reported in the previous subsections and yields  
\begin{align}
&\frac{\tra{\rho_A^n(t'+L_A/2)}}{\tra{\rho_A^n(L_A/2)}}=\\&\sum_{x=1}^{2t'} \left[\sum_{\beta=1}^{m_{\rt}} \frac{{c^{{\rt}}_{\beta}}^{n}}{{d_{\beta}^{{\rt}}}^{2(n-1)}}\right]^x\left[\sum_{\alpha=1}^{m_{\lt}}\sum_{\beta=1}^{m_{\rt}} \frac{c_{\alpha,\beta}^{n}}{\left(d^{{\lt}}_{\alpha}d^{{\rt}}_{\beta}\right)^{n-1}}\right]^{2t'-x}\!\!\!\!\!\!\!\!\!\!\!\!, \notag 
\end{align}
the sum in $x$ can be explicitly carried out but, since we are interested in the leading order, we ignore all the factors not scaling exponentially in $t'$ and obtain 
\begin{align}
&\frac{\tra{\rho_A^n(t'+L_A/2)}}{\tra{\rho_A^n(L_A/2)}} \label{eq:expression}\\
&\!\!{\simeq} \max\left(\sum_{\alpha=1}^{m_{\rt}}\frac{{c^{{\rt}}_{\alpha}}^{n}}{{d_{\alpha}^{{\rt}}}^{2(n-1)}},\sum_{\alpha=1}^{m_{\rt}}\sum_{\beta=1}^{m_{\lt}} \frac{c_{\alpha,\beta}^{n}}{\left(d^{{\rt}}_{\alpha}d^{{\lt}}_{\beta}\right)^{n-1}}\right)^{2t'}\!\!\!.\notag 
\end{align}
where we recall that $\simeq$ denotes equality at the leading order in time.
\begin{figure}[t!]
\includegraphics{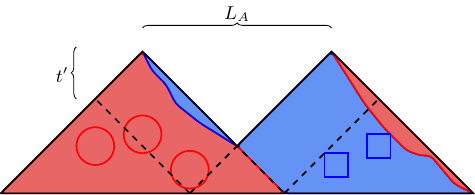}
\caption{ Dominant domain configuration in the limit $n\rightarrow 1$ assuming $s^{(r)}_n<s^{(l)}_n$}
\label{fig:domainconfig2ndphasefinal}
\end{figure}
As shown in Appendix~\ref{app:inequality}, the assumption  $s^{{\rt}}_n<s^{{\lt}}_n$ implies that the first term in Eq.~\eqref{eq:expression} is always dominant in the entanglement entropy limit $n\to1$. This means that, in this limit, the dominant domain wall configuration is the one reported in Fig.~\ref{fig:domainconfig2ndphasefinal}. This gives the following explicit result for the entanglement entropy
\begin{align}
\hspace{-.25cm}S_A(L_A/2+t')&=S_A(L_A/2)+
2t'\left(2 s^{{\rt}}_{\rm conf}\!+\!s^{{\rt}}_{\rm num}\right)\notag\\
&=4t s^{{\rt}} - (2t-L_A) s_{\rm num}^{\rt}. 
\end{align}
This result implies that the rate of growth decreases by an amount equal to the number entropy density in the sector (left or right) which has the smallest asymptotic entropy density, predicting a saturation of the exact bound in Eq.~\eqref{eq:slopesecondphase}, in the scaling limit. In other words, we find the following two entanglement velocities 
\begin{align}
 v_{E}(\zeta)\equiv& \lim_{\substack{{t,L_A\to\infty}\\ {t/L_A=\zeta}}}\frac{ S(t+1)-S(t)}{S(\infty)/L_A} \notag\\
&= \begin{cases}
		\displaystyle \frac{2s^\rt}{s^\rt+s^\lt},\quad\,\,  \zeta \le {1}/{2}\\
		\displaystyle \frac{2s^\rt-s^\rt_{\rm{num}}}{s^\rt+s^\lt}, \,\, 1/2<  \zeta  \lessapprox t_{th}/L_A 
	\end{cases}.
	\label{eq:entanglementvelocities}
\end{align}
A numerical check of this equation is reported in Fig.~\ref{fig:entgrowthnumerics} for two examples of left charged solvable states, both chosen such that $\smash{c^\lt_1=c^\lt_2=0.5}$ (the block structure of the charges is such that $\smash{m_{(\ltb/\rtb)}=2}$ and $d^{(\ltb/\rtb)}_{1,2}=2$). The data agrees with our prediction on the asymptotic value of the entanglement (see Eq.~\eqref{eq:ententropysolvabilityonesideasymptotic}) and is compatible with our results for the behaviour at times $t>L_A$ (which is not fixed by left solvability). Indeed, from Eq.~\eqref{eq:entanglementvelocities} we expect that in the case $\smash{c^\rt_1=0, c^\rt_2=1}$ (green curve) the slope in the second phase should be the same as the first one (assuming a saturation of the bound \eqref{eq:slopesecondphase}). The numerics suggests this to be the case. Instead, the blue dataset has $\smash{c^\rt_1=0.1,   c^\rt_2=0.9}$,  which implies $s^\rt_{\rm num}>0$, displaying a reduction of the slope in accordance with Eq.~\eqref{eq:entanglementvelocities}.

\section{Asymmetry dynamics and Mpemba effect}
\label{sec:asymmetrysec}

Further information on the dynamics of $\mathcal G$-symmetric dual-unitary circuits, and on the nature of the second phase of thermalisation, can be obtained by studying symmetry restoration. Namely, preparing the system in an initial state that is not an eigenstate of the charges and observing how it gradually becomes more symmetric.

This process can be conveniently characterised using the \emph{entanglement asymmetry}~\cite{ares2022entanglement}, which is the relative entropy between $\bar \rho_A(t)$ and the symmetrised state 
\begin{align}
	\bar \rho_A(t)=\sum_{\boldsymbol{\alpha},\boldsymbol{\beta}} \Pi^{{\rt}}_{\boldsymbol{\alpha}}\Pi^{{\lt}}_{\boldsymbol {\beta}} \rho_A(t) \Pi^{{\rt}}_{\boldsymbol {\alpha}}\Pi^{{\lt}}_{\boldsymbol{\beta}} ,
\end{align}
where $\Pi^{{(\rtb/\ltb)}}_{\boldsymbol{\alpha}}$, introduced in Eq.~\eqref{eq:projectors}, are the projectors in each charge sector (the latter  corresponding to the choice of  a soliton $\alpha_i$ on each leg). The entanglement asymmetry is then expressed as 
\be
\label{eq:asymmetrydefinition}
\Delta S_A(t)= \rho_A(t)\log \rho_A(t)-\bar \rho_A(t)\log \bar \rho_A(t).
\ee
Considering again a quench from the generic pair product states in Eq.~\eqref{eq:statedef} we have that the asymmetry at $t=0$ is given by  
\begin{align}
\Delta S_A(0)=- L_A \sum_{\alpha,\beta} c_{\alpha,\beta}\log(c_{\alpha,\beta}).	\label{eq:asymmetryexpression}
\end{align} 
where $\{c_{\alpha,\beta}\}$ is the probability distribution defined in Eq.~\eqref{eq:fulldistribution}. Note that this quantity is zero if and only if the initial state is in a single charge sector, i.e., $c_{\alpha,\beta} = \delta_{\alpha,\alpha_0} \delta_{\beta,\beta_0}$. For later convenience we also introduce a symbol to indicate the asymmetry density at time $t=0$
\be
\Delta s_0 \equiv  \frac{\Delta S_A(0)}{L_A} = - \sum_{\alpha,\beta} c_{\alpha,\beta}\log(c_{\alpha,\beta}), 
\ee
which is given by the Shannon entropy of $\{c_{\alpha,\beta}\}$. 

An interesting question in this setting is whether we can have instances of the quantum Mpemba effect~\cite{ares2022entanglement}. Namely, whether states that at $t=0$ are more asymmetric --- have larger asymmetry --- can reach the equilibrium value of zero asymmetry before other, initially-more-symmetric states. Our results on the entanglement dynamics suggest that this should be possible in $\mathcal G$-symmetric dual-unitary circuits evolving from non-charged-solvable states. Indeed, recent results in the context of random unitary circuits~\cite{klobas2024translation} suggest that in chaotic systems the symmetry is not restored before the thermalisation time. Moreover, because of the of the two-step thermalisation process, in $\mathcal G$-symmetric dual-unitary circuits the initial value of the asymmetry and the thermalisation time $t_{\rm th}$ (see Eq.~\eqref{eq:thermalisationtime}) can be set independently of each other. To show that this expectation is indeed correct, let us provide an explicit example.

To compute the asymmetry for $t>0$ we use the replica trick. Exploiting the cyclicity of the trace, we can write the $n$-th moment of the symmetrized density matrix as
\begin{align}
	\tra{\left(\bar \rho_A\right)^n}=\sum_{\boldsymbol{\alpha},\boldsymbol{\beta}}\tra{\left(\left(\Pi^{{\rt}}_{\boldsymbol{\alpha}}\Pi^{{\lt}}_{\boldsymbol{\beta}}\right)\rho_A(t)\right)^n}.\label{eq:symmetrizedmatrix}
\end{align}
In the early time regime $t<L_A/2$ we have three distinct contributions: one coming from the causally disconnected $2L_A-4t$ sites in the centre of $A$, which carry asymmetry $(L_A-2t) \Delta s_0$, and two separate contributions from the edges taking the form 
\begin{align}
\sum_{\boldsymbol{\alpha}}	\fineq[-0.8ex][0.5][1]{		
		\trianglediag[0][0][3][bertiniorange][pairproduct][n][d]	\foreach \i in {0,...,4}{
			\cstate[-\i+2.35][-\i+3.65]	
			\sqrstate[\i+3.65][-\i+3.65]
			\charge[\i+3.35][-\i+3.35][red]
		}
		\draw [decorate, decoration = {brace}]   (8.5,4)--++(0,-4) ;
		\node[text width=15,scale=2] at (7.5,2.3) {$2t$ sites};  },
\label{eq:leftasymmetry}
\end{align}
\begin{align}
		\sum_{\boldsymbol{\beta}}	\fineq[-0.8ex][0.5][1]{		
			\trianglediag[0][0][3][bertiniorange][pairproduct][n][d]	\foreach \i in {0,...,4}{
				\cstate[-\i+2.35][-\i+3.65]	
				\sqrstate[\i+3.65][-\i+3.65]
				\charge[-\i+2.65][-\i+3.35][blue]
			}  }.
\label{eq:rightasymmetry}
\end{align}
Note that the solitons here are chosen to be projectors and must be the same on each replica, as prescribed by Eq.~\eqref{eq:symmetrizedmatrix}.

Assuming, e.g., $s^{{\rt}}_n<s^{{\lt}}_n$, it is easy to see that the diagram in Eq.~\eqref{eq:leftasymmetry} gives the same contribution as Eq.~\eqref{eq:Sgrowthnonsolvable} at leading order. This follows from the fact that the decomposition used in Sec.~\ref{sec:genericinitialstates} is essentially the same as in that in Eq.~\eqref{eq:symmetrizedmatrix}: explicitly evaluating \eqref{eq:Sgrowthnonsolvable} in \eqref{eq:resum} (for a single edge), we get the prescription of choosing the cut to be along the rightmost diagonal, forcing the right-moving charges to be the same in each replica, as in Eq. \eqref{eq:symmetrizedmatrix}, while the left-moving ones are not, and can be re-summed to be the identity. We get, instead, a nontrivial contribution from the other edge (i.e.\ Eq.~\eqref{eq:rightasymmetry})
\begin{align}
	\sum_x \left[\sum_{\alpha,\beta}\frac{{c_{\alpha,\beta}}^n}{{d^{{\rt}}_\alpha}^{n-1}}\right]^x\left[\sum_{\beta}\frac{{c^{{\lt}}_{\beta}}^n}{{d^{{\lt}}_\beta}^{n-1}}\right]^{2t-x}=\notag\\
	\simeq \left[\max\left({\sum_{\alpha,\beta}\frac{c_{\alpha,\beta}^n}{{d^{{\rt}}_\alpha}^{n-1}},\sum_{\beta}\frac{ {c^{{\lt}}_{\beta}}^n}{{d^{{\lt}}_\beta}^{n-1}}}\right)\right]^{2t},\label{eq:asymmetryinitialphase}
\end{align}
without further information one cannot decide which contribution in Eq.~\eqref{eq:asymmetryinitialphase} is the dominant one, as both cases can be realised.

For example, to produce an example where the first term dominates, one can proceed as follows. Suppose that we have gates with chaotic blocks such that $m_\ltb=m_\rtb$ (i.e. the number of left and right solitons is the same), and  choose an initial state such that, for a certain ordering of the soliton labels we have $c_{\alpha,\beta}=c_\alpha \delta_{\alpha,\beta}$. Then  
\be
\sum_{\alpha,\beta}\frac{c_{\alpha,\beta}^n}{{d^{{\rt}}_\alpha}^{n-1}}=\sum_{\alpha}\frac{{c_{\alpha}^\rt}^n}{{d^{{\rt}}_\alpha}^{n-1}}=\exp((1-n)s^\rt_n)\,.
\ee
Since by assumption 
\be
	s^\rt_n<s^\lt_n,
\ee
we find that the first term in \eqref{eq:asymmetryinitialphase} is the largest and dominant one. Note that for our choice of the charges $c^{\lt}_\alpha=c^{\rt}_\alpha$ it is still possible to have $s^\rt_n\ne s^\lt_n$ by tuning appropriately the sizes of the charge sectors $d^\lt_\alpha\ne d^\rt_\alpha$.	

\begin{figure}
\comment{
	\begin{tikzpicture}
		\begin{axis}[grid=major,
			legend columns=2,
			legend style={at={(0.6,0.7)},anchor=south east,font=\scriptsize,draw= none, fill=none},
			xtick distance=1,
			mark size=1.3pt,	
			xlabel=$t$,
			ymin=0,
			xmin=0,
			ytick={5,4,2,1,0},
			xtick={1,1.6,2.65},
			xticklabels={$\frac{L_A}{2}$,$t_{th}^{1}$,$t_{th}^{2}$},
			yticklabels={$s^{{\lt}}_{\rm num}+s^{{\rt},1}_{\rm num}$,$s^{{\lt}}_{\rm num}+s^{{\rt},2}_{\rm num}$,$s^{{\lt}}_{\rm num}-s^{{\rt},2}_{\rm num}$,$s^{{\lt}}_{\rm num}-s^{{\rt},1}_{\rm num}$,0},
			ylabel=$ \frac{\Delta S(t)}{L_A}$,
			y label style={at={(axis description cs:.15,.55)},anchor=south,font=\large	},		
			tick label style={font=\normalsize	},	
			x label style={at={(axis description cs:.2,-0.05)},anchor=south,font=\large		},	
			]
			\def \sL {3}
			\def \sRR {2}
			\def \sR {1}
			\def \sconf {0.69314718056}	
			\addplot[dashed, red,domain = 0:1.,very thick]{(\sL+\sR-2*\sR*x)};
			\addplot[dashed, red,domain = 1:2.65,very thick]{\sL-\sR-(x-1)*(\sR/2+\sconf)};
			\addplot[dashed, blue,domain = 0:1,very thick]{(\sL+\sRR-2*\sRR*x)};
			\addplot[dashed, blue,domain = 1:1.6,very thick]{\sL-\sRR-(x-1)*(\sRR/2+\sconf)};
\end{axis}	\end{tikzpicture}}
\includegraphics[scale=0.95]{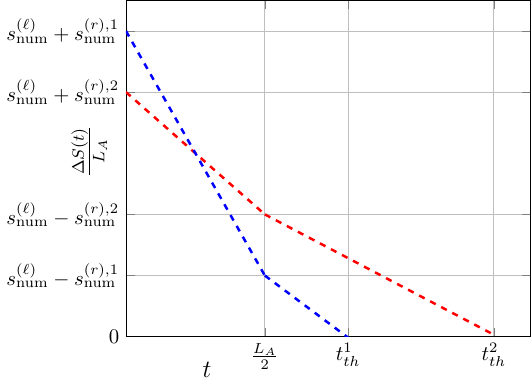}
\caption{Example of Mpemba effect choosing two initial product states and tuning them so that they have equal $s_{\rm num}^{{\lt}}>s_{\rm num}^{{\rt},1/2}$, and different $s_{\rm num}^{{\rt},1}>s_{\rm num}^{{\rt},2}$. The symbols $t_{\rm th}^{1/2}$ denote the different thermalisation times, computed according to Eq.~\eqref{eq:thermalisationtime}.}
\label{fig:Mpemba}
\end{figure}
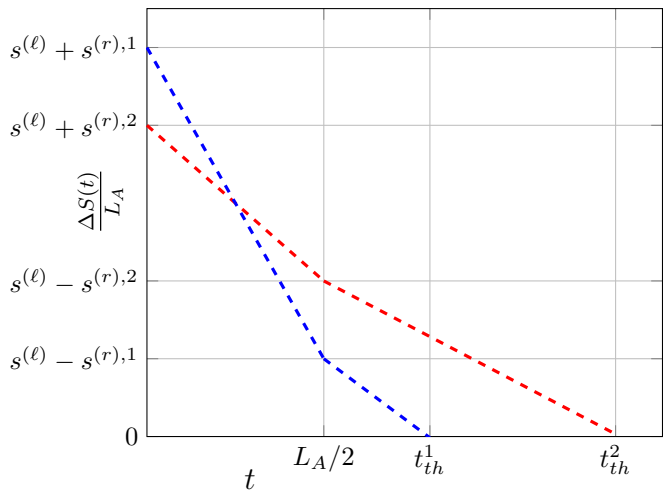

From now on, however, we consider a case where the second term dominates. Specifically, we focus on the case where both right and left symmetry sectors have fixed  size $\mathcal{D}$
\be
\!\!\!\!d^\lt_\alpha=d^\rt_\beta=\mathcal{D},\,\,\, \forall\alpha\!\in\!\{1,\ldots,m_\ltb\},\beta\!\in\!\{1,\ldots,m_\rtb\},
\ee
 which immediately implies
\begin{align}
\sum_{\alpha,\beta}\frac{c_{\alpha,\beta}^n}{\mathcal{D}^n}\le\sum_{\beta}\frac{{}{\left(\sum_{\alpha}c_{\alpha,\beta}\right)^n}}{\mathcal{D}^n}=\sum_{\beta}\frac{{c^{{\lt}}_{\beta}}^n}{\mathcal{D}^n},
\end{align}
and 
\be
s^{(\ltb/\rtb)}_{\rm conf}=\log(\mathcal{D}).
\ee
In this case we can simplify Eq.~\eqref{eq:asymmetryinitialphase} and, putting all together, find the following expression for the entanglement entropy of the symmetrised state 
\begin{align}
-\bar \rho_A(t)\log \bar \rho_A(t) \simeq & \min(2t,L_A)\left({s^{{\lt}}+s^{{\rt}}}\right) \notag\\
&+ \max(L_A-2t,0) \Delta s_0.
\end{align}
This value remains constant at later times, since the dominant configuration of domain walls at times $t\approx L_A/2$ is the same as the long times one in Fig.~\ref{fig:domainconfig}. Plugging it into Eq.~\eqref{eq:asymmetrydefinition} and combining it with the results of the previous section we finally obtain
\begin{widetext}
\be
{\Delta S_A(t)} \simeq
\begin{cases}
{2t}({s^{{\lt}}_{\rm num}-s_{\rm num}^{{\rt}}})+(L_A-2t)\Delta s_0,   & t <L_A/2 \\ 
\\
L_A (s^{{\lt}}_{\rm num}-s_{\rm num}^{{\rt}})-({2t}-{L_A})(s^{{\rt}}_{\rm num}+2\log(\mathcal{D})) & t_{\rm th}\geq t\geq L_A/2\\
 \\
 0 & t > t_{\rm th}
\end{cases}. \label{eq:asymmetrycomputed}
\ee
\end{widetext}
Note that the slope in the first phase is always negative. Indeed (see Eq.~\eqref{eq:asymmetryexpression})
\be
s^{{\lt}}_{\rm num}+s^{{\rt}}_{\rm num}\ge\Delta s_0 \ge \max(s^{{\lt}}_{\rm num},s^{{\rt}}_{\rm num}).
\ee
One can always saturate the first inequality by choosing an initial product state (i.e.\ a rank $1$ matrix $m$ in Eq.~\eqref{eq:statedef}). Indeed, this implies 
\begin{align}
	c_{\alpha,\beta}=c^{{\lt}}_\alpha c^{{\rt}}_\beta\implies \Delta s_0=s^{{\lt}}_{\rm num}+s^{{\rt}}_{\rm num},
	\label{eq:upperbounde}
\end{align}
allowing for a simpler form of \eqref{eq:asymmetrycomputed}.
 In this case, it is easy to find an explicit instance of Mpemba effect, by tuning the initial value of the asymmetry to be high without modifying the thermalisation time, which only depends on $s^{(\ltb/\rtb)}_{\rm num}$. In particular, recalling the definition of thermalisation time in Eq.~\eqref{eq:thermalisationtime}, we see that we can choose two initial product states that have the same (larger) left number entropy density and different right number entropy densities
\begin{align}
s^{{\rt},2}_{\rm num}<s^{{\rt},1}_{\rm num}<s^{\lt ,2}_{\rm num}=s^{\lt,1}_{\rm num}	,
\end{align}
and trigger a quantum Mpemba effect, see, e.g., Fig.~\ref{fig:Mpemba}.

\section{Discussion and Outlook}
\label{sec:conclusions}

In this work we introduced and characterised $\mathcal G$-symmetric dual-unitary circuits, a class of quantum circuits where it is possible  to investigate exactly the interplay between entanglement and charge fluctuations. These are dual-unitary circuits with an arbitrary number of independent $U(1)$ charges that are generically chaotic in each charge sector. We showed that in these circuits the transport of charge is always ballistic and one can define a class of low-entangled initial states, the charged solvable states, with exactly solvable entanglement dynamics. This class contains --- but is not limited to --- the conventional solvable states of dual-unitary circuits described in Ref.~\cite{piroli2020exact}. An important difference is that charged solvable states generically relax to non-trivial generalised Gibbs ensembles while conventional solvable states always relax to the infinite temperature state. We then moved on to study the entanglement growth in these systems, which, because of the ballistic transport of charge, show linear growth of all R\'enyi entropies as opposed to systems where the transport of charge is diffusive~\cite{rakovszky2019sub, huang2020dynamics, znidaric2020entanglement}. Very surprisingly, however, we found that if one breaks the solvability condition the entanglement dynamics displays a sharp qualitative change. While after quenches from charged solvable states the entanglement entropy follows the conventional linear-growth-to-relaxation pattern, quenches from states that are not charged solvable show a two step relaxation with a second phase of entanglement growth characterised by a smaller slope. Interestingly, despite these circuits are dual-unitary, non-charged solvable states show a submaximal entanglement velocity even in the early time regime. Moreover, because of their two-step relaxation they can also display instances of quantum Mpemba effect. 

A pressing question for future research is to unveil what are the conditions for this two-step relaxation to occur: Our analysis suggests that the latter results from the constraints imposed by charge conservation over timescales where the information about the environment has fully transversed the subsystem of interest. This suggests that the two-step relaxation of the entanglement entropy does not rely on dual-unitarity and occurs in generic quantum circuits with conservation laws. The special feature brought by dual-unitarity is that a similar phenomenology is also observed for R\'enyi entropies with index larger than one.

Moreover, the findings discussed above are just starting to unveil the rich structure that conservation laws impose on the non-equilibrium dynamics of $\mathcal G$-symmetric dual-unitary circuits and we expect that future research will help identifying many more such exotic phenomena. An immediate question that can be directly attacked with the techniques introduced here concern, for instance, investigating the operator spreading in these systems, e.g. by computing out-of-time ordered correlators or the operator-space entanglement of local operators. Another interesting set of questions concerns the complexity of computing the time evolution of these circuits with classical computers. Conventional dual unitary circuits are known to be hard to simulate both directly~\cite{suzuki2022computational} and by exchanging the roles of space and time~\cite{foligno2023temporal}, and it is interesting to ask whether the additional charge structure can help the classical simulation. Finally, it is appealing to wonder whether one can generalise our construction to circuits with non-abelian symmetries or where the dual unitarity condition has been weakened following the logic of Ref.~\cite{yu2023hierarchical}. 

\begin{acknowledgments}
We acknowledge financial support from the Royal Society through the University Research Fellowship No.\ 201101 (A.\ F.\ and B.\ B.). P.\ C.\ has been supported by the European Research Council under Consolidator Grant number 771536 ``NEMO''. A.\ F.\ and B.\ B.\ thank SISSA for hospitality during the preparation of this work. 
\end{acknowledgments}

\appendix
\section{Charge Propagation in Dual-Unitary Circuits}
\label{app:chargepropagation}

Assume to have a conserved charge $Q$, which can be written as the sum of one-site charge densities 
\begin{align}
	Q=\sum_{x=0}^{L} q_{x}+q_{x+1/2}.
	\label{eq:fullcharge}\end{align}
Calling $\mathds{U}$ the time evolution operator (which commutes with $Q$),  we can write the continuity equation for the charge density (assume $x$ is an integer site)
\begin{align}
	\mathds{U}q^{{}}_{x} \mathds{U}^\dagger = q^{{}}_{x}- J^{{}}_{x+1/2}+ J^{{}}_{x-1/2},\label{eq:continuity equation}
\end{align}
we make no assumption on $J_x$, apart from the constraints due to causality and the request that it cancels out once summing on all sites as in Eq. \eqref{eq:fullcharge}. In fact, without loss of generality, we can take $J_x$ to be traceless. Indeed, if $\tra{J}\neq 0$ we can always consider $J\mapsto J'=J- {\tra{J}} \mathds{1} /{d}$, which also obeys \eqref{eq:continuity equation} and is traceless. 

Graphically, we can represent Eq.~\eqref{eq:continuity equation} using the notation introduced in Sec. \ref{sec:diagrams} as
\begin{align}
	&	\fineq[-0.8ex][0.8][1]{
		\roundgate[0][0][1][topright][bertiniorange][1]
		\roundgate[1][1][1][topright][bertiniorange][1]
		\roundgate[-1][1][1][topright][bertiniorange][1]
		\cstate[-.5][-.5]
		\cstate[.5][-.5]
		\cstate[1.5][.5]
		\cstate[-1.5][.5]
		\charge[-.3][-.3][red]
	}=\label{eq:chargeprop}\\&
	=\fineq[-0.8ex][0.7][1]{
		\foreach\i in {0,...,3}
		{
			\draw[thick] (-1.5+\i,.5)--++(0,.5);
		}
		\cstate[-.5][.5]
		\cstate[.5][.5]
		\cstate[1.5][.5]
		\charge[-.5][.8][red]
		\cstate[-1.5][.5]
	}-\fineq[-0.8ex][0.7][1]{
		\foreach\i in {0,...,3}
		{
			\draw[thick] (-1.5+\i,.5)--++(0,.5);
		}
		\cstate[1.5][.5]
		\draw[fill=black] (-1,.5) ellipse (0.7 and 0.15);
		\cstate[.5][.5]
	}+\fineq[-0.8ex][0.7][1]{
		\foreach\i in {0,...,3}
		{
			\draw[thick] (-1.5+\i,.5)--++(0,.5);
		}
		\cstate[-.5][.5]
		\draw[fill=black] (1,.5) ellipse (0.7 and 0.15);
		\cstate[-1.5][.5]
	},\notag
\end{align}
where we represented the $J$ operator with a black ellipse to indicate a generic two-site operator. Tracing out the two rightmost sites (in the replica language this corresponds to the insertion of a bullet) we have 
\begin{align}
	&	\fineq[-0.8ex][0.8][1]{
		\roundgate[0][0][1][topright][bertiniorange][1]
		\roundgate[1][1][1][topright][bertiniorange][1]
		\roundgate[-1][1][1][topright][bertiniorange][1]
		\cstate[-.5][-.5]
		\cstate[.5][-.5]
		\cstate[1.5][.5]
		\cstate[1.5][1.5]
		\cstate[.5][1.5]
		\cstate[-1.5][.5]
		\charge[-.3][-.3][red]
	}=0=\label{eq:charge2}\\&
	=d^2\left(\fineq[-0.8ex][0.7][1]{
		\foreach\i in {0,...,1}
		{
			\draw[thick] (-1.5+\i,.5)--++(0,.5);
		}
		\cstate[-.5][.5]
		\charge[-.5][.8][red]
		\cstate[-1.5][.5]
	}-\fineq[-0.8ex][0.7][1]{
		\foreach\i in {0,...,1}
		{
			\draw[thick] (-1.5+\i,.5)--++(0,.5);
		}
		\draw[fill=black] (-1,.5) ellipse (0.7 and 0.15);
	}\right)+\fineq[-0.8ex][0.7][1]{
		\foreach\i in {0,...,3}
		{
			\draw[thick] (-1.5+\i,.5)--++(0,.5);
		}
		\cstate[-.5][.5]
		\draw[fill=black] (1,.5) ellipse (0.7 and 0.15);
		\cstate[-1.5][.5]
		\cstate[1.5][1]
		\cstate[.5][1]
	},\notag
\end{align}
where we used dual unitarity and unitarity to simplify the left hand side of \eqref{eq:charge2} obtaining $0$ thanks to the traceless condition on $q^\rt$. We then immediately have 
\begin{align}
	\fineq[-0.8ex][0.7][1]{
		\foreach\i in {0,...,1}
		{
			\draw[thick] (-1.5+\i,.5)--++(0,.5);
		}
		\draw[fill=black] (-1,.5) ellipse (0.7 and 0.15);
	}=\fineq[-0.8ex][0.7][1]{
		\foreach\i in {0,...,1}
		{
			\draw[thick] (-1.5+\i,.5)--++(0,.5);
		}
		\cstate[-.5][.5]
		\charge[-.5][.8][red]
		\cstate[-1.5][.5]
	},
\end{align}
meaning that $J^{{}}_{x}=-q^{{}}_{x+1/2}$, which shows to a ballistic transportation of charge at the operatorial level. We will refer from this point on to the one-site charge densities as solitons. Explicitly, they obey
\be
\mathds{U}q^{{}}_{x} \mathds{U}^\dagger = q^{{}}_{x+1}
\ee
or graphically
\be
\fineq[-0.8ex][0.8][1]{
		\roundgate[0][0][1][topright][bertiniorange][1]
		\roundgate[1][1][1][topright][bertiniorange][1]
		\roundgate[-1][1][1][topright][bertiniorange][1]
		\cstate[-.5][-.5]
		\cstate[.5][-.5]
		\cstate[1.5][.5]
		\cstate[-1.5][.5]
		\charge[-.3][-.3][red]
	}=\fineq[-0.8ex][0.7][1]{
		\foreach\i in {0,...,3}
		{
			\draw[thick] (-1.5+\i,.5)--++(0,.5);
		}
		\cstate[-.5][.5]
		\cstate[.5][.5]
		\cstate[1.5][.5]
		\charge[1.5][.8][red]
		\cstate[-1.5][.5]
	}.\label{eq:rightmoving}
\ee
Similarly charges starting on half-integer sites can be shown to move in the left direction: We will use a superscript $\lt$ for left-moving solitons and $\rt$ for right-moving ones.

This immediately implies that conserved charges with support on half integer and integer sites are separately conserved, since, for example
\begin{align}
	\mathbb{U}{Q'}\mathbb{U}^\dagger=\mathbb{U}\sum_{x=0}^L q_{x}\mathbb{U}^\dagger=
	\sum_{x=0}^L q_{x+1}=Q'.
\end{align}
Moreover, one can produce exponentially many conserved charges~\cite{bertini2020operator}. Noting that the identity operator $\mathds{1}_d$ trivially fulfils Eq.~\eqref{eq:rightmoving} we have that a charge density with support on $\ell$ sites, having on each integer site either $q^\rt$ or $\mathds{1}_d $, obeys Eq.~\eqref{eq:rightmoving} too. This gives us $2^\ell$ choices for the charge density. Ruling out the identity operator itself (this corresponds to choosing $\mathds{1}_d$ on each site) one is left with $2^\ell-1$ independent charge densities with support on integer sites; considering also the ones on half integer sites (left-moving) the total number of conserved charges is $2^{\ell+1}-2$. More generally, with $m_\ltb/m_\rtb$ solitons (including the identity), the total number of charges with support on at most $2\ell$ sites is $(m_\ltb)^\ell+(m_\rtb)^\ell -2$.

\section{Proof of Theorem \ref{th:chargeclassification}}
\label{app:chargeclassifications}

In this appendix we provide a proof of Theorem~\ref{th:chargeclassification}.

Without loss of generality we work with right-moving solitons. We begin by noting that one can simplify the leftmost gate in the diagram in Eq.~\eqref{eq:rightmoving} (multiplying both sides by its inverse) obtaining
\begin{align}
	&\left(\mathds{1}_d\otimes U\right) \left(U\otimes\mathds{1}_d\right)		\left(\sigma^{{\rt}} \otimes \mathds{1}_d^{\otimes 2}\right)\left(U^\dagger\otimes\mathds{1}_d\right)\left(\mathds{1}_d\otimes U^\dagger\right)	 = \notag \\
	&=\left(\mathds{1}_d^{\otimes 2} \otimes \sigma^{{\rt}} \right).\label{eq:th1-2}
\end{align}
Equation \eqref{eq:th1-2} is our starting point. First we note that it implies that the space of right-moving solitons forms an algebra: If $\sigma^{{\rt}}$ and $\sigma'^{{\rt}}$ both obey \eqref{eq:th1-2}, then their sum and product also  do. {Note that in general the members of this space are not hermitian (because we allow for linear combination with complex coefficient) but they are be normal operators, because by assumption the solitons $\sigma^{{\rt}}_i$ commute which each other.} We call this algebra $Q^{{\rt}}$.

This observation is enough to prove the first part  of the theorem. If $\sigma^{{\rt}}$ is a soliton we can decompose it in orthogonal subspaces (this can always be done because $\sigma^{{\rt}}$ normal) as follows 
\begin{align}
	\sigma^{{\rt}}=\sum_\mu \Pi_\mu \mu,
\end{align}
where we called $\Pi_\mu$ the projector in the eigenspace corresponding to the eigenvalue $\mu$.
Then, for every eigenvalue $\mu^*$, one can define the polynomial
\begin{align}
	P_{\mu^*}(x)=\prod_{\mu \ne \mu^*} (x-\mu),
\end{align}
such that $P_{\mu^*}(\sigma^{{\rt}})\propto \Pi_{\mu^*}$. But since  the space of solitons is closed under product and sum also $P_{\mu^*}(\sigma^{{\rt}})$ must be a soliton. Since these projectors are linearly independent this proves that we can always choose a basis where the solitons are projectors (this is true even if the charge densities do not commute).

To prove the second part of the theorem we rewrite Eq.~\eqref{eq:th1-2} as 
\begin{align} 
\left(U\otimes\mathds{1}_d\right)\left(\sigma^{{\rt}} \otimes \mathds{1}_d\otimes\mathds{1}_d\right)\left(U^\dagger\otimes\mathds{1}_d\right)	 = \notag \\
	=
	\left(\mathds{1}_d\otimes U^\dagger\right)\left(\mathds{1}_d\otimes \mathds{1}_d \otimes \sigma^{{\rt}} \right)
	\left(\mathds{1}_d\otimes U\right)
	\label{eq:th1-1}.
\end{align}
We now observe that the left hand side of Eq.~\eqref{eq:th1-1} is the identity on the last (rightmost) site , while the right hand side is the identity on the first, meaning that both sides have to be equal to \begin{align}
	\mathds{1}_d \otimes \mathscr{L}[\sigma^{{\rt}}] \otimes 	\mathds{1}_d ,
\end{align} 
where $\mathscr{L}[\bullet]$ is some generic function which is defined from the space of the conserved charge densities $\sigma^{{\rt}}$ to the space of local operators on one site.
More specifically, we can rewrite \eqref{eq:th1-1} as \begin{align}
	U	\left(\sigma^{{\rt}} \otimes \mathds{1}_d \right)U^\dagger= \mathds{1}_d \otimes \mathscr{L}[\sigma^\rt].\label{eq:th1-3}
\end{align}
\begin{widetext}
We now show that if $\sigma^\rt$ obeys Eq.~\eqref{eq:th1-2}, then also $\mathscr{L}[\sigma^\rt]$ does. Starting from Eq.~\eqref{eq:th1-3} we expand it on three sites (by attaching an identity on the left) and conjugate both sides with $\mathds{1}_d \otimes U$ to obtain
\begin{align}
(\mathds{1}_d \otimes U)		\left(U \otimes\mathds{1}_d\right)	\left(\sigma^{{\rt}} \otimes \mathds{1}_d\otimes \mathds{1}_d \right)\left(U^\dagger \otimes\mathds{1}_d\right)(\mathds{1}_d \otimes U^\dagger)= (\mathds{1}_d \otimes U)(\mathds{1}_d \otimes \mathscr{L}[\sigma^\rt]\otimes\mathds{1})(\mathds{1}_d \otimes U^\dagger),
\end{align}
which can be simplified using \eqref{eq:th1-2} on the left hand side to obtain 
\begin{align}
		\left(\mathds{1}_d\otimes \mathds{1}_d \otimes\sigma^{{\rt}} \right)=(\mathds{1}_d \otimes U)(\mathds{1}_d \otimes \mathscr{L}[\sigma^\rt]\otimes\mathds{1})(\mathds{1}_d \otimes U^\dagger). 
\end{align}
Finally, we can rewrite this equation by multiplying both sides with the same combination of gates, and then substituting Eq.~\eqref{eq:th1-3}. Specifically 
\begin{align}
\left(\mathds{1}_d \otimes\mathds{1}_d \otimes U\right)	(\mathds{1}_d \otimes U\otimes\mathds{1}_d)(\mathds{1}_d \otimes \mathscr{L}[\sigma^\rt]\otimes\mathds{1}_d\otimes\mathds{1}_d)(\mathds{1}_d \otimes U^\dagger\otimes\mathds{1}_d )\left(\mathds{1}_d \otimes\mathds{1}_d \otimes U^\dagger\right)=\notag\\
=\left(\mathds{1}_d\otimes \mathds{1}_d \otimes U \right)\left(\mathds{1}_d\otimes \mathds{1}_d \otimes\sigma^{{\rt}} \otimes \mathds{1}_d\right)\left(\mathds{1}_d\otimes \mathds{1}_d \otimes U^\dagger \right)=\left(\mathds{1}_d\otimes \mathds{1}_d \otimes  \mathds{1}_d \otimes\mathscr{L}[\sigma^\rt]\right). 
\end{align}
\end{widetext}
Ignoring all the identities on the leftmost sites, this shows that $\mathcal{L}[\sigma^\rt]$ is itself a right-moving soliton, implying that $\mathscr{L}$  is a function that goes from the space of conserved, right-moving solitons to itself. Additionally, notice that $\mathscr{L}[\sigma^\rt]$ is an \emph{endomorphism}, meaning that not only it maps the algebra  $Q^\rt$ to itself, but it is also compatible with its operations, and that 
\begin{align}
	\left(\mathscr{L}\right)^{2}[\sigma^{{\rt}}]=\sigma^{{\rt}}, 
	\label{eq:th1-4}
\end{align}
where we used Eq.~\eqref{eq:th1-2}. Finally, from Eq.~\eqref{eq:th1-3}, we see that $\mathscr{L}$ is unitary with respect to the Hilbert–Schmidt inner product 
\begin{align}
	\langle \sigma_1,\sigma_2\rangle=\tra{\sigma_1^\dagger \sigma_2}.
\end{align}
This means that $\mathscr{L}$ can be diagonalised and its eigenvalues must lie on the unit circle. In fact, Eq.~\eqref{eq:th1-4} implies that they can only be $\pm1$ and we denote the corresponding eigenspaces and their respective dimension as $Q^\pm$ and $n_\pm$. We consider the subspace of charges with eigenvalue $+1$ and call it $Q^+$. It is immediate to see that this is a sub-algebra of $Q$: if $\sigma_1^{{\rt}},\sigma_2^{{\rt}}\in Q^+$ then 
\begin{align}
	\mathscr{L}[\sigma_1^{{\rt}} \sigma_2^{{\rt}}]&=\mathscr{L}[\sigma_1^{{\rt}}]\mathscr{L}[\sigma_2^{{\rt}}]= \sigma_1^{{\rt}}\sigma_2^{{\rt}}\notag\\
	&\implies \sigma_1^{{\rt}}\sigma_2^{{\rt}} \in Q^+,
\end{align}
and similarly for the sum operation. By the theorem's assumptions all the solitons commute, so we diagonalise all of them simultaneously and find a common decomposition in eigenspaces. We use the projectors on these eigenspaces as a basis of $Q^+$ and denote its elements by $\sigma^{{\rt}}_{+,\alpha}$.

Next, we consider solitons in $Q^-$, the vector space corresponding to the negative eigenvalue of $\mathscr{L}$. This vector space is not closed under multiplication, since if $\sigma_1^-,\sigma_2^-\in Q^-$, then
\begin{align}
	\mathscr{L}[\sigma_1^- \sigma_2^-]&=\mathscr{L}[\sigma_1^-]\mathscr{L}[\sigma_2^-]= (-\sigma_1^-)(-\sigma_2^-)\notag\\
	&\implies \sigma_1^-\sigma_2^- \in Q^+.
\end{align}
Then we decompose the solitons $\sigma_{-,\alpha}^{{\rt}}\in Q^-$ as follows: first we notice that $\sigma_{-,\alpha}^{{\rt}\,2} \in Q^+$ so it can be decomposed into the orthogonal projectors as before
\begin{align}
	\left(\sigma_{-,\alpha}^{{\rt}}\right)^2=\bigoplus_\beta \mu_\beta\, \sigma^{{\rt}}_{+,\beta}.
	\label{eq:th2-5}
\end{align}
If some $\mu_\beta\neq 0$ is degenerate we decompose $\sigma_{-,\alpha}^{{\rt}}$ further as
\begin{align}
	\sigma_{-,\alpha}^{{\rt}} = \bigoplus_{\beta|\mu_\beta\ne 0} \sigma_{-,\alpha}^{{\rt}} \sigma_{+,\beta}^{{\rt}},
\end{align}
and take each $\sigma_{-,\alpha}^{{\rt}}\sigma_{+,\beta}^{{\rt}}$ as a new basis of solitons (being careful of maintainingva the linear independence).
We iterate this procedure for all the solitons in $Q^-$ in such a way to obtain a new basis, and reorder the indices of the solitons such that  \begin{align}
	\sigma^\rt_{+,\alpha}\sigma^\lt_{-,\beta}=\delta_{\alpha \beta}\sigma^\lt_{-,\beta}\qquad \forall \alpha,\beta\le n_-\\
	\sigma^\rt_{+,\alpha}\sigma^\lt_{-,\beta}=0 \qquad \forall \alpha> n_-, \beta\le n_-.
\end{align}
We then define the projectors belonging to the first group of Theorem~\ref{th:chargeclassification} as the ones in $Q^+$ orthogonal to all the solitons $\in Q^-$
\begin{align}
	\sigma^{{\rt}}_{+,\alpha}=\Pi_\alpha^{{\rt}} \qquad \alpha >n_-.
\end{align}
To find those in the second group we observe that since $\mathscr{L}[\sigma_{-,\alpha}^{{\rt}} ]=- \sigma_{-,\alpha}^{{\rt}}$, this means that a similarity transformation (which defined $\mathscr{L}$)  connects $\sigma_{-,\alpha}^{{\rt}} $  and $-\sigma_{-,\alpha}^{{\rt}} $. Thus the spectrum of $\sigma_{-,\alpha}^{{\rt}} $ must be symmetric around the origin. Then, given that ${\sigma_{-,\alpha}^{{\rt}2}\propto \sigma_{+,\alpha}^{{\rt}}}$ has only a nontrivial eigenspace, $\sigma^-_\alpha$ must be written as
\begin{align}
	\!\!\sigma_{-,\alpha}^{{\rt}}\!=\!\mu (P^{{\rt}}_{+,\alpha}-P^{{\rt}}_{-,\alpha})
	\implies
	\sigma^+_\alpha\!\propto\! P^{{\rt}}_{+,\alpha}+P^{{\rt}}_{-,\alpha} ,
\end{align}
for some orthogonal projectors $P^\pm_\alpha$. It is then immediate to see that
\begin{align}
	&\mathscr{L}[P^{{\rt}}_{+,\alpha}]=\notag\\	&=\mathscr{L}[P^{{\rt}}_{+,\alpha}-P^{{\rt}}_{-,\alpha}+P^{{\rt}}_{+,\alpha}+P^{{\rt}}_{-,\alpha}]/2=
	P^{{\rt}}_{-,\alpha},\\
	&\mathscr{L}[P^{{\rt}}_{-,\alpha}]=\notag\\	&=\mathscr{L}[P^{{\rt}}_{-,\alpha}-P^{{\rt}}_{+,\alpha}+P^{{\rt}}_{-,\alpha}+P^{{\rt}}_{+,\alpha}]/2=P^{{\rt}}_{+,\alpha},
\end{align}
proving Eq.~\eqref{eq:secondgroupP}. To prove the completeness relation in Eq.~\eqref{eq:completeness}, it is sufficient to note that in our notation the identity operator is a conserved charge (altough trivial) belonging in $Q^+$. Therefore, if the sum of all the mutually orthogonal projectors  $\Pi^{{\rt}}_\alpha,P^{{\rt}}_{\pm,\alpha}$ does not give the identity our set of charges is not a complete basis (note that, since the projectors are mutually orthogonal, their only linear combination that produces the identity is when they are all summed with coefficient $1$), and we find a contradiction.

\section{Proof of Theorems \ref{th:solvabilityclassification}-\ref{th:characterizationsolvMPS}}
\subsection{Left solvability}
\label{app:solvablestates}

Here we prove Theorem \ref{th:solvabilityclassification} following the proof of Theorem 1 in Ref.~\cite{piroli2020exact}. Without loss of generality we consider the case of left charged solvable states. The condition that the transfer matrix $\tau(\mathcal{M})$ has spectrum contained in the disc $\abs{\lambda}<1$ with only one eigenvalue on its border, corresponds to the request that the initial state $\ket{\Psi(0)}$ is normalised in the thermodynamical limit, since (assuming a periodic boundary)
\begin{align}
	\braket{\Psi^{L}_0(\mathcal{M})}=\tra{\tau(\mathcal{M})^{L}}.
\end{align} 
The left solvability condition \eqref{eq:leftsolvabilityformula} implies that the left eigenvector corresponding to this eigenvalue is the (vectorized) operator $S$, i.e., 
\begin{align}
	\bra{S}=\sum_{i,j}\mel{i}{S}{j}\bra{i,j}.
\end{align}
To see this explicitly, we trace Eq.~\eqref{eq:leftsolvabilityformula} on the physical space indices, finding:\begin{align}
\sum_{i,j}	\left(\mathcal{M}^{i,j}\right)^\dagger S\mathcal{M}^{i,j}=\sum_{\alpha}\frac{c^{{\rt}}_\alpha}{d_{\alpha}}\tra{\Pi^{{\rt}}_\alpha} S= S.\label{eq:initialass}
\end{align}
Then, we can use Theorem 3.5 in Ref.~\cite{SpectralProperties}, which states that a completely positive map $\mathcal{E}$ with a spectral radius $r$ ($r=1$ in this case), has at least one positive operator $X$ such that $\mathcal{E}[X]=rX$. This theorem can be applied to the matrix $\tau$, which can also be seen as a linear map on the space of operators
\begin{align}
	\tau\leftrightarrow \mathcal{E}_\tau[X]=\sum_{i,j} \left(\mathcal{M}^{i,j}\right)^\dagger X \mathcal{M}^{i,j}.
\end{align}
By assumption, the spectrum of $\tau$ is nondegenerate at the border of the unit circle, and by virtue of left solvability \eqref{eq:leftsolvabilityformula} we know $S$ is a fixed point, from which it follows that $S$ must be a positive operator.

We now show that a solvable MPS is always equivalent, in the thermodynamic limit, to another MPS for which the matrix $S$ (defined as the unique fixed point of the transfer matrix), is not only positive, but strictly positive. This is equivalent to Eq.~\eqref{eq:leftsolv}, because if $S$ is strictly positive it is invertible, and we can do a gauge transformation on the MPS
\begin{align}
	\mathcal{M}^{a,b}\rightarrow A \mathcal{M}^{a,b} A^{-1}\implies S\rightarrow  \left(A^{-1} \right)^* S A^{-1},\label{eq:gaugetransf}
\end{align}
which maps $S$ to the identity operator by choosing $A=\sqrt{S}$.\\
To show the strict positivity of $S$, suppose by contradiction that $S$ has  some zero eigenvalues.
Choosing a basis where $S$ is diagonal ($S=\sum_\alpha \mu_\alpha  \ketbra{\alpha}$), we have\begin{align}
0 \!=\! \sum_\alpha \mu_\alpha  \ketbra{\alpha} \!-\! \sum_{i,j} \sum_\alpha \mu_\alpha \left(\mathcal{M}^{i,j}\right)^\dagger \ketbra{\alpha} \mathcal{M}^{i,j} .
\label{eq:7}
\end{align}
Let $P$ be the projector on the kernel of $S$ --- $P=\mathds{1}_\chi-\sum_\alpha \ketbra{\alpha}$ ---  then Eq. \eqref{eq:7} implies \begin{align}
	\sum_{i,j} \sum_\alpha \mu_\alpha P\left(\mathcal{M}^{i,j}\right)^\dagger \ketbra{\alpha} \mathcal{M}^{i,j} P =0.
\end{align}
The matrices inside the sum  
\begin{align}
	P\left(\mathcal{M}^{i,j}\right)^\dagger \ketbra{\alpha} \mathcal{M}^{i,j} P, 
\end{align}
are all positive and we are requiring their sum to vanish. This can only be true if each of them is $0$
\begin{align}
\forall \alpha,i,j\qquad 	&P\left(\mathcal{M}^{i,j}\right)^\dagger \ketbra{\alpha} \mathcal{M}^{i,j} P =0\notag\implies\\&
 \mathcal{M}^{i,j} P=P\mathcal{M}^{i,j} P,
\end{align}
meaning that there is a basis in which the  matrices $M^{i,j}$ are all triangular
\begin{align}
	\mathcal{M}^{i,j}=\begin{pmatrix}
		\mathcal{A}^{i,j} & 		\mathcal{B}^{i,j}\\
				0 & \mathcal{C}^{i,j}\label{eq:MPStriangular}
	\end{pmatrix}.
\end{align}
Given our periodic boundary conditions on the MPS, the MPS is completely equivalent to one that only has the diagonal part in Eq.~\eqref{eq:MPStriangular} (i.e.\ $\mathcal{B}^{ij}$ can be set to $0$). Such MPS can be written as $\ket{\Psi}=\ket{\Psi_1}+\ket{\Psi_2}$, where $\ket{\Psi_1}$ is obtained using the matrices $\mathcal{A}^{ij}$ as MPS and $\ket{\Psi_2}$ with $\mathcal{B}^{i,j}$. Since the leading eigenvalue of $\tau$ is nondegenerate by assumption, and has support only on the block corresponding to $\mathcal{A}^{ij}$ it means that the norm of $\ket{\Psi_{2}}$ is suppressed in the thermodynamic limit, so that we can  reduce the bond dimension using the MPS 
\begin{align}
	P\mathcal{M}^{ij}P,
\end{align}
which is equivalent (in the thermodynamical limit) to the original one. In this reduced space the matrix $S$ is strictly positive, proving the theorem.

\subsection{Charged solvability}
\label{app:proofsolvabilitybothsides}

Suppose to have a state which obeys both Eqs.~\eqref{eq:leftsolvabilityformula} and \eqref{eq:rightsolvabilityformula}. Using the left-solvability we can consider an equivalent MPS which can written as in \eqref{eq:leftsolvablestateconstruction}
\begin{align}
	\Gamma(\mathcal{M})= V \left(\sqrt{C^{{\rt}} }\otimes \mathds{1}_\chi\right), 
\end{align}
where we defined
	\begin{align}
	C^{{{\rt}}}\equiv \sum_{\alpha=1}^{m_\rtb}\frac{c^{{{\rt}}}_\alpha}{d^{{{\rt}}}_\alpha} \Pi^{{{\rt}}}_\alpha.
\end{align}
Similarly we can define  
\begin{align}
	C^{{{\lt}}}\equiv \sum_{\alpha=1}^{m_\ltb}\frac{c^{{{\lt}}}_\alpha}{d^{{{\lt}}}_\alpha} \Pi^{{{\lt}}}_\alpha.
\end{align}
Then, condition \eqref{eq:rightsolvabilityformula} tells us that there is a \emph{unique} matrix $S$ in the auxiliary space such that \begin{align}
V \left(C^{{\rt}} \otimes S\right) V^\dagger= C^{{\lt}} \otimes S.
\label{eq:condition1}
\end{align}
First of all we note that, since $V$ is a unitary matrix, the operation $V \left(\bullet\right) V^\dagger$ is a similarity transformation. This means that the spectra of $C^{{\rt}}\otimes S$ and $C^{{\lt}}\otimes S$ have to match, implying that there is a matrix $U'\in SU(d)$ such that
\begin{align}
	C^{{\rt}}=U'C^{{\lt}} U'^\dagger.
\end{align}
We use $U'$ to change the basis on the odd sites (which generate left moving particles), so that $C^{{\rt}}=C^{{\lt}}\equiv C$: this proves the blocking condition in Eq.~\eqref{eq:blockingcondition}.

Repeating the reasoning of App.~\ref{app:solvablestates}, one can show that $S$ is a strictly positive operator (or that the MPS is equivalent to another MPS with lower bond dimension such that this property holds). In passing we note that there are examples where $S$ cannot be taken to be the identity matrix, differently from the solvable states of Ref.~\cite{piroli2020exact}. 

\section{Late time regime for left charged solvable states}
\subsection{Leading eigenvectors of the transfer matrix}
\label{app:transferM}

We consider transfer matrices as the one in Eq.~\eqref{eq:transfermatrsubstitution}. We start by writing the right-moving soliton on the main diagonal as a linear combination of projectors
\begin{align}
	\mathcal{T}=\sum_{\alpha=1}^{m_\rtb} c_\alpha^{{\rt}} \fineq[-0.8ex][0.750001][1]{\roundgate[0][0][1][topright][bertiniorange][1]
		\roundgate[1][1][1][topright][bertiniorange][1]
		\cstate[1.5][1.5]
		\cstate[-.6][-.6]
		\charge[-.3][-.3][red]
		\draw [<-] (-.3,-.4)--(.2,-1);
	\node at (.6,-1) {$\Pi^{{\rt}}_\alpha$};	}\frac{1}{d^{{\rt}}_\alpha}\equiv \sum_{\alpha=1}^{m_\rtb} c_\alpha^{{\rt}}\mathcal{T}_\alpha.
\label{eq:transfermatrixlsolvable}
\end{align}
In this way, we express the transfer matrix as a convex combination (with coefficients $c_\alpha^{{\rt}}\ge 0$ that sum to $1$) of other transfer matrices $\mathcal{T}_\alpha$ which have the projectors $\Pi^{{\rt}}_\alpha$ propagating along the main diagonal. 

Note that, by unfolding, these transfer matrices can be expressed as quantum channels written in Kraus form, i.e., 
\begin{align}
\mathcal{T}_\alpha\leftrightarrow\mathcal{E}_\alpha[X]&=\frac{1}{d^{{\rt}}_\alpha}
	\fineq[-0.8ex][0.750001][1]{
		\begin{scope}[shift={(0,0)},rotate around={45:(0,0)}]
		\roundgate[0][0][1][topright][bertinired]
		\roundgate[-1][-1][1][topright][bertinired]
	\end{scope}		\begin{scope}[shift={(3,0)},rotate around={45:(0,0)}]
	\roundgate[0][0][1][topright][bertiniblue]
	\roundgate[-1][-1][1][topright][bertiniblue]
\end{scope}
\draw[rounded corners=10pt] (0.707,0.5)--++(0,-2.5)--++(1.586,0)--++(0,2.5)--cycle;
\node[scale=2] at (1.5,-.8) {$X$};	
\draw[thick, rounded corners=3pt] (0,.707)--++(0,.2)--++(3,0)--++(0,-.2);	
\draw[thick, rounded corners=3pt] (0,-.707*3)--++(0,-.2)--++(3,0)--++(0,.2);
\begin{scope}[shift={(1.5,-.707*3-.2)},scale=2]
\cstate[0][0]
\node[scale=0.5] at (0,0)  {$\Pi^{{\rt}}_\alpha$};
	\end{scope}}=\notag\\
&=\sum_{i,j=1}^d K^{i,j}_\alpha X \left(K^{i,j}_\alpha\right)^\dagger,
\label{eq:quantumchannel}
\end{align}
where we introduced 
\begin{align}
 K^{i,j}_\alpha=\sqrt{\frac{\mel{i}{ \Pi^{{\rt}}_\alpha }{i}}{d^{{\rt}}_\alpha}} \fineq[-0.8ex][0.750001][1]{		\begin{scope}[shift={(0,0)},rotate around={45:(0,0)}]
 		\roundgate[0][0][1][topright][bertinired]
 		\roundgate[-1][-1][1][topright][bertinired]
 \end{scope}	
\node at (0,1) {$j$};
\node at (0,-2.3) {$i$};},
\end{align}
and used $i,j$ as indexes for an orthonormal basis on one site. Unitarity and dual-unitarity imply the channel is both trace preserving and unital (and completely positive since it is in Kraus form), thus showing its maximal eigenvalue has to be $\le 1$.

Since the original transfer matrix is a generic convex combination of these channels, it has an eigenvalue of magnitude $1$ if and only if all the channels $\mathcal{E}_{\alpha}$ have common eigenvectors with equal eigenvalue of magnitude $1$. Some are fixed by the soliton property. Indeed, all operators of the form (we denote by $x$ the number of legs of this quantum channel)
\begin{align}
X^{(\vec{\beta})}=\bigotimes_{i=1}^x \frac{\Pi^{{\lt}}_{\beta_i}}{d^{{\lt}}_{\beta_i}},
\label{eq:maximaleigenvector}
\end{align}
are eigenvectors of $\mathcal{E}_\alpha$ with eigenvalue $1$, where we denote with $\vec{\beta}$ a string of length equal to the number of legs of the transfer matrix, taking values in $\beta_i=1,\ldots ,m_{\ltb}$.

We expect these to be the only eigenvectors with this property, assuming that all the blocks in which the charge conserving gate is built upon (see Eq.~\eqref{eq:DUdecomposiiton}) are chosen independently and are generic enough. This statement is proven rigorously in \cite{folignoinprep} for blocks of dual unitaries of local dimension $d_\alpha^{(\rtb/\ltb)}=2$. 

Let us now contradict this hypothesis and suppose there exists a common, nontrivial eigenvector of all the channels $\mathcal{E}_\alpha$, starting from  $x$ sites: we want to show that this implies the existence of extra conserved charges (which cannot be built with the one-site solitions). Consider a transfer matrix as in Eq.~\eqref{eq:transfermatrixlsolvable}, with the choice ${c^{{\rt}}_\alpha}=d^{\rt}_\alpha/d$, which can be written simply as \begin{align}
	\mathcal{T}^\lt =	\frac{1}{d} \fineq[-0.8ex][0.750001][1]{\roundgate[0][0][1][topright][bertiniorange][1]
			\roundgate[1][1][1][topright][bertiniorange][1]
			\roundgate[2][2][1][topright][bertiniorange][1]
			\cstate[2.5][2.5]
			\cstate[-.6][-.6]
				}.
		\label{eq:transfermatrixl}\end{align}
By assumption, if all the reduced channels $\mathcal{E}_\alpha$ have a common eigenvector (with the same eigenvalue $\lambda$ such that $\abs{\lambda}=1$), it means that also this matrix has a nontrivial eigenvector $\ket{v}$ with a phase as eigenvalue

\begin{align}
	\ket{v}&=	\mathcal{T}^\lt\ket{v}= \frac{1}{d}\fineq[-0.8ex][0.750001][1]{\roundgate[0][0][1][topright][bertiniorange][1]
		\roundgate[1][1][1][topright][bertiniorange][1]
		\roundgate[2][2][1][topright][bertiniorange][1]
		\cstate[2.5][2.5]
		\cstate[-.6][-.6]
 \draw[black,thick,fill=jonasgreen] (.5-.1,-.5-.1) arc (-45-40:-45+40:2.4)--cycle;
		}\notag\\
		&=e^{i\phi}\fineq[-0.8ex][0.750001][1]{
		\foreach \i in {0,1,2}
		{
			\draw[thick] (\i+.5,\i-.5)--++(-.3,.3);
		}
		\draw[black,thick,fill=jonasgreen] (.5-.1,-.5-.1) arc (-45-40:-45+40:2.4)--cycle;
	}.
\end{align}
Given the unitarity of the folded gates (which are a tensor product of two unitary gates so are unitary), and the fact that the norm squared of the bullet state is $d$, this implies
\begin{align}
\fineq[-0.8ex][0.750001][1]{\roundgate[0][0][1][topright][bertiniorange][1]
		\roundgate[1][1][1][topright][bertiniorange][1]
		\roundgate[2][2][1][topright][bertiniorange][1]
		\cstate[-.6][-.6]
		\draw[black,thick,fill=jonasgreen] (.5-.1,-.5-.1) arc (-45-40:-45+40:2.4)--cycle;
	}=e^{i\phi}\fineq[-0.8ex][0.750001][1]{
		\foreach \i in {0,1,2}
		{
			\draw[thick] (\i+.5,\i-.5)--++(-.3,.3);
		}
		\draw[black,thick,fill=jonasgreen] (.5-.1,-.5-.1) arc (-45-40:-45+40:2.4)--cycle;
		\draw[thick] (2.5,2.5)--++(.5,.5);
				\cstate[2.5][2.5]
	}.\label{eq:eigenvaluetransferM}
\end{align} 
Now consider a charge density $q$ built as follows
\begin{align}
	\hspace{-.5cm}\fineq[-0.8ex][0.750001][1]{	
		\draw[black,thick,fill=jonasgreen] (.5-.1,-.5-.1) arc (-45-40:-45+40:2.4)--cycle;
		\foreach \i in {0,1,2}
		{
			\draw[thick] (\i+.5,\i-.5)--++(-.25,.25);
		}		
		\roundgate[0][0][1][topright][bertiniorange][1]
		\roundgate[1][1][1][topright][bertiniorange][1]
		\roundgate[-1][1][1][topright][bertiniorange][1]
		\cstate[-.5][-.5]
		\cstate[-1.5][.5]		}\hspace{-.25cm}=
		\fineq[-0.8ex][0.750001][1]{	
			\draw[thick,black,fill=pink] (.5,-.5) rectangle (3.5,0);
		\foreach \i in {0,...,4}
			{			
				\draw[thick] (\i/1.6+.75,0)--++(0,.3);
			}		
		}\equiv q.
\label{eq:newchargedensity}
\end{align}
Using \eqref{eq:eigenvaluetransferM}, it is straightforward to show that this charge density moves ballistically, i.e., it obeys
\begin{align}
&\fineq[-0.8ex][0.750001][1]{	
		\draw[thick,black,fill=pink] (2,-1) rectangle ++(5,.5);
		\foreach \i in {0,...,3}
		{
			\ifnumgreater{\i}{0}{
		\roundgate[2*\i][0][1][topright][bertiniorange][1]}
	{}
		\roundgate[2*\i+1][1][1][topright][bertiniorange][1]
		}	
		\cstate[1.5][-.5]
		\cstate[.5][.5]
		\cstate[7.5][.5]
	}=\label{eq:newballisticcharge}\\
&e^{2i\phi} \fineq[-0.8ex][0.750001][1]{	
	\draw[thick,black,fill=pink] (2,-1) rectangle ++(5,.5);
	\foreach \i in {0,...,7}
	{
		\draw[thick](2.5+\i,-.5)--++(0,.3);
		\ifnumgreater{\i}{4}{
			\cstate[2.5+\i][-.5];}{}
	}
	}.
\notag
\end{align}
However, since the matrix in pink is finite dimensional and the unitary matrices applied to it on the left side of Eq. \eqref{eq:newballisticcharge} are a similarity transformation, the spectrum of the matrix cannot be changed after the transformation, implying that ${\phi}/{\pi}\in \mathds{Q}$.

Finally, we note that it is possible to renormalise the gates to make all the charges like those in Eq.~\eqref{eq:newballisticcharge} one-site and two-site charge density, and, moreover, turn the phase in \eqref{eq:newballisticcharge} to $1$. A blocked $n$-sites dual unitary gate is constructed as  
\begin{align}
\fineq[-0.8ex][0.750001][1]{	
	\foreach \i in {0,...,2}
	{
		\foreach \j in {0,...,2}
		{
			\roundgate[-\i+\j][\i+\j][1][topright][bertinired][-1]	
		}	
	}		  }=\fineq[-0.8ex][0.750001][1]{	
\begin{scope}[scale=3]
	\roundgate[0][0][1][topright][violet][-1]	
	\draw[thick] (.25,.1)--++(.25,.25);
	\draw[thick] (.1,.25)--++(.25,.25);
	\draw[thick] (-.25,.1)--++(-.25,.25);
	\draw[thick] (-.1,.25)--++(-.25,.25);
	\draw[thick] (.25,-.1)--++(.25,-.25);
	\draw[thick] (.1,-.25)--++(.25,-.25);
	\draw[thick] (-.25,-.1)--++(-.25,-.25);
	\draw[thick] (-.1,-.25)--++(-.25,-.25);
\end{scope},
}		  \label{eq:renormalizedgate}
\end{align}
and, crucially, it continues to be dual unitary. By grouping enough sites together, we can make it such that a single time step of the renormalised circuit commutes with the new charge density. More precisely, if the new gate is formed by $r^2$ old gates, then a new time-step will give a phase $e^{2ir\phi}$, however, since ${\phi}/{\pi}\in \mathds{Q}$, it must be possible to find an $r$ such that this phase is $1$ and we have a ``standard'' conserved charge.

\subsection{Bound between left/right entanglement entropies} \label{app:boundsRL}
In this appendix we compute the R\'enyi entropies of the state in Eq.~\eqref{eq:GGEforMPS}. We begin by expressing the latter in formulae as follows 
\be
\rho_A(\infty)=	\rho^{GGE}_\ltb\otimes\rho^{GGE}_\rtb
\label{eq:GGEexpr}
\ee
where the tensor product is between even and odd sites and we defined 
\begin{align}
&\!\!\!\!\rho^{GGE}_\rtb=\left(\sum_{\alpha} \frac{c_\alpha^{\rt}}{d_\alpha^{\rt }}\Pi^{\rt}_\alpha\right)^{\otimes L_A},\\
&\!\!\!\!\rho^{GGE}_\ltb =\sum_{\vec{\alpha}}\frac{1}{d^{\lt}_{\vec{\alpha}}}\!\! \left[\bigotimes_{i=1}^{L_A} \Pi^{\lt}_{\alpha_i} \mel{\mcirc_{\chi,1}}{N^{\alpha_1}N^{\alpha_2}\cdots}{S}\right]\!\!,
\end{align}
and in the second equation we set 
\begin{align}
&{N}^\alpha\equiv \sum_{i,j=1}^{d} \mathcal{M}^{i,j}\otimes \left(\mathcal{M}^{i,j}\right)^* \mel{i}{\Pi_\alpha^{\lt}}{i}, \\
& d^{\lt}_{\vec{\alpha}}\equiv\prod_i d^{\lt }_{\alpha_i}. 
\end{align}
Considering traces of powers of $\rho_A(\infty)$ we have 
\be
\tr[\rho_A(\infty)^n] = \tr[\rho^{GGE\,\, n}_\ltb] \tr[\rho^{GGE\,\, n}_\rtb], 
\ee
where 
\be
\tr[\rho^{GGE\,\, n}_\rtb] =\left(\sum_{\alpha} \frac{c_\alpha^{\rt\, n}}{d_\alpha^{\rt\, n-1}}\right)^{L_A}, 
\ee
and 
\begin{align}
\tr[\rho^{GGE\,\, n}_\ltb]=\sum_{\vec{\alpha}}  \frac{\mel{\mcirc_{\chi,1}}{N^{\alpha_1}N^{\alpha_2}\cdots}{S}^n}{d^{{\lt\,n-1}}_{\vec{\alpha}}}.
\end{align}
The latter expression can be rewritten as
\begin{align}
\tr[\rho^{GGE\,\, n}_\ltb]= \mel{\mcirc_{\chi,1}}{{}^{\otimes n}N(n)^{L_A}}{S}^{\otimes n},
\label{eq:MPSrenyientropy}
\end{align}
where we defined 
\be
N(n)\equiv \sum_{\alpha=1}^{m_{\rtb}}  \frac{(N^{\alpha})^{\otimes n}}{d^{\lt n-1}_\alpha}. 
\ee
This expression can be evaluated efficiently for integer values of $n$ using a transfer matrix approach: the expression is determined by the leading eigenvector of $N(n)$. In this way, however, we are in general unable to provide an efficient analytic continuation for $n\rightarrow1$.

The entanglement entropy is split in configurational and number entropy contributions: defining 
\begin{align}
	p(\vec{\alpha})=\mel{\mcirc_{\chi,1}}{N^{\alpha_1}N^{\alpha_2}\ldots}{S},
\end{align}
then it is clear we can write
\begin{align}
&S^{\lt}\equiv	\lim_{n\rightarrow1} \frac{\tr[\rho^{GGE\,\, n}_\ltb]}{1-n}=\sum_{\vec{\alpha}} S_{\rm num}^\lt+ S_{\rm conf}^\lt\\
&	S_{\rm num}^\lt=\sum_{\vec{\alpha}}-p(\vec{\alpha})\log(p(\vec{\alpha}))\notag\\
&		S_{conf}^\lt=\sum_{\vec{\alpha}}-p(\vec{\alpha})\log(d^\lt_{\vec{\alpha}})\qquad d^\lt_{\vec{\alpha}}=\prod_{i=1}^{L_A} d^\lt_{\alpha_i}\notag.	
\end{align}
For a generic MPS the classical probability distribution $p(\vec{\alpha})$ is not the product of independent distribution at each site, making its Shannon entropy (which is the number entropy, in our language) hard to evaluate explicitly. In the case of bond dimension one, instead, it simplifies in the product of $L_A$ independent distributions with $p(\alpha)=c^\lt_\alpha$. The Shannon entropy becomes then additive, and, since we choose a translational invariant states, one has
\begin{align}
	S_{\rm num}=L_A\sum_\alpha c^\lt_\alpha\log(c^\lt_\alpha)\,.
\end{align}

We now prove that for a left charged solvable state one has 
\begin{align}
	S^{\lt}\ge L_A s^\rt 
	\label{eq:bound},
\end{align}
even when $p(\vec{\alpha})$ is not factorised.

We start by considering the following pure state\begin{align}
	\rho=	\fineq[-0.8ex][0.50001][1]{
		\trianglediag[0][0][3][bertiniorange][MPS][1][d]	
		\foreach \i in {1,...,6}
		{
			\MPSinitialstate[2*\i-3][-1][bertiniorange][topright][1]
		}\draw[thick,red,dashed] (3,-2)--++(0,6);
		\draw[very thick,dashed] (0,-1.05)--++(11,0);
		\draw[decorate,decoration={brace}] (3.6,4.4)--++(4.5,-4.5);
		\node[scale=2] at (7,2.5) {$x$};	
		\node[scale=2.5] at (0,4) {$B$};
	\node[scale=2.5] at (6,4) {$\bar B$}; },\label{eq:examplestate}\end{align}where the bottom line is built with our left-solvable MPS and is understood to be infinite in the right direction.
We consider its entanglement across the bipartition $B\bar{B}$ shown with a red dashed line. 
Tracing out $B$, and simplifying the transfer matrices with the leading eigenvector, we find\begin{align}
\rho_{\bar{B}}=	\fineq[-0.8ex][0.50001][1]{
		\trianglediag[0][0][3][bertiniorange][MPS][1][d]	
		\foreach \i in {1,...,6}
		{
			\MPSinitialstate[2*\i-3][-1][bertiniorange][topright][1]
		}\draw[thick,red,dashed] (3,-2)--++(0,6);
		\draw[very thick,dashed] (0,-1.05)--++(11,0);
		\draw[decorate,decoration={brace}] (3.6,4.4)--++(4.5,-4.5);
		\node[scale=2] at (7,2.5) {$x$};	
		\cstate[-2][-1.05]
	\foreach \i in {0,...,4}
	{\cstate[\i-1.5][\i-.5]}}
\end{align}
and then, combining solvability, charge conservation and dual unitarity, we can simplify it as 
\begin{align}
\fineq[-0.8ex][0.50001][1]{
	\foreach \i in {6}
	{
		\MPSinitialstate[2*\i-3][-1][bertiniorange][topright][1]
	}	\cstate[8][-1.05]
	\foreach \i in {0,...,4}
	{\draw[thick] (8-\i-1.5,\i-.5)--++(.5,.5);
		\cstate[8-\i-1.5][\i-.5]
		\charge[8-\i-1.5+.3][\i-.5+.3][red]
	}		
	\draw[decorate,decoration={brace}] (3.4,4.6)--++(4.5,-4.5);
	\node[scale=2] at (7,2.5) {$x$};	
},
\end{align} 
so that, apart from a $o(1)$ contribution from the initial MPS, its R\'enyi entropies are given by  
\begin{align}
	S^{(n)}[\rho_B]=x s^{{\rt}}_n.
\end{align}
Let us now trace Eq.~\eqref{eq:examplestate} from the side of $\bar{B}$. We want to apply on each of the $x$ open leg on the triangle the following quantum channel
 \begin{align}
	\mathcal{E}^\lt[X]= \sum_{\beta=1}^{m_\ltb} \frac{\Pi^\lt_\beta}{d^\lt_\beta} \tra{X \Pi^\lt_\beta},\label{eq:quantumchannelprojectors}
\end{align}
which is unital (preserves the identity) and can be written in Kraus form (it is the same channel considered in the previous subsection, i.e.,  it corresponds to the transfer matrix in Eq.~\eqref{eq:transfermatrixl}). 

On the left open leg of the auxiliary MPS  space we instead apply the unital quantum channel
\begin{align}
	\mathcal{E}^\chi[X]=\frac{\mathds{1}_\chi}{\chi} \tra{X}.
\end{align}
After the application of these channels, the reduced density matrix can be represented as
\begin{widetext}
\begin{align}
	\left({\mathcal{E}^\lt}^{\otimes x} \otimes \mathcal{E}^\chi\right)[\rho_B]
	=\sum_{\vec{\beta}}\frac{1}{d^\lt_{\vec{\beta}}}	\fineq[-0.8ex][.8][1]{
		\trianglediag[0][0][3][bertiniorange][MPS][1][d]	\draw[thick,red,dashed] (3,-2)--++(0,6);
		\foreach \i in {0,...,4}
		{	
			\node at (\i-2.1,\i-.7) {$\Pi^\lt_{\pgfmathparse{\i+1}\beta_{\pgfmathprintnumber[precision=1]{\pgfmathresult}}}$};
			\draw[thick,->] (\i-2.3,\i-.5)--++(0,.7);
			\draw[thick,->] (\i-1.95,\i-.57)--++(.55,-.1);
			\cstate[\i-1.6][\i-.4]
			\cstate[\i+3.6][3.6-\i]
			\charge[\i-1.3][\i-.7][blue]
			\draw[thick](\i-2,\i)--++(-.5,.5);
			\charge[\i-2.3][\i+.3][blue]
			\cstate[\i-1.6-.4][\i]
		}
		\cstate[-2.][-1.05]
		\cstate[8][-1.05][][black]
	},\label{eq:disegno2}
\end{align}\end{widetext} 
where we substituted the leading eigenvector of the MPS transfer matrix on the right, as per Eq.~\eqref{eq:leadingMPSTmat}, and the sum is over all strings of left charges $\vec{\beta}$ of length $x$. Moving all the charges to the base of the triangle, it is immediate to see that we recover $\rho^{GGE}_\ltb$, defined on the $x$ left-sites. Since a unital quantum channel can only increase the R\'enyi entropies~\cite{FrankLieb2013}, this shows that the R\'enyi entropies of \eqref{eq:disegno2}, and thus those of $\rho^{GGE}_\ltb$, are always larger than $S^{(n)}[\rho_B]=S^{(n)}[\rho_{\bar B}]=x s^\rt_n$. This proves Eqs.~\eqref{eq:bound} and \eqref{eq:bound2} in the main text.

\subsection{Bound on the rate of growth of entanglement in the second phase }
\label{app:boundsecondphase}

To prove the bound in Eq.~\eqref{eq:slopesecondphase} we begin by observing that 
\be
\tr[{\rho}_A(t)^n] = \tr[\widetilde{\rho}_A(t)^n],
\label{eq:identity}
\ee
where we noted that ${\rho}_A(t)$ in Eq.~\eqref{eq:rdmls} can be written as 
\begin{align}
	\rho_A(t)	=\!\!\!\!\!\!\fineq[-0.8ex][0.55][1]{
		\MPSinitialstate[1][-1][bertiniorange][topright][1]
		\roundgate[2][0][1][topright][bertiniorange][1]
		\MPSinitialstate[3][-1][bertiniorange][topright][1]
		\MPSinitialstate[5][-1][bertiniorange][topright][1]
		\cstate[5.5][-.5]	\cstate[4.5][-.5]
		\draw[very thick, dashed] (5,-1.05)	--++(2,0);	
		\foreach\i in {0,...,4}
		{
			\roundgate[-\i][\i][1][topright][bertiniorange][1]
			\roundgate[-\i+1][\i+1][1][topright][bertiniorange][1]
		}
		\roundgate[-5][5][1][topright][bertiniorange][1]
		\roundgate[-4][6][1][topright][bertiniorange][1]
		\roundgate[-6][6][1][topright][bertiniorange][1]
		\cstate[0][-1.05][][white]
		\draw[dashed, color=red, rounded corners=5pt](-4.5,2.5)--++(3,3)--++(-1,1)--++(-3,-3)--cycle;
		\foreach \i in {0,...,6}
		{
			\cstate[-\i+3.5][\i-.5]
			{
				\cstate[-\i-.55][\i-.55]}{}
			\charge[-\i-.3][\i-.3][red]
		}
		\draw [decorate, decoration = {brace,mirror}]   (2.2,1.8)--++(-4.5,4.5) ;
		\node[scale=1.5] at (1.5,4.5){$2t-L_A$};
		\node[scale=1.5, red] at (-5,2.5){$\mathcal{T}$};
	}\hspace{-2cm},\label{eq:rdmlswithoutMPSTM}
\end{align}
and we defined
\begin{align}
	&\widetilde{\rho}_A(t)=\!\!\!\!\!\!\!\!\!\!\!	\fineq[-0.8ex][0.50001][1]{
	\foreach \i in {0,1,-1}
{		
	\MPSinitialstate[3-2*\i][-1][bertiniorange][topright][1]
}	
	\draw[very thick, dashed] (6.75,-1.05)--++(-6,0);	\roundgate[2][0][1][topright][bertiniorange][1]
		\foreach\i in {0,...,4}
		{
			\roundgate[-\i][\i][1][topright][bertiniorange][1]
			\roundgate[-\i+1][\i+1][1][topright][bertiniorange][1]
		}
		\roundgate[-5][5][1][topright][bertiniorange][1]
		\roundgate[-4][6][1][topright][bertiniorange][1]
		\roundgate[-6][6][1][topright][bertiniorange][1]
		\foreach \i in {0,...,6}
		{
			\charge[-\i-.3][\i-.3][red]
		}
		\draw [decorate, decoration = {brace,mirror}]   (2.2,1.8)--++(-4.5,4.5) ;
		\node[scale=2] at (3,3.5){$2t-L_A$};
		\cstate[-6.5][6.5]
		\cstate[-5.5][6.5]
		\cstate[-4.5][6.5]
		\cstate[-3.5][6.5]
	}\hspace{-1cm}. \label{eq:rhotilde}
\end{align}
The identity follows from the fact that the above diagrams with all legs open corresponds to a pure state and the white bullets correspond to a trace. Therefore, Eq.~\eqref{eq:identity} is just stating that two reduced density matrices obtained by tracing out two complementary subsystems of a pure state have the same spectrum. 

Next, we apply the quantum channel in Eq.~\eqref{eq:quantumchannelprojectors}, but defined with right-moving solitons, on the open legs on the right of the matrix in Eq.~\eqref{eq:rhotilde}, i.e., 
\begin{align}
	&\left({\mathcal{E}^\rt}\right)^{\otimes 2t-L_A}[\tilde{\rho}_A]= \notag\\
	&\sum_{\vec{\alpha}} \frac{1}{d^{{\rt}}_{\vec{\alpha}}}	\fineq[-0.8ex][0.45][1]{
		\foreach \i in {0,1,-1}
		{		
			\MPSinitialstate[3-2*\i][-1][bertiniorange][topright][1]
		}	\draw[very thick, dashed] (6.75,-1.05)--++(-6,0);	
			\MPSinitialstate[1][-1][bertiniorange][topright][1]
			\roundgate[2][0][1][topright][bertiniorange][1]
			\MPSinitialstate[3][-1][bertiniorange][topright][1]
			\foreach\i in {0,...,4}
			{
				\roundgate[-\i][\i][1][topright][bertiniorange][1]
				\roundgate[-\i+1][\i+1][1][topright][bertiniorange][1]
			}
			\roundgate[-5][5][1][topright][bertiniorange][1]
			\roundgate[-4][6][1][topright][bertiniorange][1]
			\roundgate[-6][6][1][topright][bertiniorange][1]
				\foreach \i in {0,...,6}
			{
				\charge[-\i-.3][\i-.3][red]
				\ifnumgreater{\i}{1}{
				\draw[thick] (-\i+4,\i)--++(.5,.5);
				\cstate[-\i+4][\i]
				\charge[-\i+4.3][\i+.3][red]
				\charge[-\i+3.35][\i-.65][red]
				\cstate[-\i+3.6][\i-.4]
		}{}
			}
			\draw[thick,<-](-1.7,6.5)--(-3,7);
			\draw[thick,<-](-2.7,5.55)--(-3,6.8);
			\node[scale=2] at (-3,7.5){$\Pi^{{\rt}}_{\alpha_1}$};
			\cstate[-6.5][6.5]
			\cstate[-5.5][6.5]
			\cstate[-4.5][6.5]
			\cstate[-3.5][6.5]
			\draw[thick,<-] (-6.3,5.6)--++(0,-4);
			\node[scale=2] at (-6.3,1.1) {$\sum_\beta \Pi^{{\rt}}_\beta \frac{c^{{\rt}}_\beta}{d^{{\rt}}_\beta}$};
			\draw[very thick] (-.5,-1.05)--++(-.5,0);
			\cstate[0][-1.05]
			\cstate[-.5][-1.05]
		}\hspace{-1cm},\label{eq:rmdlstilde1}
\end{align}
where we set
\be		
d^\rt_{\vec{\alpha}}\equiv \prod_i	d^\rt_{{\alpha}_i}
\ee
and the sum is over strings of $2t-L_A$ projectors on $\alpha_i$ (the charges on right side). In the above equation we also reported the explicit form of the solitons on the leftmost diagonal (which are a linear combination of projectors instead).

The data processing inequality generalised for R\'enyi entropies, see Ref.~\cite{FrankLieb2013}, implies that a unital quantum channel can only increase the entropies, so the entropies of \eqref{eq:rmdlstilde1} will provide an upper bound on the desired ones. We can simplify the diagram using unitarity, dual unitarity and left charged solvability of the initial state to obtain
\begin{align}
	\sum_{\vec{\alpha}} \frac{1}{d^{{\rt}}_{\vec{\alpha}}}	\fineq[-0.8ex][0.50001][1]{
		\foreach \i in {0,...,6}
		{
				\draw[thick] (-\i+4,\i)--++(.5,.5);
				\cstate[-\i+4][\i]
				\charge[-\i+4.3][\i+.3][red]
			\ifnumgreater{\i}{1}{
				\draw[thick] (-\i+3.6,\i-.4)--++(-.5,-.5);
				\charge[-\i+3.35][\i-.65][red]
				\cstate[-\i+3.6][\i-.4]
			}{}
		}
	\foreach \i in {7,8}
	{
			\draw[thick] (-\i+3.6,\i-.4)--++(-.5,-.5);
			\charge[-\i+3.35][\i-.65][red]
			\cstate[-\i+3.6][\i-.4]
	}
	\draw[dashed,color=red] (-3.6,4.8)--++(2.3,2.3)--++(4.8,-4.8)--++(-2.3,-2.3)--cycle;
		\draw[thick,<-](-1.5,6.3)--(1,7);
		\draw[thick,<-](-2.65,5.1)--(-2.7,3.);
		\node[scale=2] at (1.6,7.2)
		{$\Pi^{{\rt}}_{\alpha_1}$};
		\node[scale=2] at (-2.7,2.){$\frac{c^{{\rt}}_{\alpha_1}}{d^{{\rt}}_{\alpha_1}}\Pi^{{\rt}}_{\alpha_1}$};
			\draw[dashed,color=blue] (1.3,-0.1)--++(2.3,2.3)--++(2,-2)--++(-2.3,-2.3)--cycle;
			\draw[dashed,color=blue] (1.3-7,-0.1+7)--++(2.3,2.3)--++(2,-2)--++(-2.3,-2.3)--cycle;
			\draw[<-,thick] (3.35+1,.65)--++(0,4);
			\node[scale=2] at (4.35,5) {$\sum_\beta \Pi^{{\rt}}_\beta \frac{c^{{\rt}}_\beta}{d^{{\rt}}_\beta}$};
		\draw[very thick,dashed](-4,-2.05)--++(7,0);
			\foreach \i in {0,1,-1}
			{		
				\MPSinitialstate[-2*\i-1][-2][bertiniorange][topright][1]
				\cstate[-4][-2.05]
			}
	}.\label{eq:qchannelrhotilde}
\end{align}

From \eqref{eq:qchannelrhotilde} we can exactly compute the entanglement entropy, finding (we report only the expression for the Von Neumann entropy and neglect the initial entanglement of the state)
\begin{align}
&	(2t-L_A) \sum_\alpha \left(2c^{{\rt}}_\alpha \log d_\alpha^{{\rt}} -c^{{\rt}}_\alpha \log(c_\alpha)\right)+\\
&	L_A\sum_\alpha \left(c^{{\rt}}_\alpha \log d_\alpha^{{\rt}}-c^{{\rt}}_\alpha \log(c_\alpha)\right)= S[\mathcal{E}[\tilde{\rho}_A]]\ge S[\rho_A], \notag
\end{align}
which proves Eq. \eqref{eq:slopesecondphase}.

\section{Simplification of Eq. \eqref{eq:expression}}
\label{app:inequality}

In the limit $n\rightarrow1$, the two quantities in Eq. \eqref{eq:expression} can be written as (taking the log and dividing the result by ${1}/{(1-n)}$)
\begin{align}
&	s^{{\rt}}_{\rm num}+2s^{{\rt}}_{\rm conf},\notag\\
&	\sum_{\alpha,\beta} -c_{\alpha,\beta}\log(c_{\alpha,\beta})+s^{{\lt}}_{\rm conf}+s^{{\rt}}_{\rm conf}.
\end{align}
We use a chain of inequalities based on the fact that, by assumption,
\begin{align}
s^{{\rt}}=s^{{\rt}}_{\rm num}+s^{{\rt}}_{\rm conf}<s^{{\lt}}=s^{{\lt}}_{\rm num}+s^{{\lt}}_{\rm conf}.
\end{align}
The inequalities are as follows
\begin{align}
	&s^{{\rt}}_{\rm num}+2s^{{\rt}}_{\rm conf}<s^{{\lt}}_{\rm num}+s^{{\lt}}_{\rm conf}+s^{{\rt}}_{\rm conf}<\notag \\
	&<s^{{\lt}}_{\rm conf}+s^{{\rt}}_{\rm conf}-\sum_{\alpha,\beta} c_{\alpha,\beta}\log(c_{\alpha,\beta})\label{eq:chaininequalities}.
\end{align}
Here in the last step, we used the fact that, since 
\begin{align}
	c^{{\rt}}_{\alpha}=\sum_{\beta} c_{\alpha,\beta} \implies c^{{\rt}}_{\alpha}\succ c_{\alpha,\beta},
\end{align}
the (classical) distribution of $c^{{\rt}}_\alpha$ majorises the one $c_{\alpha,\beta}$, meaning that its corresponding entropy has to be strictly less, i.e. 
\begin{align}
	\sum_{\alpha,\beta} -c_{\alpha,\beta}\log(c_{\alpha,\beta})\ge 	s^\lt_{\rm conf}.
\end{align}
The chain in Eq.~\eqref{eq:chaininequalities} proves the desired inequality.

\section{Numerical methods}
\label{app:numericsdetails}

To produce the plots in Figs.~\ref{plot:entanglementgrowth} and \ref{fig:entgrowthnumerics} we contracted exactly the relevant tensor networks for different values of $t$. In particular, for Fig.~\ref{plot:entanglementgrowth}, we constructed the matrix
\begin{align}
		\fineq[-0.8ex][0.750001][1]{
			\trianglediag[0][0][3][bertiniorange][pairproduct][1][d]
			\foreach \i in {0,...,3}
			{
				\cstate[\i-.5][\i+.5]
					\draw [decorate, decoration = {brace}]   (4,4)--++(3.5,-3.5) ;
			}
			\cstate[-1.5][-.5]
			\node[scale=1.5] at (6.5,3) {$2t$ legs};
		},
\end{align} 
whose eigenvalues (for a bipartition across the dashed line) correspond to the entanglement spectrum of the system partitioned in two semi-infinite lines (i.e.\ the entanglement growth at a single edge of a subsystem). The initial pair product state is chosen by assigning randomly its vector elements and normalising the result. To produce, the data in Fig.~\ref{fig:entgrowthnumerics} we instead built the reduced density matrix in Eq.~\eqref{eq:rdmls}. The eigenvalue decomposition and the tensor contractions were carried out using the C++ ITensor library~\cite{itensor}.

\bibliography{bibliography.bib}

\end{document}